\documentclass[12pt]{article}

\usepackage{epsfig,epsf}
\usepackage{graphics}
\usepackage{cite}
\usepackage{geometry}
\usepackage{xcolor}
\usepackage{rotating}
\usepackage{url}
\usepackage{microtype}
\usepackage{listings}
\usepackage[breaklinks=false]{hyperref}

\usepackage{amsmath}
\usepackage{amsthm}
\usepackage{amsfonts}
\usepackage{amssymb}

\usepackage{slashed}
\usepackage{braket}

\newcommand{\Tprod}[1]{{\mathrm T}\lbrack #1 \rbrack}

\newcounter{MBQ}

\definecolor{darkgreen}{rgb}{0,0.4,0}

\numberwithin{equation}{section}

\begin{document}

\allowdisplaybreaks
\thispagestyle{empty}

\begin{flushright}
{\small
TUM-HEP-1113/17\\
IPPP/17/87\\
FTUAM-17-27\\
IFT-UAM/CSIC-17-111\\[0.1cm]
%arXiv:1711.10429 [hep-ph] \\[0.1cm]
\today
}
\end{flushright}

\vskip1.2cm
\begin{center}
\textbf{\Large\boldmath
Non-resonant and electroweak NNLO correction\\[0.15cm]
to the $e^+ e^-$ top anti-top threshold}
\\
\vspace{1.2cm}
{\sc M.~Beneke}$^a$, {\sc A.~Maier}$^b$, {\sc T.~Rauh}$^{b}$
and {\sc P.~Ruiz-Femen\'ia}$^{c}$
\\[0.5cm]
\vspace*{0.1cm} $^a$\,{\it
Physik Department T31,\\
James-Franck-Stra\ss{}e~1,
Technische Universit\"at M\"unchen,\\
85748 Garching, Germany}\\[0.3cm]
$^b$\,{\it
IPPP, Department of Physics,
University of Durham,\\
DH1 3LE, United Kingdom}\\[0.3cm]
$^c$\,{\it
Departamento de F\'isica Te\'orica and Instituto de F\'isica Te\'orica UAM-CSIC,
Universidad Aut\'onoma de Madrid, E-28049 Madrid, Spain}

\def\thefootnote{\arabic{footnote}}
\setcounter{footnote}{0}

\vskip1.8cm
\textbf{Abstract}\\
\vspace{1\baselineskip}
\parbox{0.9\textwidth}{
We determine the NNLO electroweak correction to the
$e^+ e^-\to b\bar{b}W^+W^- X$ production cross section near
the top-pair production threshold. The calculation includes
non-resonant production of the final state as well as electroweak
effects in resonant top anti-top pair production with
non-relativistic resummation, and elevates the theoretical prediction
to NNNLO QCD plus NNLO electroweak accuracy. We then
study the impact of the new contributions on the top-pair threshold scan
at a future lepton collider.
}

\end{center}

%{\small \tableofcontents}

\newpage
\setcounter{page}{1}

%%%%%%%%%%%%%%%%%%%%%%%%%%%%%%%%%%%%%%%%%%%%%%%%%%%%%%%%%%%%%%%%%%%%%%%%%
%%%%%%%%%%%%%%%%%%%%%%%%%%     Introduction      %%%%%%%%%%%%%%%%%%%%%%%%
%%%%%%%%%%%%%%%%%%%%%%%%%%%%%%%%%%%%%%%%%%%%%%%%%%%%%%%%%%%%%%%%%%%%%%%%%

\section{Introduction\label{sec:intro}}

The precision study of the top pair production
threshold is among the main motivations for the construction of
a high-energy $e^+e^-$ collider~\cite{Fujii:2015jha}.
About 100~fb$^{-1}$ of integrated luminosity spread over ten
center-of-mass energies
distributed around $\sqrt{s}\approx 345\mbox{ GeV}$ can provide a
measurement of the top-quark $\overline{\text{MS}}$ mass with
an experimental uncertainty of about $50\mbox{ MeV}$~\cite{
Seidel:2013sqa,Horiguchi:2013wra,Simon:2016pwp}.
This must be compared to the ultimate precision possible at the LHC,
which is constrained to $\mathcal{O}(1~\mbox{GeV})$ due to the limited
understanding of the relation between the $\overline{\text{MS}}$ mass
and the mass parameter in the calculation and simulation of the final
state from which the top mass is directly reconstructed. There has
been some progress in the quantification of this relation when the mass
is reconstructed from two-jettiness in $e^+ e^-$
collisions in the boosted top regime~\cite{Butenschoen:2016lpz},
but the extension of this approach to hadron collider processes requires
the consideration of additional effects \cite{Andreassen:2017ugs}.
In addition, the top width, the strong coupling constant and the top Yukawa
coupling can be extracted  from the threshold scan to varying degree
of accuracy.

The threshold region is defined as the kinematic regime where the top
quarks have a small three-velocity $v=(\sqrt{s}/m_t-2)^{1/2}$ of the
order of the strong coupling constant $\alpha_s$. Thus, the top quarks
are non-relativistic and are subject to the colour Coulomb
interaction, that would facilitate the formation of toponium bound
states if the top quarks were stable. The sizeable top decay width
caused by the electroweak interaction also prevents hadronization.
Therefore, the top threshold dynamics is governed by the colour Coulomb
interaction, which must be treated non-perturbatively, while the
strong coupling $\alpha_s \ll 1$ is still small. This interplay between
the strong Coulomb attraction and the large top decay width has first been
realized in~\cite{Bigi:1986jk,Fadin:1987wz}.

A significant effort has since been invested into providing
high-precision predictions for top pair production near threshold.
The major focus has naturally been the strong interaction effects,
which have now been computed to next-to-next-to-next-to-leading
order (NNNLO) accuracy~\cite{Beneke:2015kwa} in an expansion where
$\alpha_s\ll1$ and $v \ll 1$, but $\alpha_s/v = {\cal O}(1)$. The effective
field theory formalism and ingredients that underlie this calculation
are summarized in \cite{Beneke:2013jia}, to which we refer for
more details on the QCD aspects of the calculation.\footnote{
See~\cite{Marquard:2014pea,Luke:1997ys,Beneke:2005hg,
Beneke:2013PartII,Beneke:2008cr,Wuester:2003,Kniehl:2001ju,Beneke:2014qea,
Anzai:2009tm,Smirnov:2009fh,Lee:2016cgz,Beneke:2013kia} for the
computation of specific NNNLO ingredients.}
The NNNLO QCD result has finally
settled the issue of the poor convergence of the perturbative
expansion up to NNLO~\cite{Hoang:2000yr}. The NNNLO corrections are
well behaved and the remaining scale uncertainty of the QCD result is at
the level of $\pm3\%$. Similarly, it has been observed that the RG-improved
prediction at the (almost) next-to-next-to-leading logarithmic
order~\cite{Hoang:2013uda} stabilizes the scale uncertainty at the
level of~$\pm5\%$.

In the present work we are concerned with electroweak effects and
non-resonant production of the
observable final state $b\bar b W^+ W^- +X$ of the decayed top anti-top
pair. An analysis of various electroweak
effects~\cite{Beneke:2015lwa} has demonstrated that they
are as large as $10\%$. Thus, the full NNLO non-resonant and
electroweak contributions must be included to salvage the precision of the
prediction. Even more importantly, as will be discussed below,
they are required to obtain a well-defined result,
since the pure QCD cross section by itself contains
divergences proportional to the top-quark decay
width~\cite{Beneke:2008cr}, which are cancelled only once
the non-resonant production is included~\cite{Beneke:2010mp,Jantzen:2013gpa}.

The main result of this work is the NNLO calculation of all
electroweak and non-resonant effects. We also provide an implementation
of initial-state radiation in a scheme consistent with Coulomb
resummation and the inclusion of ${\cal O}(\alpha)$ electromagnetic
corrections, following a similar treatment as for the $W^+ W^-$
threshold \cite{Beneke:2007zg,Actis:2008rb}. To define the
precise meaning of ``NNLO'' for electroweak effects, we note that
they introduce the electromagnetic ($\alpha_{\rm em}$), SU(2) electroweak
($\alpha_{\rm EW}$) and top-quark Yukawa ($\lambda_t$) coupling.
For the purpose of power counting we do not distinguish between
$\alpha_{\rm em}$ and $\alpha_{\rm EW}$ and count
\begin{equation}
\alpha_{\rm EW} \sim \alpha_t\equiv
\frac{\lambda_t^2}{4\pi} \sim \alpha_s^2 \sim v^2,
\label{eq:parametercounting}
\end{equation}
that is, an electroweak coupling counts as two powers of the strong
coupling, which is consistent with counting $\Gamma_t\sim m_t\alpha_{\rm EW}
\sim m_t v^2$, which is always adopted in the pure QCD calculation.
The pure QCD calculation up to NNNLO then accounts for all terms in
the total cross section $\sigma$ of the form
\begin{equation}
  \label{eq:PNRQCD_power_counting}
  \sigma_{\rm QCD\,only} \sim \alpha_{\rm EW}^2 v \sum_{k=0}^\infty
  \bigg(\frac{\alpha_s}{v}\bigg)^{\!k}\times
  \begin{cases}
    1 & \text{LO}\\
    \alpha_s,v & \text{NLO}\\
    \alpha_s^2,\alpha_s v, v^2 & \text{NNLO}\\
    \alpha_s^3,\alpha_s^2 v,\alpha_s v^2, v^3 & \text{NNNLO}
  \end{cases}\,,
\end{equation}
where the global factor $\alpha_{\rm EW}^2 v$ accounts for the phase-space
suppression of the cross section near the threshold and the electroweak
production in $e^+ e^-$ collisions. The electromagnetic,
electroweak, Yukawa and non-resonant terms are of the parametric form
\begin{eqnarray}
  \sigma &\sim& \alpha_{\rm EW}^2 v \sum_{k=0}^\infty
  \bigg(\frac{\alpha_s}{v}\bigg)^{\!k}\times
  \begin{cases}
    {\displaystyle \frac{\alpha_{\rm em}}{v}} & \hskip-1.5cm \text{NLO}
    \\[0.3cm]
    {\displaystyle \left(\frac{\alpha_{\rm em}}{v}\right)^{\!2}},
    {\displaystyle \frac{\alpha_{\rm em}}{v}}
    \times \{\alpha_s, v\}, \alpha_{\rm EW}, \sqrt{\alpha_{\rm EW}\alpha_t},
    \alpha_t
    & \hskip-1.5cm \text{NNLO}
    \\[0.3cm]
    {\displaystyle \left(\frac{\alpha_{\rm em}}{v}\right)^{\!3}},
    {\displaystyle \left(\frac{\alpha_{\rm em}}{v}\right)^{\!2}}
    \times \{\alpha_s, v\},
    {\displaystyle \frac{\alpha_{\rm em}}{v}}
    \times \{\alpha_s^2, \alpha_s v,v^2,\sqrt{\alpha_{\rm EW}\alpha_t}\},
    & %\text{NNNLO}
    \\[0.2cm]
    \hskip0.1cm
    \alpha_t\times \{{\displaystyle \frac{\alpha_{\rm em}}{v}},\alpha_s, v\},
    \ldots  & \hskip-1.5cm \text{NNNLO}
  \end{cases}
\nonumber\\[0.1cm]
&&+\,\alpha_{\rm EW}^2\times
  \begin{cases}
    \alpha_{\rm EW} & \text{NLO}\\
    \alpha_{\rm EW} \alpha_s & \text{NNLO}\\
    \ldots & \text{NNNLO}
  \end{cases}\,,
  \label{eq:EW_power_counting}
\end{eqnarray}
where the first line refers to resonant and the second to non-resonant
production. We note the absence of phase-space suppression and Coulomb
resummation for the non-resonant part.
The non-resonant contribution is known at NLO~\cite{Beneke:2010mp},
but only partial results are available at NNLO~\cite{Jantzen:2013gpa,
Ruiz-Femenia:2014ava,Penin:2011gg}. On the resonant side, the
$(\alpha_{\rm em}/v)^k$ terms arise from the QED Coulomb potential.\footnote{
We do not distinguish $\alpha_{\rm em}$ and $\alpha_{\rm EW}$ in the
other terms.}
These as well as all Yukawa coupling effects have already been
included up to NNNLO in~\cite{Beneke:2015lwa}. This result together with
the NLO non-resonant and the NNNLO QCD calculation has been
made available in the
\texttt{QQbar\_threshold} code~\cite{Beneke:2016kkb}.
The NNLO non-resonant and the remaining NNLO electroweak
contributions are computed in this work, thus elevating the precision
at the top-pair threshold to complete NNNLO QCD+Yukawa and NNLO
EW+non-resonant. The ellipses in
(\ref{eq:EW_power_counting}) denote third-order electroweak and
non-resonant terms that remain unknown.

The outline of the paper is as follows.
In Section~\ref{sec:setup} we describe how the calculation is split into
resonant and non-resonant contributions, such that no double-counting
occurs and the divergences are cancelled consistently. We also discuss
the implementation of an invariant mass cut.
For the practical calculation we split the total cross section into three
separately finite parts, which are computed, each within its own
computational scheme, in Sections~\ref{sec:PartI}, \ref{sec:PartII}
and \ref{sec:PartIII}, respectively. Section~\ref{sec:checks_and_imp}
describes a consistency check we performed for our results and the
comparison with some previous results.
In Section~\ref{sec:results} we analyze the importance of the
various contributions for the threshold scan including initial-state
radiation. We conclude in Section~\ref{sec:conclusions}. Several
appendices collect technical results, in particular the
implementation of the new results into the \texttt{QQbar\_threshold} code.

%%%%%%%%%%%%%%%%%%%%%%%%%%%%%%%%%%%%%%%%%%%%%%%%%%%%%%%%%%%%%%%%%%%%%%%%%%%
%%%%%%%%%%%%%%%%%%%%%%%%%%%%    Setup      %%%%%%%%%%%%%%%%%%%%%%%%%%%%%%%%
%%%%%%%%%%%%%%%%%%%%%%%%%%%%%%%%%%%%%%%%%%%%%%%%%%%%%%%%%%%%%%%%%%%%%%%%%%%

\section{Setup of the computation\label{sec:setup}}

\subsection{Resonant and non-resonant separation in unstable
particle EFT}
\label{sec:resnonres}

Precision calculations of top pair production near threshold are most
conveniently done in potential non-relativistic effective field
theory (PNREFT)~\cite{Pineda:1997bj,Beneke:1999qg},
which describes the dynamics of slowly moving particles with three-momentum
$m_t v$ coupled to ultrasoft radiation/massless particles with
energy $m_t v^2$ after hard and soft effects have been integrated
out. The computation contains uncancelled divergences proportional
to the top-quark width, which start at NNLO in dimensional
regularization.

The top-pair production cross section is thus an ill-defined
quantity. Instead one must consider the final state of the
decay products $b\bar{b}W^+W^-+X$. The narrow-width approximation is
not applicable since the top width is not small compared to the
top kinetic energy $E=\sqrt{s}-2 m_t\sim m_t v^2$.\footnote{We assume
$|V_{tb}|^2=1$. Despite the $W$-boson lifetime being of similar
size as the top lifetime, the $W$ decay width can be dropped
(expanded out) in the propagators, because the $W$ bosons are
always hard. Thus, it is justified to treat the $W$ bosons as
stable particles.}
The above final state can also be produced non-resonantly,
i.e. without an intermediate non-relativistic top pair. The resonant and
non-resonant production mechanisms cannot be distinguished physically
and must be summed. Only the sum is well-defined and finite-width
divergences must cancel~\cite{Beneke:2008cr}. This cancellation has already
been demonstrated up to NNLO~\cite{Jantzen:2013gpa}, and will be
reproduced in the computation of the full NNLO correction in this
paper.

To account for the non-resonant production mechanism, one must embed
the effective theory framework for the QCD result~\cite{Beneke:2013jia}
into Unstable Particle Effective Theory~\cite{Beneke:2003xh,Beneke:2004km}.
The complete NNLO cross section can be written as the sum
of a resonant and a non-resonant contribution
\begin{equation}
 \sigma^\text{NNLO}(s)=\sigma_\text{res}^\text{NNLO}(s)
 +\sigma_\text{non-res}^\text{NNLO}(s).
\label{eq:master}
\end{equation}
The resonant contribution has the form
\begin{equation}
 \sigma_\text{res}^\text{NNLO}(s)\sim
\text{Im}\left[\sum_{k,l} C^{(k)} C^{(l)} \int d^4 x \,
\braket{e^- e^+ |\Tprod{i {\cal O}^{(k)\dagger}(0)\,i{\cal O}^{(l)}(x)}
|e^- e^+}\right].
\label{eq:resstructure}
\end{equation}
It is understood that the imaginary part refers only to discontinuities of
the forward amplitude that correspond to a $b\bar b W^+ W^- X$ final
state.\footnote{This includes cutting nearly on-shell top lines
in the effective theory, since the effective top propagator contains
the top width and the top is assumed to decay exclusively into
$bW^+ X$.}
The production operators ${\cal O}^{(l)}$ annihilate the incoming
$e^+e^-$ states and produce a nearly on-shell top and anti-top quark
with small relative velocity. The matrix element is evaluated within
PNREFT, appropriately generalized from QCD to account for
electroweak effects and top decay. In addition one must consider the
interactions of the energetic initial-state electrons.
The $C^{(l)}$ are the hard matching coefficients of the production operators.
They also receive electroweak corrections and furthermore acquire an
imaginary part from diagrams involving cuts corresponding to $\bar{t}bW^+$
and $t\bar{b}W^-$ final states. The imaginary part arises, for example,
from the interference of the process $e^+ e^- \to W W^*$, where the
off-shell $W$ decays to $\bar t b$ with the process $e^+ e^-\to t\bar t$,
where the on-shell $t$ decays to $W b$. In unstable particle theory
this contribution appears in the resonant term, since the separation
into resonant and non-resonant is done strictly on the basis of the
virtuality of the top propagators, which in this example is small
for both $t$ and $\bar t$.

The non-resonant part takes the form
\begin{equation}
 \sigma_\text{non-res}^\text{NNLO}(s)\sim \sum_{k} \,
\text{Im}\left[C_{4 e}^{(k)}\right]
\braket{e^- e^+|i {\cal O}_{4e}^{(k)}(0)|e^- e^+}.
\label{eq:nonresstructure}
\end{equation}
It originates from cuts over hard propagators that correspond to
the physical final state $b\bar{b}W^+W^-X$. Hard cuts over
the $t\bar{t}$ final state are not possible kinematically near threshold.
Thus, the leading corrections are from $\bar{t}bW^+$ and
$t\bar{b}W^-$ cuts and are of the order $\alpha_{\rm EW}^3$, which
constitutes a NLO contribution to the cross section
$\sigma^\text{LO}\sim \alpha_{\rm EW}^2v$. The non-resonant
term arises from expanding the full-theory diagrams in $E$.
Since both $E$ and $\alpha_{\rm EW}$ count as two orders in the
expansion, the NNLO contribution is given by the QCD $\mathcal{O}(\alpha_s)$
corrections to the process $e^+ e^-\to \bar{t}bW^+ + t\bar{b}W^-$,
computed directly at the threshold $\sqrt{s}=2 m_t$,
while actual $b\bar{b}W^+W^-$
cuts as well as electroweak and $E/m_t$ corrections are of the order
$\alpha_{\rm EW}^4$ and only contribute at NNNLO. The construction
implies that the poles of internal top propagators in the non-resonant
contribution are not regulated by a finite-width prescription, since
any width terms would have to be expanded out. This leads to
singularities at phase-space boundaries $(p_b+p_{W^+})^2\rightarrow m_t^2$,
which must be regulated
dimensionally. The $1/\epsilon$ poles cancel exactly the finite-width
divergences that appear in the resonant contribution. The computation
of the  QCD correction to the process $e^+ e^-\to \bar{t}bW^+ + t\bar{b}W^-$
with this specific prescription, required for consistency with the
resonant PNREFT calculation in dimensional regularization, is the
major result of the present work.

%%%%%%%%%%%%%%%%%%%%%%%%%%%%%%%%%%%%%%%%%%%%%%%%%%%%%%%%%%%%%%%%%%%%%%%
%%%%%%%%%%%%%%%%%%%%%     Organization      %%%%%%%%%%%%%%%%%%%%%%%%%%%
%%%%%%%%%%%%%%%%%%%%%%%%%%%%%%%%%%%%%%%%%%%%%%%%%%%%%%%%%%%%%%%%%%%%%%%

\subsection{Organization of the computation
\label{sec:organization}}

We now discuss the structure of the phase-space endpoint divergences
in more detail. The clarification of their diagrammatic origin allows
us to divide the sum of resonant and non-resonant
NNLO contributions into several separately divergence-free parts, and
this separation determines the organization of the actual calculation.
The cross sections of the processes $e^+e^-\rightarrow \bar{t}W^+b$
and $e^+e^-\rightarrow tW^-\bar{b}$ are equal by CP symmetry, hence we
shall only consider the final state $\bar{t}W^+b$ below and
multiply the result by two in the end.

%%%%%%%%%%%%%%%%%%%%%%%%%%%%%%%%%%%%%%%%%%%%%%%%%%%%%%%%%%%%%%%%%%%%%%%%
\begin{figure}[t]
\begin{center}
\includegraphics[width=\textwidth]{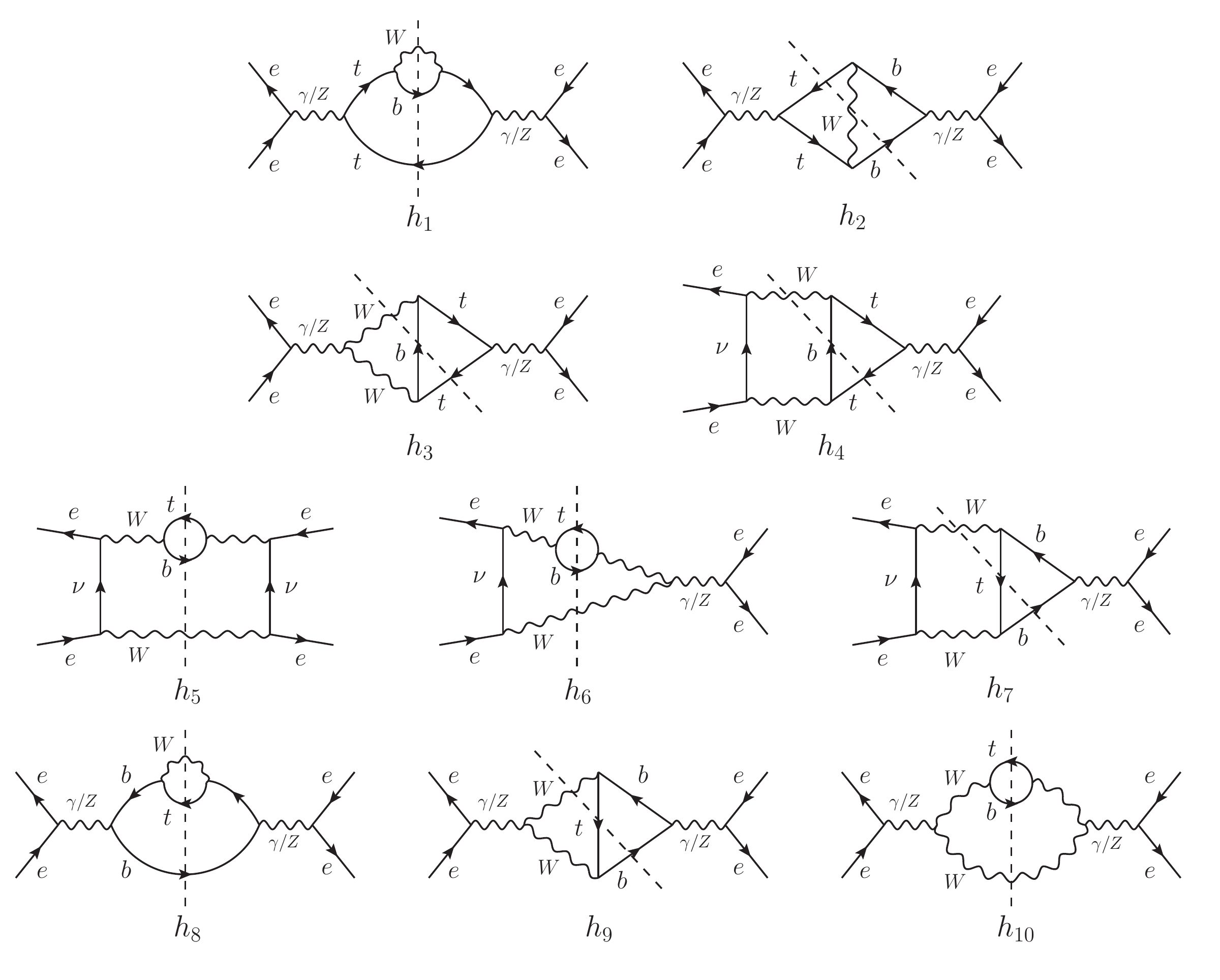}
\vskip-0.5cm
\caption{\label{fig:NonresNLO} NLO non-resonant diagrams.
Symmetric diagrams and diagrams with $tW^-\bar{b}$ cuts are not displayed.}
\end{center}
\end{figure}
%%%%%%%%%%%%%%%%%%%%%%%%%%%%%%%%%%%%%%%%%%%%%%%%%%%%%%%%%%%%%%%%%%%%%%%%

In unitary gauge the NLO non-resonant contribution is given by the diagrams
shown in Figure~\ref{fig:NonresNLO}~\cite{Beneke:2010mp}. At NNLO real and
virtual gluon corrections must be considered. While this appears to be a
standard NLO QCD correction computation to a $2\to 3$ process, existing
automation tools can nevertheless not be employed due to the endpoint
divergences, which are present in addition to the usual UV and IR
singularities.

To illustrate this issue, we consider the phase-space integral of a virtual
diagram such as $h_{ix}$ below,
where the integrand $f_{ix}$ is a Lorentz scalar, i.e.
it only depends on scalar products of its arguments. This allows us to
define
\begin{eqnarray}
\int_y^1dt\,g_{ix}(t) &  \equiv &
\int d\text{LIPS}_{e^+e^-\rightarrow\bar{t}W^+b}\,f_{ix}(p_{e^+},p_{e^-},
p_{\bar{t}},p_{W^+},p_b)\,\theta\!\left((p_{W^+}+p_b)^2-ym_t^2\right)
\nonumber \\[-0.2cm]
 & = & \frac{m_t^2}{2\pi}\int\limits_y^1dt
\int d\text{LIPS}_{e^+e^-\rightarrow t\bar{t}}
\int d\text{LIPS}_{t\rightarrow W^+b}\,
f_{ix}(p_{e^+},p_{e^-},p_{\bar{t}},p_{W^+},p_b),\quad
\label{eq:PSI}
\end{eqnarray}
where
\begin{equation}
d\text{LIPS}_{i_1\dots i_n\rightarrow f_1\dots f_m} =
\delta^{(d)}\!\left(\sum\limits_{i=1}^n p_{i_i}-\sum\limits_{i=1}^m p_{f_i}
\right)\,
\prod\limits_{i=1}^m\,\frac{d^{d-1}\mathbf{p}_{f_i}}{(2\pi)^{d-1}2p_{f_i}^0}
\end{equation}
is the $d$-dimensional Lorentz-invariant phase space for the
process $i_1(p_{i_1})\dots i_n(p_{i_n})\to f_1(p_{f_1})\dots f_m(p_{f_m})$ and
$t\equiv(p_{W^+}+p_b)^2/m_t^2$. The Heaviside function accounts for
the optional cut on the invariant mass of the top quark as will be discussed
in Section~\ref{sec:nonres}. Since the bottom quark mass can be safely
neglected for this calculation, for the total cross section $y=m_W^2/m_t^2$.
The real corrections can be brought into the same form as~\eqref{eq:PSI}
with the variable $t^*\equiv(p_{W^+}+p_b+p_g)^2/m_t^2$ instead of $t$.

The endpoint divergences originate from the region $t\rightarrow1$, where
the integrand becomes singular due to negative powers of
$(1-t)=(m_t^2-(p_{W^+}+p_b)^2)/m_t^2$, which stem from top-quark propagators
becoming resonant. In~\cite{Jantzen:2013gpa} the leading terms in an
expansion around $t=1$ of the integrands $g_{ix}(t)$
were obtained using the expansion by regions
approach~\cite{Beneke:1997zp,Jantzen:2011nz}. The remaining $t$-integration
for the expanded result is trivial,
\begin{equation}
\int_y^1dt\,(1-t)^{-a-b\epsilon}=
\frac{(1-y)^{1-a-b\epsilon}}{1-a-b\epsilon}.
\label{eq:tintegral}
\end{equation}
The divergent integrals with $a\geq1$ are regulated dimensionally
by the $b\epsilon$ in the exponent, which is inherited from the $d-1$
dimensional phase-space integral. At NNLO endpoint-divergent
integrals with $a=1,3/2,2$ are present, but only those with $a=1$
manifest as $1/\epsilon$ poles. This is related to the well-known
property of dimensional regularization, that it renders some
power-like divergent integrals finite for $\epsilon\rightarrow0$.

%%%%%%%%%%%%%%%%%%%%%%%%%%%%%%%%%%%%%%%%%%%%%%%%%%%%%%%%%%%%%%%%%%%%%%%%%%
\begin{figure}[t]
 \begin{center}
  $\vcenter{\hbox{\includegraphics[width=0.23\textwidth]{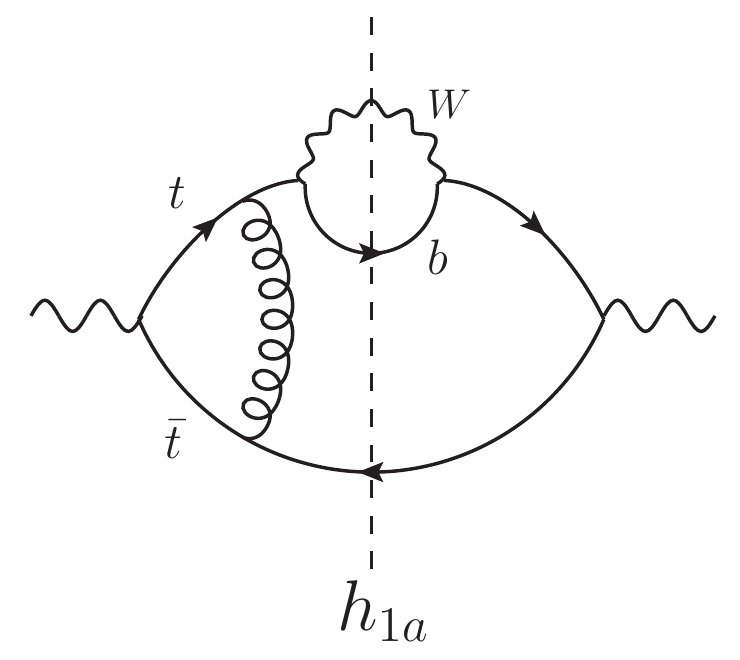}}}$
  $\vcenter{\hbox{\includegraphics[width=0.23\textwidth]{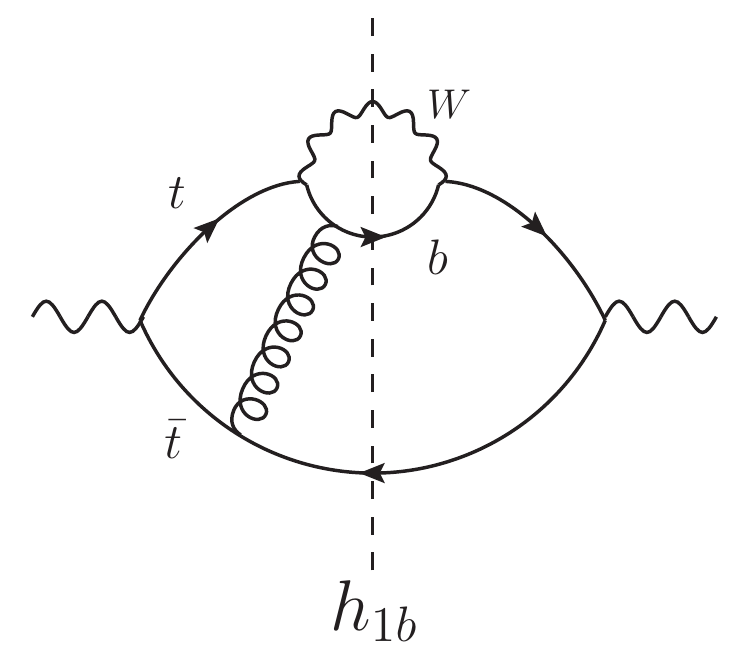}}}$
  $\vcenter{\hbox{\includegraphics[width=0.23\textwidth]{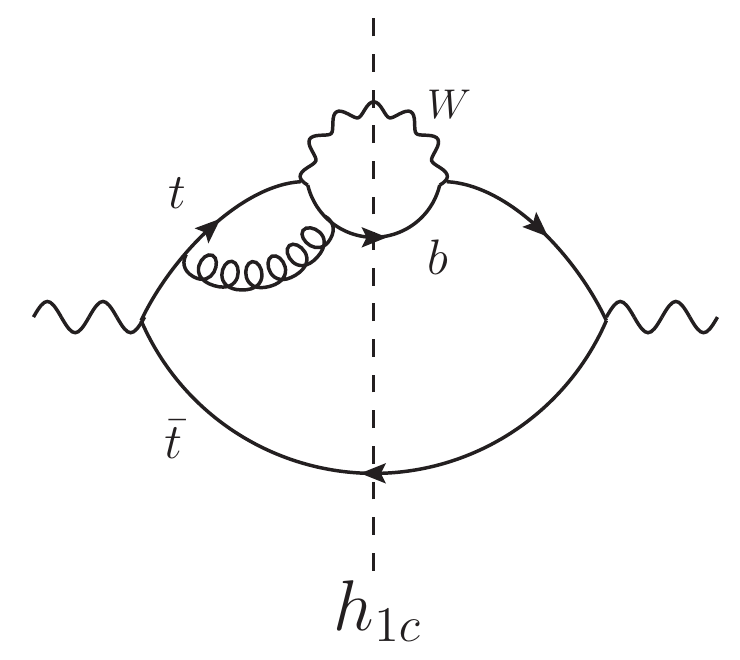}}}$\\
  \vspace*{0.3cm}
  $\vcenter{\hbox{\includegraphics[width=0.23\textwidth]{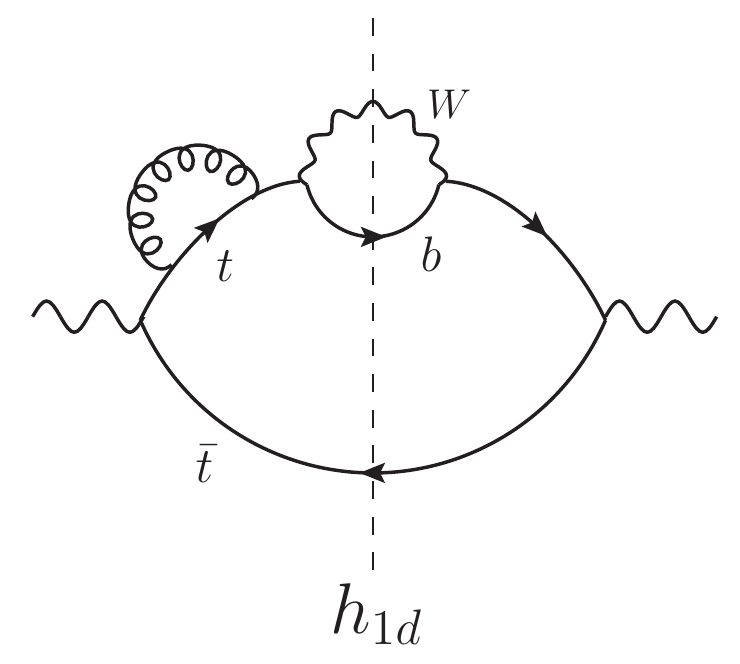}}}$
  $\vcenter{\hbox{\includegraphics[width=0.23\textwidth]{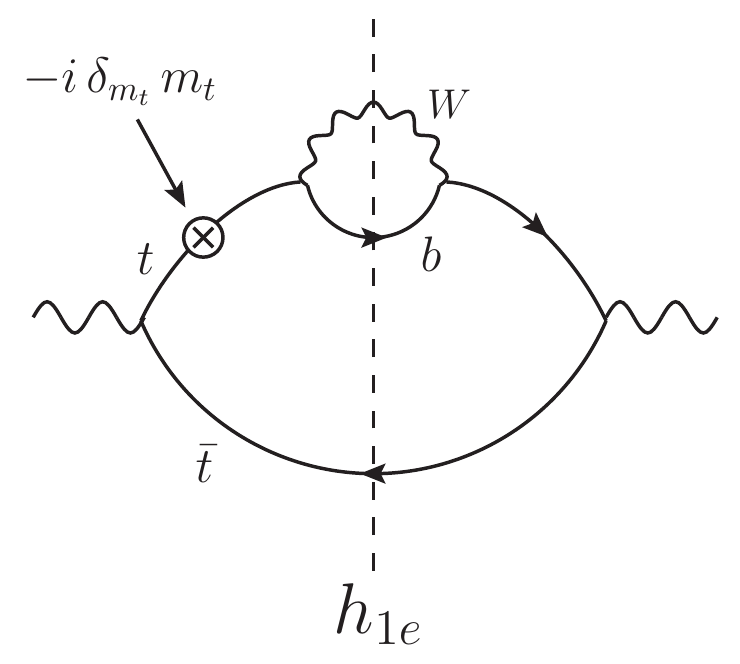}}}$
  $\vcenter{\hbox{\includegraphics[width=0.23\textwidth]{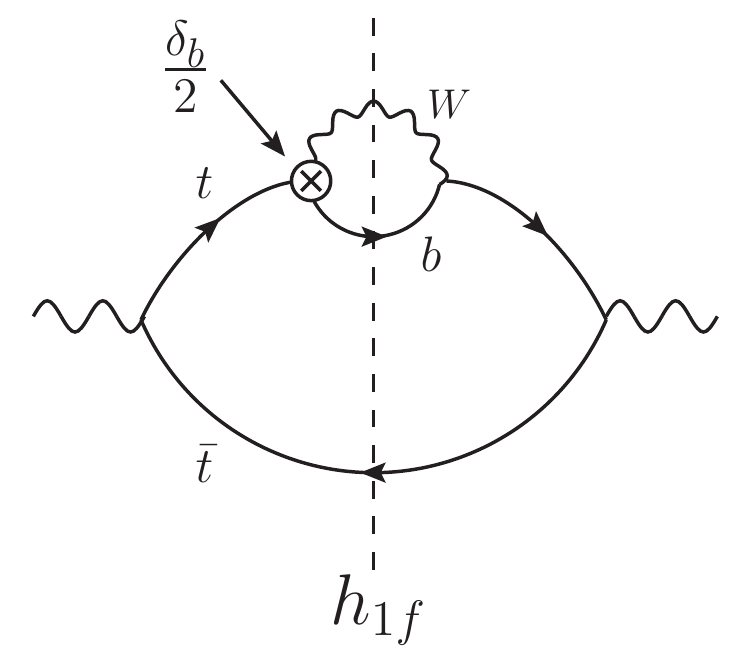}}}$
  $\vcenter{\hbox{\includegraphics[width=0.23\textwidth]{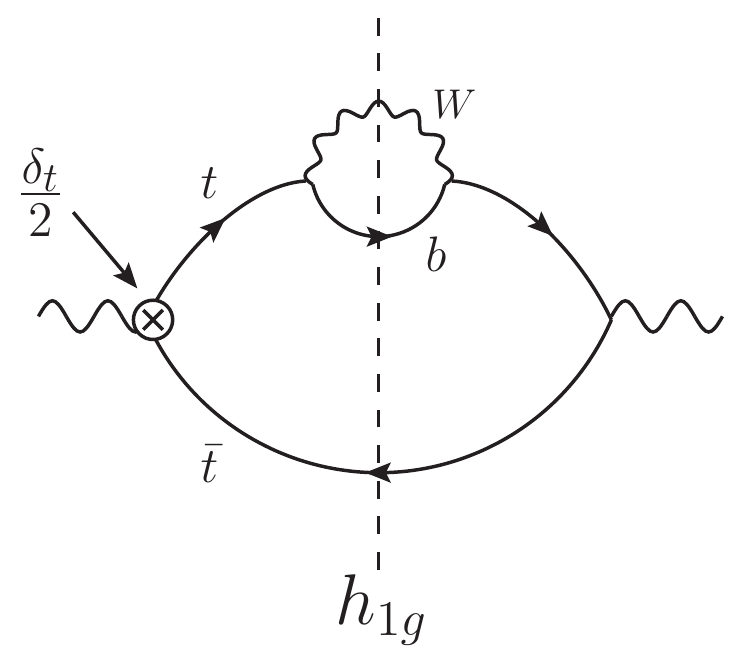}}}$\\
  \vspace*{0.2cm}
  $\vcenter{\hbox{\includegraphics[width=0.23\textwidth]{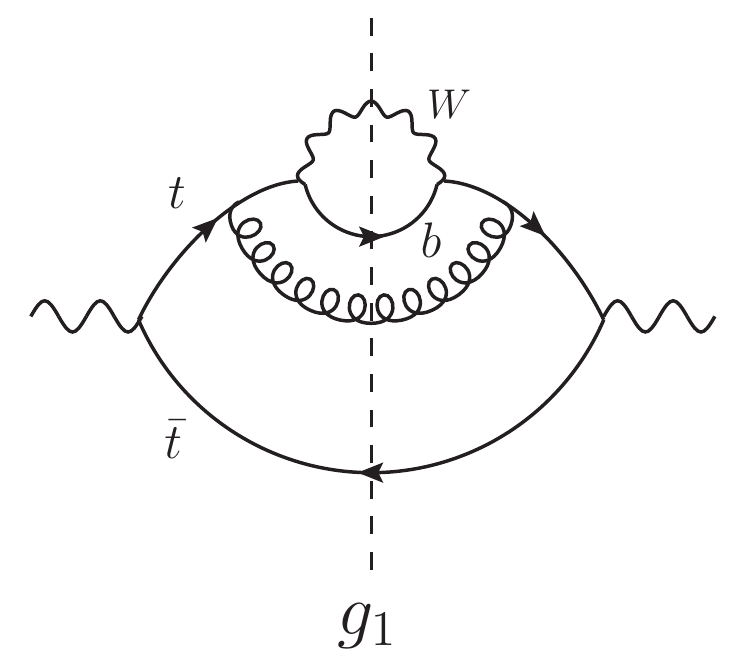}}}$
  $\vcenter{\hbox{\includegraphics[width=0.23\textwidth]{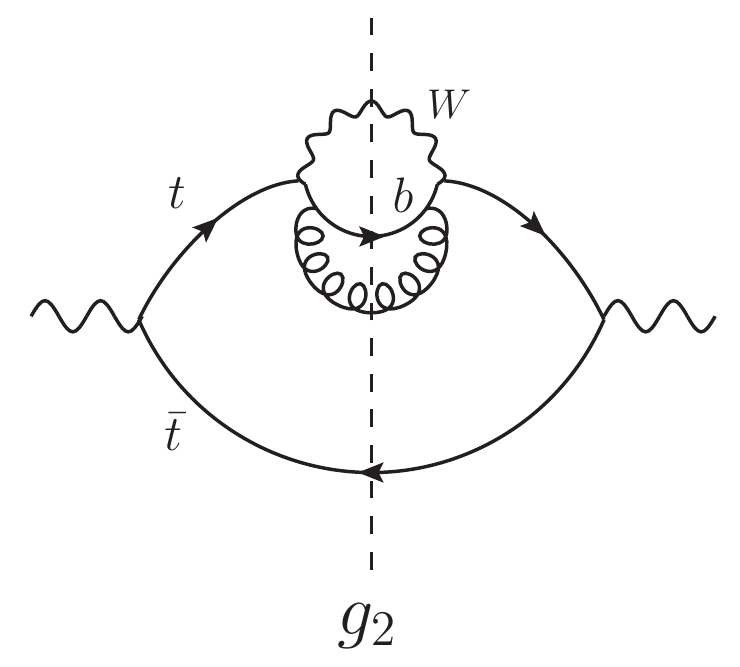}}}$
  $\vcenter{\hbox{\includegraphics[width=0.23\textwidth]{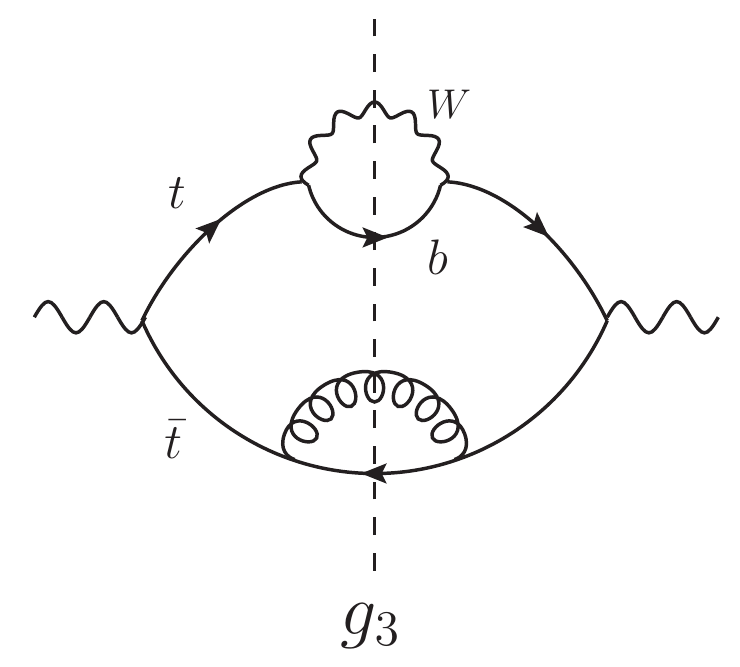}}}$\\
  \vspace*{0.3cm}
  $\vcenter{\hbox{\includegraphics[width=0.23\textwidth]{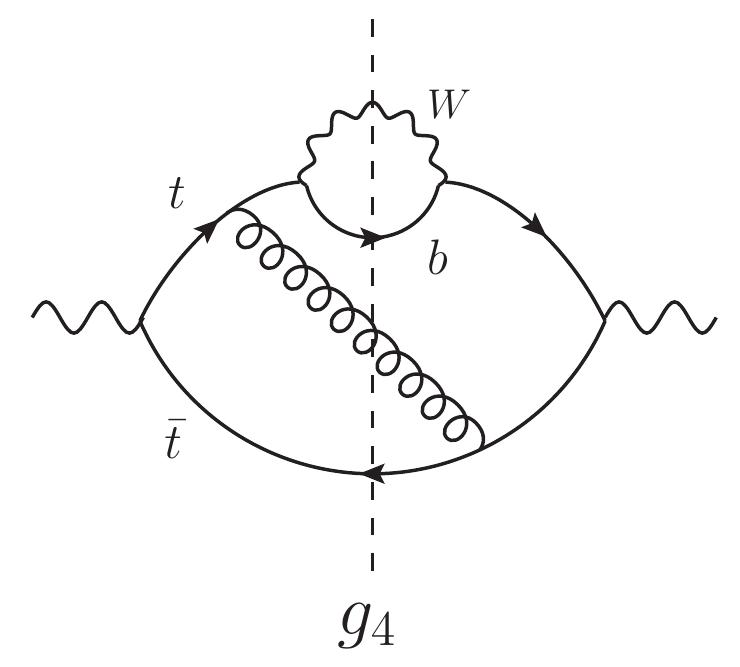}}}$
  $\vcenter{\hbox{\includegraphics[width=0.23\textwidth]{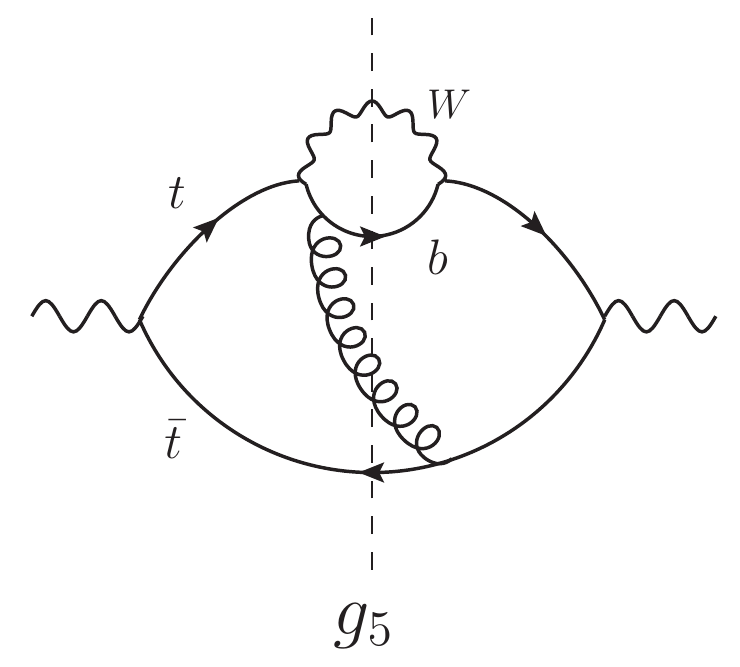}}}$
  $\vcenter{\hbox{\includegraphics[width=0.23\textwidth]{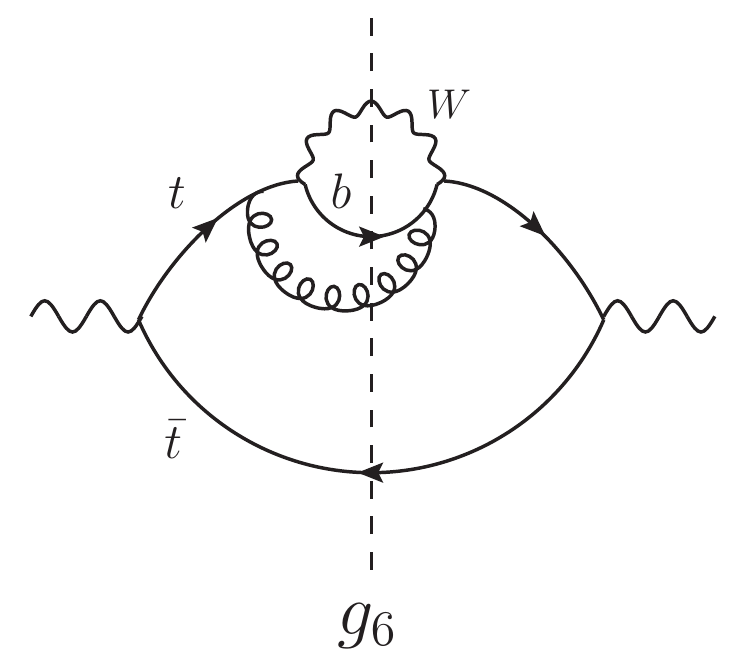}}}$
  \caption{\label{fig:squared}Gluon corrections to the tree-level diagram
  $h_1$. This set of endpoint divergent diagrams is UV and IR finite and
  will be denoted as the squared contribution. Symmetric diagrams and
  diagrams with $tW^-\bar{b}(g)$ cuts are not displayed.}
 \end{center}
\end{figure}
%%%%%%%%%%%%%%%%%%%%%%%%%%%%%%%%%%%%%%%%%%%%%%%%%%%%%%%%%%%%%%%%%%%%%%%%%%

It is obvious from~\eqref{eq:tintegral} that the integrands $g_{ix}(t)$ must
not be expanded in $\epsilon$, because it would spoil the dimensional
regularization of the endpoint divergences. This implies that the loop
integrals in the virtual corrections cannot be expanded in $\epsilon$,
since even the tree-level phase-space integration is divergent.
Expressions for scalar one-loop integrals in general $d$ dimensions with up to
four external legs were obtained recently~\cite{Bluemlein:2015sia}, but
a simpler strategy is to take the results for the endpoint divergent
terms from~\cite{Jantzen:2013gpa} as subtractions to the complete integrand.
The integrals \eqref{eq:PSI} are decomposed as follows:
\begin{equation}
\int_y^1dt\,g_{ix}(t)=\int\limits_y^1dt\left[g_{ix}(t)-
\!\sum_{a=1,\frac32,2}\sum_{b}\frac{\hat{g}_{ix}^{(a,b)}}{(1-t)^{a+b\epsilon}}
\right]+
\!\sum_{a=1,\frac32,2}\sum_{b}\frac{\hat{g}_{ix}^{(a,b)}(1-y)^{1-a-b\epsilon}}
{1-a-b\epsilon},
\label{eq:EPsubtraction}
\end{equation}
where the required coefficients $\hat{g}_{ix}^{(a,b)}$ of the series
expansion in $(1-t)$ are available up to order $\mathcal{O}(\epsilon^0)$
from~\cite{Jantzen:2013gpa}. This renders the $t$-integration on the
right-hand side finite and allows us to expand the subtracted expression in
the square bracket in $\epsilon$. Thus, the integral can be performed
numerically. Additionally, we require the $\mathcal{O}(\epsilon)$
contributions to $\hat{g}_{ix}^{(1,b)}$, because the coefficients with $a=1$
are multiplied with a $1/\epsilon$ pole in~\eqref{eq:EPsubtraction}.

\begin{figure}[t]
 \begin{center}
  $\vcenter{\hbox{\includegraphics[width=0.28\textwidth]{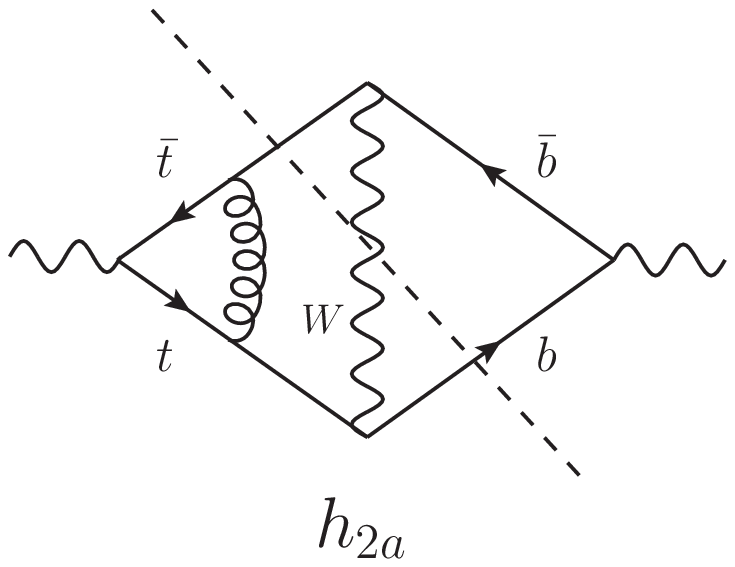}}}$
  $\vcenter{\hbox{\includegraphics[width=0.28\textwidth]{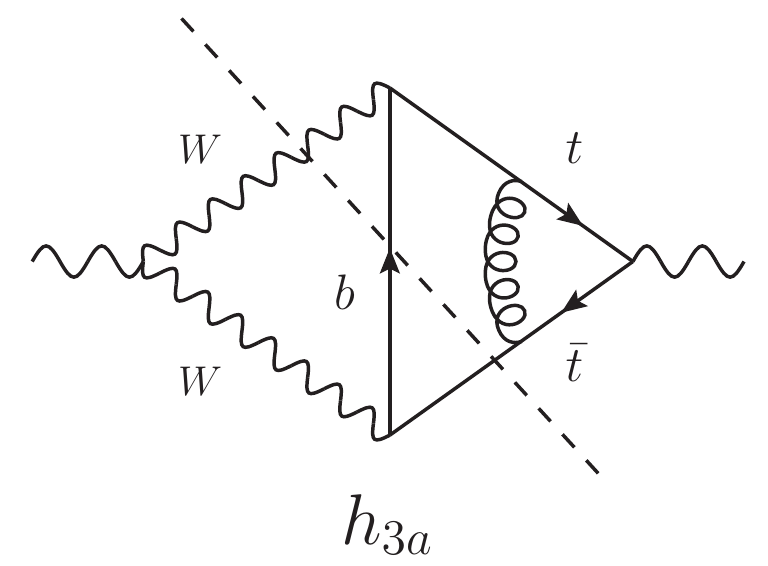}}}$
   $\vcenter{\hbox{\includegraphics[width=0.28\textwidth]{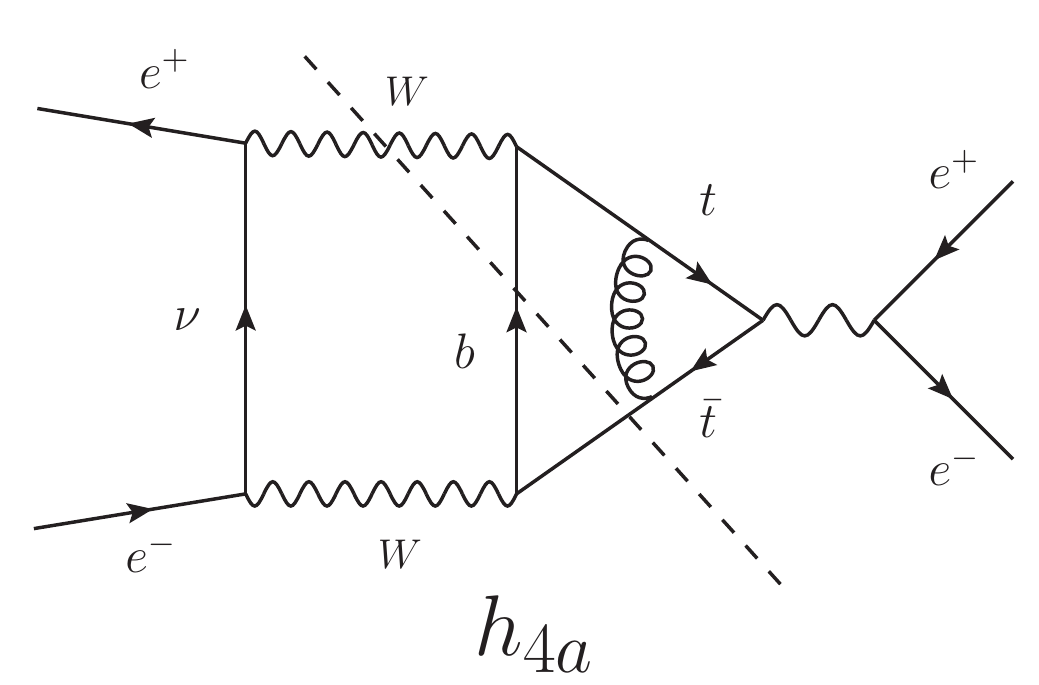}}}$
  \caption{\label{fig:interference} Additional endpoint singular diagrams
  for the NNLO non-resonant part. This set of endpoint divergent
  diagrams also contains UV divergences and will be denoted as the
  interference contribution. Symmetric diagrams and diagrams with
  $tW^-\bar{b}$ cuts are not displayed.}
 \end{center}
\end{figure}

In total the NNLO non-resonant correction requires the evaluation
of the order of 100 diagrams obtained by attaching one gluon to the
diagrams in Figure~\ref{fig:NonresNLO} in all possible ways. Fortunately
only about 15\% of those contain endpoint divergences. They have been
identified in~\cite{Jantzen:2013gpa} and are shown in
Figures~\ref{fig:squared} and \ref{fig:interference}. They are computed
manually by applying the subtractions~\eqref{eq:EPsubtraction}.
The remaining large number of finite diagrams is computed in an automated
fashion using suitably edited~\texttt{MadGraph} code. This latter contribution
will be referred to as the automated part $\sigma_\text{aut}$.

%%%%%%%%%%%%%%%%%%%%%%%%%%%%%%%%%%%%%%%%%%%%%%%%%%%%%%%%%%%%%%%%%%%%%%%%%%
%%%%%%%%%%%%%%%%%%%%%%%%%%%%%%%%%%%%%%%%%%%%%%%%%%%%%%%%%%%%%%%%%%%%%%%%%%

%%%%%%%%%%%%%%%%%%%%%%%%%%%%%%%%%%%%%%%%%%%%%%%%%%%%%%%%%%%%%%%%%%%%%%%%%%
\begin{figure}[t]
\begin{center}
\includegraphics[width=0.85\textwidth]{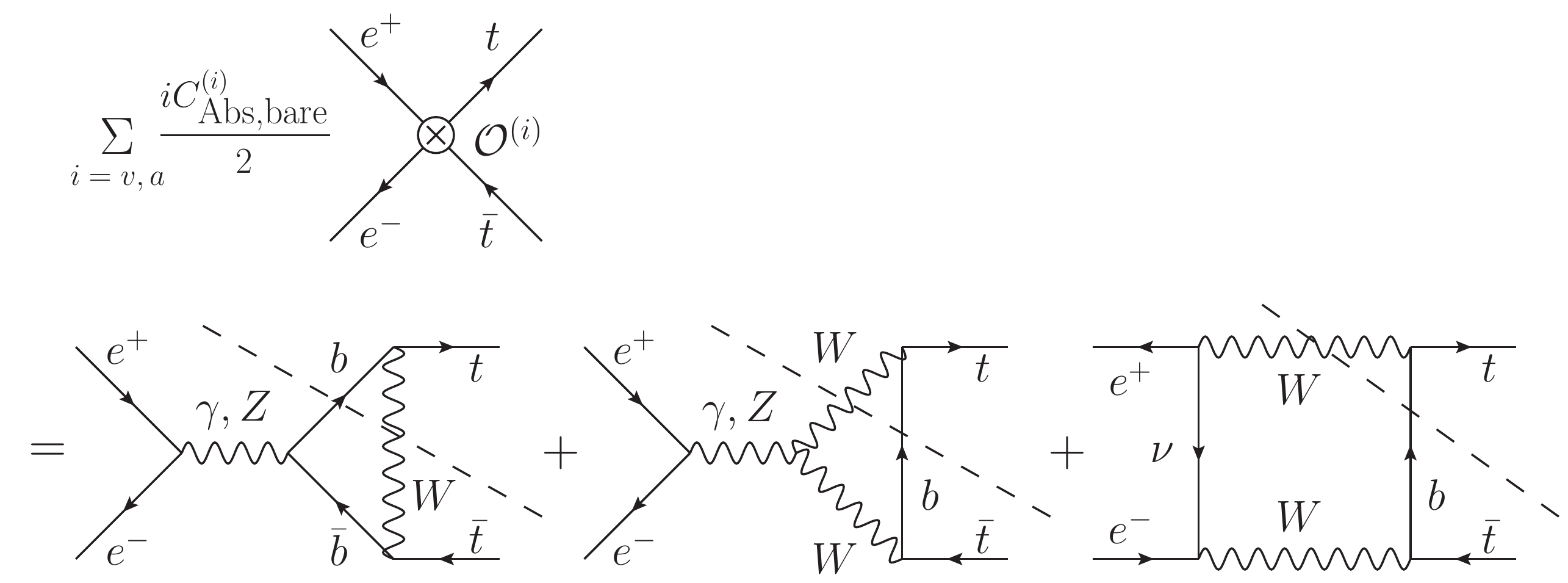}
\vskip0.8cm
\includegraphics[width=0.71\textwidth]{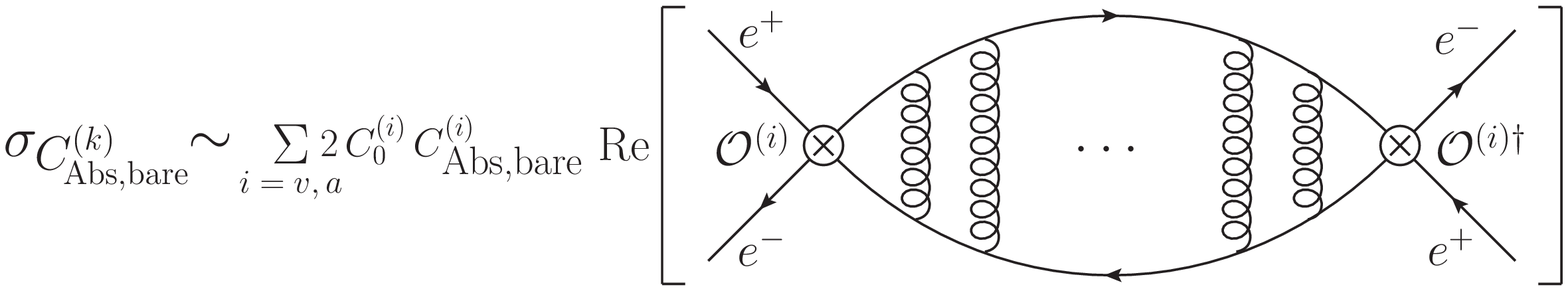}
\vskip0.2cm
\caption{\label{fig:Cabs} The middle panel shows the diagrams accounting
  for the bare absorptive contribution to the matching coefficients $C^{(k)}$.
  The lower panel sketches the respective contribution to the cross section,
  where the ladder exchanges of gluons cause the Coulomb singularities
  $(\alpha_s/v)^k$ and are, therefore, of the same order.
  Only the diagram with a single gluon exchange contains a $1/\epsilon$
  pole, which implies a scheme dependence of the
  finite part. We obtain the same diagrams (up to symmetry) as in
  Figure~\ref{fig:interference} by restoring the full theory graphs in place
  of the effective operators, i.e. by replacing $C_\text{Abs,bare}^{(i)}$
  times the insertion of $\mathcal{O}^{(i)}$ with the diagrams in the upper
  panel and replacing $C_0^{(i)}$ times the insertion of the insertion of
  $\mathcal{O}^{(i)\dagger}$ with $s$-channel photon and $Z$ boson
  exchange. There is no
  double counting, because the two contributions account for different
  momentum regions. When both are summed, the $1/\epsilon$ pole and the scheme
  dependence cancel, see Section~\ref{sec:interference}.}
 \end{center}
\end{figure}
%%%%%%%%%%%%%%%%%%%%%%%%%%%%%%%%%%%%%%%%%%%%%%%%%%%%%%%%%%%%%%%%%%%%%%%%%%

The endpoint divergent diagrams are divided into two parts. The first is
given by the QCD corrections to the diagram $h_1$, shown in
Figure~\ref{fig:squared}, and is denoted as the squared contribution
$\sigma_\text{sq}$.
It is UV and IR finite, because it includes the complete virtual, real and
counterterm contributions to $h_1$. The remaining endpoint divergent
diagrams, shown in Figure~\ref{fig:interference}, are referred to as the
interference contribution $\sigma_\text{int}$. In addition to the
endpoint divergences, the interference part contains UV divergences, which are
cancelled by endpoint-finite counterterm contributions contained in the
automated part. We disentangle the two types of divergences by performing the
subtraction~\eqref{eq:EPsubtraction} and obtain
\begin{equation}
\sigma_\text{int} = \sigma_\text{int}^{(\text{EP div})} +
\sigma_\text{int}^{(\text{EP fin})},
\label{eq:splitinterference}
\end{equation}
where
\begin{equation}
\sigma_\text{int}^{(\text{EP div})} \sim
\sum_{i=2}^4\,\hat{g}_{ia}^{(1,2)}\,\frac{(1-y)^{-2\epsilon}}{-2\epsilon}
\label{eq:sigmaintEPdiv}
\end{equation}
is endpoint-divergent but UV-finite, and
\begin{equation}
\sigma_\text{int}^{(\text{EP fin})} \sim
\sum_{i=2}^4\int_y^1dt\left[g_{ia}(t)-\frac{\hat{g}_{ia}^{(1,2)}}
{(1-t)^{1+2\epsilon}}\right]
\label{eq:sigmaintEPfin}
\end{equation}
is endpoint-finite but UV-divergent. In total, this allows us to split the
non-resonant contribution into the following parts
\begin{equation}
 \sigma_\text{non-res}^\text{NNLO} = \sigma_\text{sq} +
\sigma_\text{int}^{(\text{EP div})} +
\left[\sigma_\text{int}^{(\text{EP fin})} + \sigma_\text{aut}\right].
 \label{eq:splitnonres}
\end{equation}
Only the first two terms contain endpoint divergences. The third term,
enclosed in square brackets, is finite. The endpoint divergences cancel with
the resonant contribution. Specifically, the endpoint divergence
$\sigma_\text{int}^{(\text{EP div})}$ of the interference contribution is
compensated by $\sigma_{C^{(k)}_{\text{Abs,bare}}}$
from the bare absorptive parts $C^{(k)}_{\text{Abs,bare}}$ of
the hard matching coefficients $C^{(k)}$ appearing
in~\eqref{eq:resstructure}. The $C^{(k)}_{\text{Abs,bare}}$ are given by the
diagrams in the upper and middle panel of Figure~\ref{fig:Cabs},
which have a direct correspondence to the diagrams $h_{ia}$ in
Figure~\ref{fig:interference}. Following this observation
we split the resonant contribution into two parts,
\begin{equation}
 \sigma_\text{res}^\text{NNLO} = \sigma_{C^{(k)}_{\text{Abs,bare}}} +
\sigma_\text{res, rest},
  \label{eq:splitres}
\end{equation}
where the remainder $\sigma_\text{res, rest}$ contains various contributions
described in detail in Section~\ref{sec:PartI}. Here, we only
point out that $\sigma_\text{res, rest}$ cancels the endpoint divergence of
the squared contribution. Thus, we can now split the cross section into three
separately finite parts
\begin{equation}
 \sigma^\text{NNLO} = \underbrace{\Big[\sigma_\text{sq}+
\sigma_\text{res, rest}\Big]}_{(\text{I})}
  + \underbrace{\left[\sigma_\text{int}^{(\text{EP div})}+
\sigma_{C^{(k)}_{\text{Abs,bare}}}\right]}_{(\text{II})}
  + \underbrace{\left[\sigma_\text{int}^{(\text{EP fin})} +
\sigma_\text{aut}\right]}_{(\text{III})}.
\label{eq:splitfull}
\end{equation}
The finiteness allows us to evaluate each of the parts $(\text{I})$,
$(\text{II})$ and $(\text{III})$ in a different computational scheme. They
will be computed in Sections~\ref{sec:PartI}, \ref{sec:PartII}
and~\ref{sec:PartIII}, respectively. An overview over the divergences that 
appear in the individual parts $(\text{I})$, $(\text{II})$ and $(\text{III})$ 
is given in Table~\ref{tab:divergences}. The computations are performed in 
the top-quark mass pole scheme. The results are then converted to an IR 
renormalon-free mass scheme for the numerical evaluations performed in 
Section~\ref{sec:results}. The bottom-quark mass is neglected in all 
contributions except $\sigma_\text{aut}$, where the default value 
$m_b=4.7$\,GeV of \texttt{MadGraph} is used unless indicated otherwise.

%%%%%%%%%%%%%%%%%%%%%%%%%%%%%%%%%%%%%%%%%%%%%%%%%%%%%%%%%%%%%%%%%%%%%%%%%%
\begin{table}[t]
\centering
\begin{tabular}{ |c|c|c ccc |}
  \multicolumn{3}{c}{}                                                                & UV finite  & IR finite  & \multicolumn{1}{c}{EP finite} \\ \hline
  \multicolumn{1}{|c}{(I)} & \multicolumn{2}{c}{}                                     & \checkmark & \checkmark & \checkmark \\ \cline{2-6}
  & \multicolumn{1}{|c}{$\sigma_\text{sq}$} & \multicolumn{1}{c}{}                    & \checkmark & \checkmark & -- \\ \cline{3-6}
  & & \multicolumn{1}{|c}{}                                                           &            &            & \\[-0.55cm]
  & & \multicolumn{1}{|c}{$\sigma_\text{sq}^{(h_{1a},\dots,h_{1g})}$}                 & \checkmark & --         & -- \\
  & & \multicolumn{1}{|c}{$\sigma_\text{sq}^{(g_1,\dots,g_6)}$}                       & \checkmark & --         & $\star$ \\ \cline{2-6}
  & \multicolumn{1}{|c}{$\sigma_\text{res, rest}$} & \multicolumn{1}{c}{}             & \checkmark & \checkmark & -- \\ \cline{3-6}
  & & \multicolumn{1}{|c}{$\sigma_\text{QCD}$}                                        & \checkmark & \checkmark & -- \\
  & & \multicolumn{1}{|c}{$\sigma_\text{P-wave}$}                                     & \checkmark & \checkmark & -- \\
  & & \multicolumn{1}{|c}{$\sigma_H$}                                                 & \checkmark & \checkmark & \checkmark \\
  & & \multicolumn{1}{|c}{$\sigma_{\delta V_\text{QED}}$}                             & \checkmark & \checkmark & \checkmark \\
  & & \multicolumn{1}{|c}{$\sigma_{\Gamma}$}                                          & \checkmark & \checkmark & -- \\
  & & \multicolumn{1}{|c}{$\sigma_{C_{\text{EW}}^{(k)}}$}                             & \checkmark & \checkmark & \checkmark \\
  & & \multicolumn{1}{|c}{$\sigma_{C^{(k)}_{\text{Abs,}Z_t}}$}                        & \checkmark & \checkmark & -- \\
  & & \multicolumn{1}{|c}{$\sigma_{\text{IS}}^\text{conv}$}                           & \checkmark & \checkmark & \checkmark \\ \hline \hline
  \multicolumn{1}{|c}{(II)} & \multicolumn{2}{c}{}                                    & \checkmark & \checkmark & \checkmark \\ \cline{2-6}
  & \multicolumn{1}{|c}{} & \multicolumn{1}{c}{}                                      &            &            & \\[-0.55cm]
  & \multicolumn{1}{|c}{$\sigma_\text{int}^{(\text{EP div})}$} & \multicolumn{1}{c}{} & \checkmark & \checkmark & -- \\ \cline{2-6}
  & \multicolumn{1}{|c}{$\sigma_{C^{(k)}_{\text{Abs,bare}}}$} & \multicolumn{1}{c}{}  & \checkmark & \checkmark & -- \\[0.15cm] \hline \hline
  \multicolumn{1}{|c}{(III)} & \multicolumn{2}{c}{}                                   & \checkmark & \checkmark & \checkmark \\ \cline{2-6}
  & \multicolumn{1}{|c}{} & \multicolumn{1}{c}{}                                      &            &            & \\[-0.55cm]
  & \multicolumn{1}{|c}{$\sigma_\text{int}^{(\text{EP fin})}$} & \multicolumn{1}{c}{} & -- & \checkmark & \checkmark \\ \cline{2-6}
  & \multicolumn{1}{|c}{$\sigma_{\text{aut}}$} & \multicolumn{1}{c}{}                 & -- & \checkmark & \checkmark \\ \hline
\end{tabular}
\vskip0.2cm
\caption{Overview over the divergences that appear in the various 
contributions to the cross section. The contributions 
$\sigma_\text{sq}^{(h_{1a},\dots,h_{1g})}$ and 
$\sigma_\text{sq}^{(g_1,\dots,g_6)}$ correspond to the virtual plus 
counterterm and real contributions to $\sigma_\text{sq}$, respectively.
With $\,\star\,$ we indicate contributions that are endpoint divergent by 
power counting, but finite in dimensional regularization. The contributions 
that make up $\sigma_\text{res, rest}$ are defined in 
Section~\ref{sec:PartI}. \label{tab:divergences}}
\end{table}
%%%%%%%%%%%%%%%%%%%%%%%%%%%%%%%%%%%%%%%%%%%%%%%%%%%%%%%%%%%%%%%%%%%%%%%%%%

%%%%%%%%%%%%%%%%%%%%%%%%%%%%%%%%%%%%%%%%%%%%%%%%%%%%%%%%%%%%%%%%%%%%%%%%
%%%%%%%%%%%%%%%%%%%%%%  Invariant mass cut      %%%%%%%%%%%%%%%%%%%%%%%%
%%%%%%%%%%%%%%%%%%%%%%%%%%%%%%%%%%%%%%%%%%%%%%%%%%%%%%%%%%%%%%%%%%%%%%%%

\subsection{Implementation of a ``top invariant mass cut''
\label{sec:nonres}}

The main result of this work is the non-resonant NNLO correction to the full
cross section $\sigma(e^+e^-\rightarrow b\bar{b}W^+W^-X)$, but we also
present results with loose cuts on the invariant mass of the top and
anti-top quark
\begin{equation}
(m_t-\Delta M_t)^2\leq p_{t,\bar{t}}^2\leq (m_t+\Delta M_t)^2,
\label{eq:definitioncut}
\end{equation}
where $p_t$ ($p_{\bar{t}}$) denotes the momentum of the (anti-) top
quark. The implementation of cuts in the effective field theory framework
has been discussed in \cite{Actis:2008rb} and depends on the scaling
of the cut parameter $\Delta M_t$ with respect to the power counting
parameters of the EFT. The cut is termed ``loose'' when
$\Delta M_t\gg \Gamma_t$. Thus, a loose cut never affects the resonant
contribution to the cross section, where the off-shellness of the tops is
parametrically of order $m_t \Gamma_t$.

Since the physical final state is $b\bar{b}W^+W^-X$ without reference
to whether it was produced through an intermediate top or anti-top,
it is necessary to {\em define} what is meant by the (anti-) top
momentum. An invariant mass cut of the form (\ref{eq:definitioncut})
was already implemented in the NLO non-resonant calculation
\cite{Beneke:2010mp}, but since at this order the partonic final state is
always $b\bar{b}W^+W^-$, one simply defines $p_t = p_{W^+}+p_{b}$,
$p_{\bar t} = p_{W^-}+p_{\bar b}$. At NNLO the final state may contain
an additional gluon and a definition of the observable is required
at the hadronic and the partonic level. Any sensible definition of an
observable called ``top momentum'' should be equal up to an amount
of order $\Gamma_t$ to the momentum of the top quark in the hypothetical
limit that the top quark were a stable particle. It should also lend itself
to simple, infrared-finite theoretical computations. On the other hand,
the assignment of top momentum to the final state of a non-resonantly
produced $b\bar{b}W^+W^-X$ event is rather arbitrary and a matter of
definition subject to the previous two requirements. Because non-resonant
production is a sub-leading effect, different definitions differ only
by small amounts.

In the following we describe an algorithm that satisfies these
requirements. The algorithm is most likely not optimal and serves only as
a proof of concept. In the first step we cluster any hadronic or
partonic event into the objects $W^+$, $W^-$, $b$-jet, $\bar b$-jet
and other jets. For the purpose of this discussion energetic
leptons and photons are also among the ``other jets''. We require that
the event contains exactly one $W^+$, one $W^-$ and at least one
$b$- and one $\bar b$-jet.\footnote{We assume here that the charges of the
$W$ bosons and bottom jets have been reconstructed. In practice, this
will be inefficient although a Monte Carlo study for the top
forward-backward asymmetry at the ILC concluded that the discrimination
between bottom and anti-bottom jets can be achieved with a purity of 80\% at
about 60\% efficiency~\cite{Devetak:2010na}. We also do not discuss
the non-trivial combinatorial problem of reconstructing the $W$ bosons
from their hadronic decay in the presence of additional jets, since in the
logic of our discussion the $W$ bosons are considered as stable particles.}
Any jet algorithm can be used to define these objects. In a second step we
group the above pre-defined objects into exactly two clusters. The momenta
of these two clusters define the (anti-) top momentum. This fulfills the
above-mentioned requirement, since close to threshold in an event with
a resonant top and anti-top, the momentum of any other particle can be
at most of order~$\Gamma_t$.

To implement the second step, assume that the event contains
$N$ other jets and let ${\cal S}=\{p_{Ji}, i=1\ldots N\}$ be the set
of jet momenta. A partitioning of ${\cal S}$ consists of two disjoint
sets ${\cal A}$, $\bar{\cal A}$ such that ${\cal A} \cup \bar{\cal A}
= {\cal S}$. The momentum of the ``top cluster'' and the ``anti-top''
cluster for a given partitioning are defined as
\begin{equation}
p_{t{\cal A}} = p_{W^+}+p_{b} +\sum_{i\in {\cal A}} p_{Ji},
\qquad
p_{\bar t\bar {\cal A}} = p_{W^-}+p_{\bar b} +\sum_{i\in \bar{\cal A}} p_{Ji}.
\end{equation}
If there is more than one $b$-jet ($\bar b$-jet) in the event,
$p_b$ ($p_{\bar b}$) refers to the most energetic $b$-jet ($\bar b$-jet),
and the remaining ones are considered to be part of the set ${\cal S}$.
We then define $p_t$ and $p_{\bar t}$ by the value of
$p_{t{\cal A}}$ and $p_{\bar t\bar {\cal A}}$, respectively, of the
partitioning ${\cal A}$, which minimizes the product
\begin{equation}
\chi\equiv \big|(p_{t{\cal A}}^2-m_t^2) (p_{\bar t\bar {\cal A}}^2-m_t^2)
\big|\,.
\label{mincondition}
\end{equation}
An event passes the cut (\ref{eq:definitioncut}), if the so-defined
top momenta satisfy the inequality (\ref{eq:definitioncut}).

We now apply this definition to the partonic NNLO calculation. The partonic
final states are $t\bar t$ only at LO, $t\bar t$ and
$t W^-\bar b$, $\bar t W^+b$ at NLO, and at NNLO the previous and
$t W^-\bar bg$, $\bar t W^+bg$. Here $t$ ($\bar t$) means a final state
that can originate from on-shell (anti-) top decay at the given order,
with invariant mass within $m_t^2$ by an amount of order $m_t \Gamma_t$.
Consider first the $t\bar t$ final state. One might think that
the $t$ ($\bar t$) decay products are automatically clustered into
the correct parent $p_t$ ($p_{\bar t}$) and hence the event always passes the
loose cut as desired. However, this is not always true, since an
energetic gluon from top decay may be radiated collinear to the
$\bar b$ from anti-top decay, in which case the jet algorithm merges
it into the $\bar b$-jet rather than the $b$-jet. In this case the
loose-cut condition (\ref{eq:definitioncut}) might not be satisfied, and the
event is missed even though both tops were produced resonantly.\footnote{If
the gluon is not energetic but ultrasoft with momentum of order $\Gamma_t$,
the misassignment is irrelevant and the loose-cut condition remains
satisfied.} Since our calculation does not include the kinematics of
resonant top decay, we cannot correct it for this misassignment. However,
the probability for such a misassignment is at most of order
\begin{equation}
\frac{\alpha_s(m_t)}{\pi}\times\frac{\pi R^2}{4\pi} \
\approx 0.1\%,
\end{equation}
where $R$ is the half-opening angle of the $\bar b$-jet. The first
factor represents the suppression of energetic, large-angle gluon
radiation and the second the jet area on the unit sphere. The numerical
estimate is obtained for $R\approx 0.3$. We can therefore safely
neglect this error.

Next we consider the non-resonant final state $t \,W^-\bar b$. (With obvious
modifications the following discussion applies to the CP-conjugate final
state.) To NNLO accuracy, on-shell top decay must be taken in NLO, and
may contain an additional gluon. Whenever there is no gluon or the gluon
is merged with the $b$- or $\bar b$-jet, the set of partitionings is empty
and the definition of $p_t$ and $p_{\bar t}$ is the sum of the appropriate
$W$ and $b$-jet momenta. The loose cut is passed except when the gluon
is misassigned to the $\bar b$-jet as above, but in this case the
probability for this to occur is even further suppressed due to the
suppression of non-resonant production in the first place. If, on the
other hand, the jet algorithm returns an additional (gluon) jet, there
are two partitionings, one where the gluon jet momentum is (correctly)
added to the top decay, i.e. to $p_{W^+}+p_b$, and the other, where it
is not. The first, correct, possibility will almost always minimize
$\chi$ in (\ref{mincondition}) and then satisfy the loose-cut condition,
whenever the invariant mass of the non-resonant $W^-\bar b$ pair is
larger than $y m_t^2$, where $y\equiv (m_t-\Delta M_t)^2/m_t^2$. Hence,
imposing the cut yields a single Heaviside function
$\theta((p_{W^-}+p_{\bar b})^2-y m_t^2)$ in the phase-space
integral, as in (\ref{eq:PSI}). The other partitioning where the additional
gluon jet is incorrectly combined with the non-resonant $W^-\bar b$
to form the anti-top momentum can minimize $\chi$ only if the invariant
mass of the $W^-$, $\bar b$-jet and gluon jet accidentally adds up to
$m_t^2$ within an amount $m_t\Gamma_t$. This possibility is suppressed
by the NNLO probability for the process to happen in the first place
times the small phase-space fraction where the kinematic requirements
for misassignment are satisfied, and hence can be neglected at NNLO.

Finally we discuss the  $t \,W^-\bar bg$ final state, which appears
at NNLO in the non-resonant part. At NNLO, it is sufficient to consider
the resonant top quark decay in the tree approximation. Hence, the
discussion of the $W^+ \,W^-b \bar b g$ final state from above can be
repeated, except that now the partitioning that minimizes $\chi$ with
overwhelming probability is the correct combination of the gluon jet
momentum with the non-resonant $W^-\bar b$ pair. Hence, up to a negligible
error, the loose cut (\ref{eq:definitioncut}) is implemented in
the real-emission phase-space of the NNLO non-resonant contribution as
the Heaviside function $\theta((p_{W^-}+p_{\bar b}+p_g)^2-y m_t^2)$.

We have not implemented other cuts, but it is in principle straightforward
to do so as long as they are loose. A general cut is a function $c(p_i)$ of
the external momenta, which evaluates to one if the event passes the cut
and to zero otherwise. We define the complementary cut as
$\bar{c}(p_i)=1-c(p_i)$. Assuming that $c(p_i)$ is loose, the
complementary-cut cross section $\sigma(\bar{c})$
is purely non-resonant and free of endpoint divergences. It can therefore
be computed with automated NLO parton level event generators such
as \texttt{MadGraph}~\cite{Alwall:2014hca}. The non-resonant contribution with the original cut is then given by subtracting
$\sigma(\bar{c})$ from the total $W^+ W^- b\bar b X$ cross section,
where the cancellation of divergences between the resonant and non-resonant
parts has already been taken care of.
This approach will also be exploited in Section~\ref{sec:checks} to perform
a powerful check of our computation.

A generalization to arbitrary cuts would affect the resonant contributions
and is beyond the scope of this work. Recently, first results of
an implementation of the fully differential cross section with NLL
accuracy near threshold matched to the NLO fixed order result have been
presented~\cite{Reuter:2016ohp}, but the generalization of this method
to NNLO accuracy as discussed here is not straightforward.

%%%%%%%%%%%%%%%%%%%%%%%%%%%%%%%%%%%%%%%%%%%%%%%%%%%%%%%%%%%%%%%%%%%%%%
%%%%%%%%%%%%%%%%%%%%%%%%%    Part I      %%%%%%%%%%%%%%%%%%%%%%%%%%%%%
%%%%%%%%%%%%%%%%%%%%%%%%%%%%%%%%%%%%%%%%%%%%%%%%%%%%%%%%%%%%%%%%%%%%%%

\section{Part (I) \label{sec:PartI}}

The scheme for part~(I) as defined in (\ref{eq:splitfull}) is fixed by the
existing QCD results for $\sigma_\text{res}^\text{NNLO}$. The resonant
QCD cross section factorizes into a leptonic tensor $L$ and the correlation
function of two top-quark currents, $\Pi(q^2)$. The former is evaluated in
four dimensions; the latter completely in $d$ dimensions in the naive
dimensional regularization scheme (NDR). The squared contribution contained
in part~(I) factorizes into the same leptonic tensor $L$ and a
hadronic tensor $H$, and the same conventions must be applied. We compute
this part in Section~\ref{sec:squared}. The electroweak NNLO corrections to
the resonant part must also abide by this scheme (except for
$\sigma_{C^{(k)}_{\text{Abs,bare}}}$ contained in part~(II)), and we
consider them first.

%%%%%%%%%%%%%%%%%%%%%%%%%%%%%%%%%%%%%%%%%%%%%%%%%%%%%%%%%%%%%%%%%%%%%%
%%%%%%%%%%%%%%%%%%%%%%%%     Resonant       %%%%%%%%%%%%%%%%%%%%%%%%%%
%%%%%%%%%%%%%%%%%%%%%%%%%%%%%%%%%%%%%%%%%%%%%%%%%%%%%%%%%%%%%%%%%%%%%%

\subsection{Resonant electroweak effects \label{sec:res}}

Electroweak corrections to the resonant cross section are computed in
the non-relativistic EFT framework extended from QCD to the full Standard
Model. We consider them to NNLO  in the counting
scheme~(\ref{eq:parametercounting}).

For ease of reference and to set up notation, we briefly recapitulate
the well-known expressions for the LO cross section~\cite{Beneke:2016kkb}
\begin{equation}
 \sigma^\text{LO}=\sigma_\text{res}^\text{LO} =
 \sigma_0\,\frac{24\pi N_c}{s}\,\left[C_0^{(v)^2}+C_0^{(a)^2}\right]\,
 \text{Im}\left[G_0(E + i\Gamma)\right],
 \label{eq:sigmaLO}
\end{equation}
where $\sigma_0=4\pi\alpha^2/(3s)$ is the high-energy limit of the
photon-mediated muon pair production cross section at leading order,
$E=\sqrt{s}-2m_t$ and $\Gamma$ is the on-shell top-quark width as defined
below. At LO the top pair is produced via s-channel exchange of a photon or
$Z$ boson. The couplings of the fermions to the $Z$ boson are given by
\begin{equation}
v_f^Z\equiv v_f=\frac{T_3^f-2e_fs_w^2}{2s_wc_w},
\hspace{1cm}
a_f^Z\equiv a_f=\frac{T_3^f}{2s_wc_w},
\label{eq:Zffcouplings}
\end{equation}
where $T_3^f$ is the third component of the weak isospin of fermion $f$,
$e_f$ is the fermion electric charge measured in units of the positron
charge, and $s_w$ and $c_w$ are the sine and cosine of the Weinberg angle,
respectively. The S- and P-wave production
operators are given by\footnote{Note $\gamma_k=-\gamma^k$ and $k$ is summed
from 1 to 3.}
\begin{eqnarray}
\label{eq:Ok1st}
\mathcal{O}^{(v)} & = & \frac{4\pi\alpha}{s}\,
\bar{e}_{c_2}W_{c_2}\gamma_k W_{c_1}^\dagger e_{c_1}\,
\psi^\dagger\sigma^k\chi,\\
 \mathcal{O}^{(a)} & = & \frac{4\pi\alpha}{s}\,
\bar{e}_{c_2}W_{c_2}\gamma_k\gamma^5 W_{c_1}^\dagger e_{c_1}\,
\psi^\dagger\sigma^k\chi,\\
 \mathcal{O}_{\text{P-wave}}^{(v)} & = & \frac{4\pi\alpha}{s}\,
\bar{e}_{c_2}W_{c_2}\gamma_k W_{c_1}^\dagger e_{c_1}\,
\psi^\dagger\frac{\left[\sigma^k,(-i)
\boldsymbol{\sigma}\cdot\mathbf{D}\right]}{2m_t}\chi,\\
 \mathcal{O}_{\text{P-wave}}^{(a)} & = & \frac{4\pi\alpha}{s}\,
\bar{e}_{c_2}W_{c_2}\gamma_k\gamma^5 W_{c_1}^\dagger e_{c_1}\,
\psi^\dagger\frac{\left[\sigma^k,(-i)
\boldsymbol{\sigma}\cdot\mathbf{D}\right]}{2m_t}\chi
\label{eq:Ok}
\end{eqnarray}
with leading-order matching coefficients
\begin{eqnarray}
&&C_0^{(v)}  =  e_ee_t+v_ev_t\frac{s}{s-m_Z^2},
\hspace*{1cm} C_0^{(a)}  =  -a_ev_t\frac{s}{s-m_Z^2},\\
&&C_{0,\text{P-wave}}^{(v)}  =  -v_ea_t\frac{s}{s-m_Z^2},
\hspace*{1.3cm} C_{0,\text{P-wave}}^{(a)} =  a_ea_t\frac{s}{s-m_Z^2}.
\label{eq:CkLO}
\end{eqnarray}
Here $\psi$ $(\chi)$ is the non-relativistic top (anti-top) field
and $e_{c_i}$ denotes the effective field (as defined in soft-collinear
effective theory (SCET)) of an energetic electron moving in the light-like
direction $n_i^\mu$. In the present context, the directions $n_1$ and $n_2$
are set by the electron and positron beams, respectively. The collinear
electromagnetic Wilson lines
\begin{equation}
W_{c_i}(x)=\text{P}\exp\left[i e\int\limits_{-\infty}^0dt \,
\bar{n}_i\cdot A_{c_i}(x+\bar{n}_it)\right]
\end{equation}
have been introduced to make the operators invariant under
collinear gauge transformations in SCET, as well as the light-like
vectors $\bar{n}_i$ with $n_i\cdot \bar{n}_i=2$.
The factor of $4\pi\alpha/s$ has been absorbed into the operators to
render the coefficients dimensionless and of order one. The P-wave
production operators and their Wilson coefficients will be required below.
Note that because the cross section is constructed as an expansion
in $E$, the energy-dependence of the $s$-channel photon and
$Z$ boson propagators could be expanded around $s=4 m_t^2$, in which
case the short-distance matching coefficients would be truly
energy-independent. However, we apply a convention where we keep the full
$s$-dependence in the $s$-channel propagators, which therefore appears in
(\ref{eq:Ok1st}) to (\ref{eq:CkLO}).

The renormalization scheme for the electroweak parameters adopted here
is the $(m_W,m_Z,\alpha(m_Z))$ scheme. The Weinberg angle is then
given by $c_w^2=m_W^2/m_Z^2$ ($s_w^2=1-c_w^2$). The electromagnetic
coupling $\alpha_{\rm em}$ from now on is denoted by $\alpha$, where
$\alpha$ refers to the scale dependent electromagnetic coupling
$\alpha(\mu_\alpha)$ defined through the photon vacuum polarization,
which interpolates between the fine-structure constant
$\alpha_0=\alpha(0)$ and the input parameter $\alpha(m_Z)$.

In (\ref{eq:sigmaLO}) $G_0(E+i\Gamma)$ denotes
the non-relativistic zero-distance Coulomb Green function
in dimensional regularization \cite{Beneke:1999zr,Eiras:1999xx},
\begin{equation}
G_0(E)=\frac{m_t^2\alpha_s C_F}{4\pi}
\left[\frac{1}{4\epsilon}+L_\lambda+\frac12-\frac{1}{2\lambda}-
\hat\psi(1-\lambda)+\mathcal{O}(\epsilon)\right],
\label{eq:G0}
\end{equation}
which describes the propagation of the top-anti-top pair at LO in the
non-relativistic EFT. It is expressed through the variable
\begin{equation}
\lambda=\frac{\alpha_s C_F}{2\sqrt{-E/m}}
\end{equation}
and $\hat\psi(x) = \gamma_E+\psi(x)$, where $\psi$ is the logarithmic
derivative of the gamma function. Furthermore, the logarithm
$L_\lambda=\ln(\lambda\mu/(m_t\alpha_sC_F))=-\frac{1}{2}\ln(-4 m E/\mu^2)$
appears.\footnote{
When the $1/\epsilon$ pole is related to a finite-width divergence,
we set $\mu=\mu_w$ and write $G_0^{(w)}(E)$ and $L_\lambda^{(w)}$ to
distinguish the finite-width scale-dependence of the resonant contribution
from the $\mu_r$ scale-dependence due to the strong coupling,
cf.~\cite{Beneke:2013kia,Beneke:2013PartII}.}

After separating $\sigma_{C^{(k)}_{\text{Abs,bare}}}$ from the NNLO
resonant contributions as explained around ~\eqref{eq:splitres},
the remaining parts are
\begin{equation}
\sigma_\text{res, rest} =
\sigma_\text{QCD} + \sigma_\text{P-wave} + \sigma_{H}
 + \sigma_{\delta V_\text{QED}} + \sigma_{\Gamma} +
 \sigma_{C_{\text{EW}}^{(k)}} + \sigma_{C^{(k)}_{\text{Abs,}Z_t}}
 +\sigma_{\text{IS}}^\text{conv}.
\label{eq:resrest}
\end{equation}
The pure QCD S-wave contribution $\sigma_\text{QCD}$ has been obtained
in the formalism employed here in~\cite{Beneke:1999qg}. Top-pair production
in a P-wave state $\sigma_\text{P-wave}$ was computed in~\cite{Beneke:2013kia}.
Higgs contributions $\sigma_{H}$ that only involve the top Yukawa coupling
have been computed already up to NNNLO~\cite{Beneke:2015lwa,Eiras:2006xm};
similarly the effect $\sigma_{\delta V_\text{QED}}$ of the LO QED Coulomb
potential $\delta V_\text{QED}=-4\pi\alpha e_t^2/
\mathbf{q}^2$~\cite{Beneke:2015lwa}. At NNLO, top decay introduces
additional contributions to the bilinear part of the PNREFT Lagrangian, which
contribute $\sigma_{\Gamma}$ to the resonant cross section
(Section \ref{sec:resGF}). While there are no electroweak contributions to
the non-relativistic potential at NNLO (Section \ref{sec:resPotnew}),
there are electroweak corrections to the hard matching coefficients
$C^{(k)}$. The contribution $\sigma_{C_{\text{EW}}^{(k)}}$ from the real part
of the hard matching coefficients is given in Section~\ref{sec:resCV}.
Contrary to the QCD case the electroweak hard matching coefficients contain
an imaginary part from cuts over all possible final states. The
$\bar{t}W^+b$ ($tW^-\bar{b}$) cuts contribute to the
$e^+ e^- \to b\bar b W^+ W^-$ cross section~\cite{Hoang:2004tg}. The
imaginary part is split into a bare contribution
$\sigma_{C^{(k)}_{\text{Abs,bare}}}$ (Section~\ref{sec:resAbs}) and a
contribution from field renormalization $\sigma_{C^{(k)}_{\text{Abs,}Z_t}}$
(Section~\ref{sec:resCZt}), because the two parts are treated in different
schemes. Partial results for the mixed-QCD-electroweak corrections to
the hard matching
coefficients~$C^{(k)}$ are available~\cite{Eiras:2006xm,Kiyo:2008mh},
but they only contribute at NNNLO and will not be considered here.
Finally, we consider effects from initial-state radiation (ISR),
$\sigma_{\text{IS}}^\text{conv}$ (Section~\ref{sec:resRad}).

\subsubsection{\boldmath Finite-width corrections
to the NNLO Green function, $\sigma_{\Gamma}$
\label{sec:resGF}}

Additional terms appear in the PNREFT Lagrangian due to the instability
of the top quark and its coupling to photons.
The coupling of the top quarks to ultrasoft photons must be multipole
expanded in the spatial component, just like the interactions with the
ultrasoft gluons. Only the leading term
\begin{equation}
\mathcal{L}_{\text{us}}^{(\gamma)} =
\psi^\dagger\left[e_teA_0^{(\gamma)}(t,\mathbf{0})+\dots\right]\psi
+\chi^\dagger\left[e_teA_0^{(\gamma)}(t,\mathbf{0})+\dots\right]\chi,
\label{eq:Lusoft}
\end{equation}
is relevant at NNLO. However, its contribution vanishes, because
the multipole-expanded field can only resolve the net electric
charge of the top anti-top pair, which is zero. In analogy
to QCD, the couplings in the Lagrangian~\eqref{eq:Lusoft} can be
removed by a field transformation involving an ultrasoft Wilson
line (cf. \cite{Beneke:2010da}).
The generalization of the bilinear part of the PNREFT Lagrangian
is~\cite{Beneke:2003xh,Beneke:2004km}
\begin{eqnarray}
 \mathcal{L}_{\text{bilinear}} & = & \psi^\dagger\left[i\partial^0
+\frac{\boldsymbol{\partial}^2}{2m_t}-\frac{\Delta}{2}
+\frac{(\boldsymbol{\partial}^2-m_t\Delta)^2}{8m_t^3}+\dots\right]\psi
\nonumber \\
 & + & \chi^\dagger\left[i\partial^0-\frac{\boldsymbol{\partial}^2}{2m_t}
+\frac{\Delta}{2}-\frac{(\boldsymbol{\partial}^2-m_t\Delta)^2}{8m_t^3}
+\dots\right]\chi,
 \label{eq:Lbilinear}
\end{eqnarray}
where $\psi(\chi)$ is the non-relativistic top (anti-top) field and $\Delta$
is a hard matching coefficient. It can be determined by matching the top
propagator in the effective theory to the full theory. In the pole mass
scheme we obtain
\begin{equation}
\Delta=-i\Gamma,
\label{eq:Delta}
\end{equation}
where $\Gamma$ is the pole width of the top quark defined through the
gauge-invariant position of the pole of the top propagator
\begin{equation}
M_\star^2=m^2-im\Gamma
\label{eq:complexpole}
\end{equation}
in the complex $p^2$ plane. We note that with this convention
\eqref{eq:Lbilinear} contains
the term $-\Gamma^2/(8m_t)(\psi^\dagger\psi-\chi^\dagger\chi)$, which
has the form of a mass shift. It can be absorbed into the definition
of the pole scheme by adding $-\Gamma^2/4$ to the right hand side
of~\eqref{eq:complexpole}, which completes the square and defines a
different convention used e.g. in~\cite{Kniehl:2008cj}. Electroweak
corrections to the top-pair production cross section near threshold have
also been considered in~\cite{Hoang:2010gu}. The absence of the
$\Gamma^2$ correction to the Green function in~\cite{Hoang:2010gu}
implies that this different convention is also adopted there. Thus, one
must be careful to account for this difference in the definition of the
top pole mass when comparing their results to ours.

The term $(i\Gamma/2)(\psi^\dagger\psi-\chi^\dagger\chi)$
in~\eqref{eq:Lbilinear} belongs to the LO Lagrangian and must be treated
non-perturbatively. It leads to the replacement $E\rightarrow E+i\Gamma$,
which \textit{defines} the QCD contribution, and makes the argument of the
Green function in~\eqref{eq:sigmaLO} complex. The two remaining terms in
\eqref{eq:Lbilinear} that contain the width are of NNLO and
can be treated perturbatively. Only two simple single
insertions are required. We denote the correction to the Green function
$G_0(E)$ from the terms
$(X/2)(\psi^\dagger\psi-\chi^\dagger\chi)$ and
$(X/2)(\psi^\dagger\boldsymbol{\partial}^2\psi-\chi^\dagger
\boldsymbol{\partial}^2\chi)$  by $\delta_X G(E)$ and
$\delta_{X\boldsymbol{\partial}^2}G(E)$, respectively. They are
given by
\begin{eqnarray}
\delta_X G(E)&=&
XG_0'(E)=\frac{X}{m\alpha_s^2C_F^2}
\frac{m^2\alpha_sC_F}{4\pi}
\left[\lambda+2\lambda^2+2\lambda^3\psi_1(1-\lambda)\right],\quad
\label{eq:delta_X_G}\\
\delta_{X\boldsymbol{\partial}^2} G(E)&=&
-mX\frac{m^2\alpha_sC_F}{4\pi}\bigg[-\frac{3}{4\lambda}+\frac{1}{2\epsilon}
+\frac12+2L_\lambda^{(w)}-2\hat{\psi}(1-\lambda)
\nonumber\\
&& \,+\frac{\lambda}{2}\psi_1(1-\lambda)\bigg],
\label{eq:delta_XD2_G}
\end{eqnarray}
where $\psi_1(x) = \psi'(x)$ is the first derivative of the polygamma
function. The corresponding NNLO contribution to the
Green function is
\begin{equation}
\delta_{2,\Gamma}G(E)=
\delta_X G(E)|_{X=-\frac{\Gamma^2}{4m_t}}
+\delta_{X\boldsymbol{\partial}^2}
G(E)|_{X=\frac{i\Gamma}{2m_t^2}}.
\label{eq:delta2GammaG}
\end{equation}

In the implementation of the cross section in the \texttt{QQbar\_threshold}
code~\cite{Beneke:2016kkb} the top width is treated as a
parameter. This implies that higher-order corrections to the tree-level
width $\Gamma_0$ are also treated non-perturbatively through the replacement
$E\rightarrow E+i\Gamma$. A subtlety arises at electroweak NNLO
when this result is combined with the non-resonant contribution,
which is computed in dimensional regularization. The pole part of
the NNLO non-resonant contribution is proportional to $\Gamma_0$ with a
finite part that follows from expanding diagrams up to
${\cal O}(\epsilon^0)$. For consistency, the tree-level contribution
to the width in~\eqref{eq:Lbilinear} must be treated as a $d$-dimensional
hard matching coefficient. Hence the  $\mathcal{O}(\epsilon)$ terms
in the $d$-dimensional tree-level expression of the top width contribute
finite terms to the resonant part from their multiplication with
the finite-width $1/\epsilon$ poles. These finite parts are not included
when $\Gamma$ is treated as a four-dimensional numerical parameter, and must
be added separately.\footnote{QCD corrections to the width, however, are
only needed in four dimensions where they are known up to
NNLO~\cite{Czarnecki:1998qc,Chetyrkin:1999ju}.}

The LO pole width, which is required in $d$ dimensions, is given by
\begin{equation}
 \Gamma_0^{(d)} =\frac{m_t\alpha}{16s_w^2}
\frac{(1-x_W)^2(1+2(1-\epsilon)x_W)}{x_W}\frac{\sqrt{\pi}}
{2\Gamma(3/2-\epsilon)}
\left(\frac{4\mu_w^2e^{\gamma_E}}{m_t^2(1-x_W)^2}\right)^\epsilon,
 \label{eq:Gamma0d}
\end{equation}
where $x_W=m_W^2/m_t^2$ and $\mu_w$ denotes the scale related to the 
finite-width divergences as discussed 
in~\cite{Beneke:2013kia,Beneke:2013PartII}. The contribution from the 
$\mathcal{O}(\epsilon)$ terms of~\eqref{eq:Gamma0d}, which multiply the 
finite-width divergence contained in~\eqref{eq:delta2GammaG} and the one in 
the pure QCD result, to be added to the cross section is
\begin{eqnarray}
\delta_{\Gamma_\epsilon/\epsilon}\sigma & = &
\sigma_0\,\frac{m_t\Gamma_0\alpha_sC_FN_c}{s}
\left[\frac{2(1+x_W)}{1+2x_W}+\ln\frac{\mu_w^2}{m_t^2}-2\ln(1-x_W)\right]
\nonumber\\
&& \times\left[C_0^{(v)^2}+C_0^{(a)^2}+C_{0,\text{P-wave}}^{(v)^2}
+C_{0,\text{P-wave}}^{(a)^2}\right],
 \label{eq:Oepswidthterms}
\end{eqnarray}
where $\Gamma_0$ is the $\epsilon\rightarrow0$ limit of \eqref{eq:Gamma0d}.
On the whole, we obtain
\begin{equation}
\sigma_{\Gamma}=\sigma_0\frac{24\pi N_c}{s}\left[C_0^{(v)^2}+C_0^{(a)^2}\right]
\text{Im}\left[\delta_{2,\Gamma}G(E + i\Gamma)\right]
+\delta_{\Gamma_\epsilon/\epsilon}\sigma.
\label{eq:resEWwidth}
\end{equation}

In the numerical evaluation we resum the perturbative corrections to the
would-be toponium bound-state poles. Due to the instability of the top quarks,
we are dealing with a non-Hermitian Lagrangian, cf.~\eqref{eq:Lbilinear}.
The implications have been discussed in~\cite{Sternheim:1972zz}.
The positions of the would-be toponium poles are the complex eigenvalues
\begin{equation}
 \mathcal{E}_n = E_n - \frac{i \Gamma_n}{2}
\end{equation}
of the non-Hermitian Hamiltonian, where $E_n$ is the bound-state energy
assuming stable top quarks and
$\Gamma_n = 2\Gamma + \delta\Gamma_n$ is the total inclusive width of
the bound state. In accordance with the earlier discussion, the top-quark
width $\Gamma$ is treated as a parameter. The corrections $\delta\Gamma_n$
describe the effects of time dilatation on the top decays due to the residual
movement of the top quarks and the annihilation of the would-be toponium state
through strong (e.g. $t\bar{t}\to ggg$) or electroweak
(e.g. $t\bar{t}\to l^+l^-$) interactions.

The eigenstates of a non-Hermitian operator do not form an orthogonal
basis of the Hilbert space~\cite{Sternheim:1972zz}. This implies, that the
completeness relation must be modified. We consider the sets of
right and left
eigenstates\footnote{We do not distinguish between bound states and
continuum states, since this is irrelevant for the discussion.}
\begin{equation}
 H \Ket{n} = \mathcal{E}_n \Ket{n}, \hspace{2cm} H^\dagger \Ket{\tilde{m}} =
\tilde{\mathcal{E}}_n \Ket{\tilde{m}}
\end{equation}
with $\tilde{\mathcal{E}}_n = \mathcal{E}_n^*$. Assuming that the Hamiltonian
transforms as
\begin{equation}
 T H T^{-1} = H^\dagger
 \label{eq:Ttrafo}
\end{equation}
under time reversal, and that the eigenvalues are non-degenerate,
the eigenstates can be normalized such that they form a bi-orthogonal
set~\cite{Sternheim:1972zz}
\begin{equation}
 \Braket{\tilde{m}|n}=\delta_{mn},
 \label{eq:biorthogonal_set}
\end{equation}
which implies that the completeness
relation takes the form
\begin{equation}
 \mathbf{1} = \sum\limits_n \Ket{n}\Bra{\tilde{n}}.
 \label{eq:completeness_relation}
\end{equation}
The property~\eqref{eq:Ttrafo} implies that the state $\Ket{\tilde{n}}$ is
exponentially growing at the same rate as $\Ket{n}$ is decaying, which
facilitates the normalization~\eqref{eq:biorthogonal_set}. After
applying~\eqref{eq:completeness_relation} the Green function takes the
following form near the poles
\begin{equation}
G(E) \,\, \mathop{=}\limits^{E\to \mathcal{E}_n} \,\,
\frac{\psi_n(\mathbf{0})\psi_{\tilde{n}}^*(\mathbf{0})}
{\mathcal{E}_n-E}+\text{regular},
\label{eq:GF_poles_exact}
\end{equation}
which generalizes the expression for the QCD result~\cite{Beneke:2013PartII}.
The pole position and residue of \eqref{eq:GF_poles_exact} have the following
perturbative expansion
\begin{eqnarray}
\mathcal{E}_n & = & \sum\limits_{k=0}^\infty \mathcal{E}_n^{(k)},\\
\psi_n(\mathbf{0})\psi_{\tilde{n}}^*(\mathbf{0})
& = & \left|\psi_n^{(0)}(\mathbf{0})\right|^2
\left(1+\sum\limits_{k=1}^\infty F_n^{(k)}\right),
\end{eqnarray}
with the LO expressions
\begin{eqnarray}
\mathcal{E}_n^{(0)} & = & E_n^{(0)} - i\Gamma
= -\frac{m_t\alpha_s^2C_F^2}{4n^2} -i\Gamma,\\
|\psi_n^{(0)}(\mathbf{0})|^2 & = &
\frac{1}{\pi}\left(\frac{m_t\alpha_sC_F}{2n}\right)^3,
\end{eqnarray}
and $\mathcal{E}_n^{(k)}=E_n^{(k)}-i\delta\Gamma_n^{(k)}/2$. At LO we have
made use of the relation $\psi_{\tilde{n}}^{(0)}(\mathbf{x})
=\psi_n^{(0)}(\mathbf{x})$. This holds, because the non-Hermitian part of
the LO Hamiltonian
\begin{equation}
 (H_0-H_0^\dagger)/2 = -i\Gamma
\end{equation}
is proportional to the identity operator and thus only affects the
eigenvalues $\mathcal{E}_n^{(0)}$ but not the eigenstates $\Ket{n}^{(0)}$.

The results for the non-relativistic Green function in perturbation theory
do not take the form~\eqref{eq:GF_poles_exact}, but contain higher-order poles
\begin{eqnarray}
G(E) \,\, \mathop{=}\limits^{E\to \mathcal{E}_n^{(0)}} & &
\frac{|\psi_n^{(0)}(\mathbf{0})|^2}{\mathcal{E}_n^{(0)}-E}
\Bigg[1+\left(F_n^{(1)}
-\frac{\mathcal{E}_n^{(1)}}{\mathcal{E}_n^{(0)}-E}\right)
\nonumber \\
& &+\left(F_n^{(2)}-\frac{\mathcal{E}_n^{(2)}+F_n^{(1)}\mathcal{E}_n^{(1)}}{\mathcal{E}_n^{(0)}-E}+\frac{\mathcal{E}_n^{(1)\,2}}{(\mathcal{E}_n^{(0)}-E)^2}\right)+\dots\Bigg]+\text{regular}.
 \label{eq:GF_poles_expanded}
\end{eqnarray}
This allows us to read off the NNLO correction to the bound state parameters
from the contribution~\eqref{eq:delta2GammaG} to the Green function
\begin{eqnarray}
\label{eq:delGammaE2}
\delta_\Gamma E_n^{(2)} & = & \frac{\Gamma^2}{4m_t},\\
\label{eq:delGamma2}
\delta_\Gamma\Gamma_n^{(2)} & = & -\frac{\Gamma\alpha_s^2C_F^2}{4n^2},\\
\label{eq:delGammaF2}
\delta_\Gamma F_n^{(2)} & = & -\frac{3i\Gamma}{2m_t}.
\end{eqnarray}
As discussed above, the $\Gamma^2$ term in \eqref{eq:Lbilinear} has the form
of a mass shift and therefore leads to an $n$-independent
correction~\eqref{eq:delGammaE2} to the position of the pole, while it does
not affect the residue. The $i\Gamma\partial^2$ term in
\eqref{eq:Lbilinear} accounts for time dilatation, which reduces the total
width of the would-be toponium resonance by~\eqref{eq:delGamma2}.
Since it is non-Hermitian, it also makes the residues complex
due to~\eqref{eq:delGammaF2}.

The corrections to the bound states from QCD effects as well as the
QED Coulomb and Higgs potentials can be found in~\cite{Beneke:2013PartII}
and~\cite{Beneke:2015lwa}, respectively. Using this input we can resum
the higher-order poles in the expanded Green function by the replacement
\begin{equation}
G(E) \to G(E) + \sum\limits_n \left[
\frac{\psi_n(\mathbf{0})\psi_{\tilde{n}}^*(\mathbf{0})}{\mathcal{E}_n-E}
-\left(\left.\frac{\psi_n(\mathbf{0})\psi_{\tilde{n}}^*(\mathbf{0})}
{\mathcal{E}_n-E}\right|_\text{expanded}\right)\right],
\end{equation}
where the expanded term has the form~\eqref{eq:GF_poles_expanded}.
In the actual implementation \cite{Beneke:2016kkb}
we apply the pole resummation procedure
to $G(E)$ alone in the electroweak contributions, but to the
entire current correlation function of vector and axial vector
currents in the pure QCD contributions.

\subsubsection{Mixed QCD-electroweak NNLO corrections in PNREFT
\label{sec:resPotnew}}

\begin{figure}
 \begin{center}
  \includegraphics[height=2.4cm]{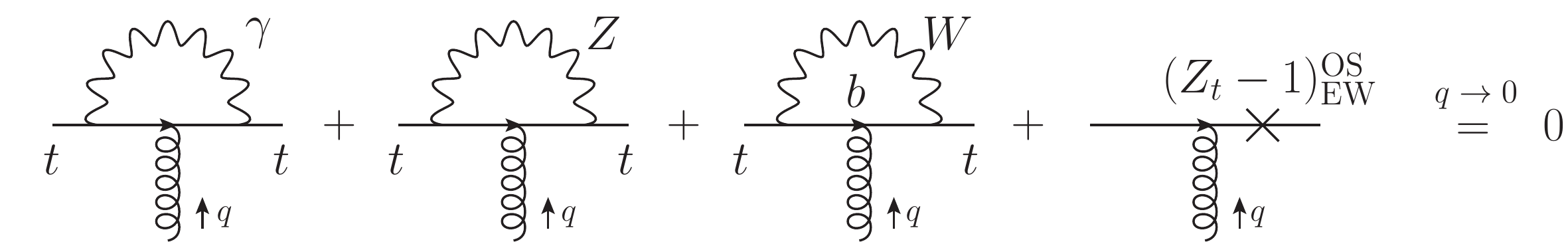}\\[0.4cm]
  \includegraphics[height=2.4cm]{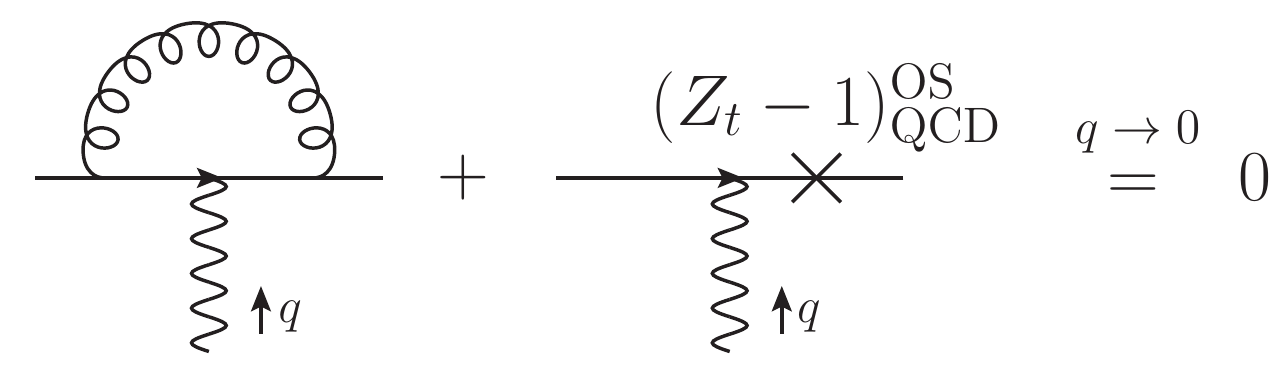}
\caption{\label{fig:cancellation}Cancellation of the electroweak
$gt\bar t$ vertex correction and the QCD $\gamma t\bar{t}$ vertex
correction in the on-shell scheme. The vector current
itself is conserved and therefore not renormalized.}
\end{center}
\end{figure}

In addition to the kinetic terms \eqref{eq:Lbilinear} the PNREFT
Lagrangian contains potential interactions. We now show that there are
no mixed QCD-electroweak corrections at NNLO.
The construction of PNREFT proceeds by
first integrating out fluctuations at the hard scale, which yields
NREFT, and then integrating out fluctuations at the soft scale.
In the first step, one must consider electroweak corrections to the
hard matching coefficients of the QCD vertex, and vice versa.
The relevant diagrams are shown in Figure~\ref{fig:cancellation}.
The loop momenta in Figure~\ref{fig:cancellation} are hard, while
the momenta of the external particles can be either soft, potential
or ultrasoft and must be expanded out of the loop integral.
Therefore the vertices are effectively evaluated at zero
external momentum, and the corresponding contribution is exactly
cancelled by the on-shell external field renormalization factor.

%%%%%%%%%%%%%%%%%%%%%%%%%%%%%%%%%%%%%%%%%%%%%%%%%%%%%%%%%%%%%%%%%%%%%%%%
%%%%%%%%%%%%%%%%%%%%%%%%%%%%%%%%%%%%%%%%%%%%%%%%%%%%%%%%%%%%%%%%%%%%%%%%

\begin{figure}
\begin{center}
\includegraphics[width=0.75\textwidth]
{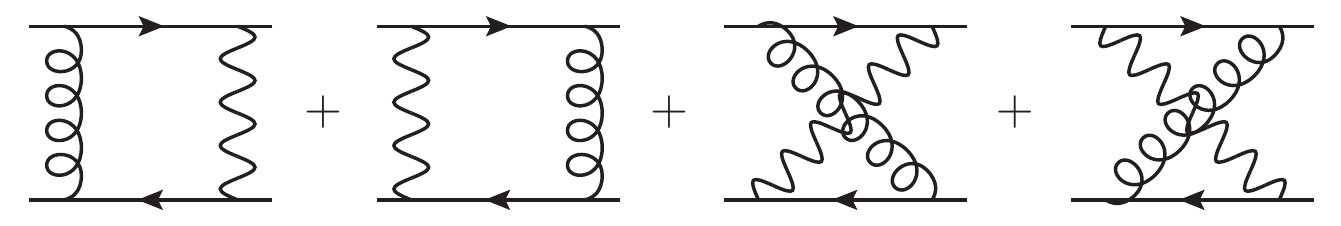}
\caption{\label{fig:mixed_QCD_QED_potential}
Contributions to the $1/\mathbf{q}^2$ potential at order
$\alpha_s\alpha$.}
\end{center}
\end{figure}

The potentials are determined in the matching procedure between
NREFT and PNREFT. The diagrams that contribute to the $1/\mathbf{q}^2$
potential at order $\alpha_s\alpha$ are shown in
Figure~\ref{fig:mixed_QCD_QED_potential}, where the momenta of the
external top quarks are potential and the loop momentum is soft.
The contributions of the first and second diagram are identical,
and are exactly opposite to those of the third and fourth diagram,
which implies that the sum of the diagrams in Figure~\ref{fig:mixed_QCD_QED_potential}
vanishes. We have not drawn the four diagrams that involve soft
vertex corrections to the tree-level potentials, because these
corrections are scaleless and vanish in dimensional regularization.
Last, but not least there are no contributions from insertions
of the one-loop corrections to the hard matching coefficients
in the tree-level potential, because these coefficients vanish
as argued above. We conclude that no mixed QCD-electroweak
potentials appear at NNLO.

%%%%%%%%%%%%%%%%%%%%%%%%%%%%%%%%%%%%%%%%%%%%%%%%%%%%%%%%%%%%%%%%%%%%%%%%
%%%%%%%%%%%%%%%%%%%%%%%%%%%%%%%%%%%%%%%%%%%%%%%%%%%%%%%%%%%%%%%%%%%%%%%%

\begin{figure}
 \begin{center}
  \includegraphics[width=0.4\textwidth]{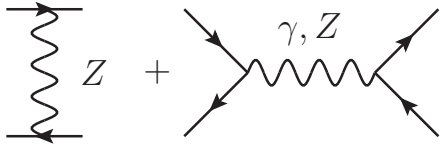}
\caption{\label{fig:localEWpotentials}
Contributions to the $1/m_t^2$ potential at order $\alpha_{\rm EW}$.}
\end{center}
\end{figure}

Furthermore, we demonstrate that the potential does not receive any
pure electroweak NNLO corrections from the exchange of Z bosons.
We count the mass of the Z bosons as hard and therefore have to
integrate out the Z boson in the hard matching to NREFT. This
implies, that all interactions that are mediated by the Z boson
in the full theory become local in PNREFT. The Z-boson exchange
potential corresponds to the full-theory diagrams shown in
Figure~\ref{fig:localEWpotentials} and is proportional to
$\alpha_{\rm EW}/m_t^2$ in momentum space. Thus, it is suppressed
by $(\alpha_{\rm EW}/\alpha_s)\times(\mathbf{q}^2/m_t^2)\sim v^3$ compared
to the LO colour Coulomb potential and only contributes at NNNLO.

%%%%%%%%%%%%%%%%%%%%%%%%%%%%%%%%%%%%%%%%%%%%%%%%%%%%%%%%%%%%%%%%%%%%%%%%
%%%%%%%%%%%%%%%%%%%%%%%%%%%%%%%%%%%%%%%%%%%%%%%%%%%%%%%%%%%%%%%%%%%%%%%%

Finally, we comment on the so-called `jet-jet' interactions
that were considered in~\cite{Melnikov:1993np}. These are
corrections involving gluon emission from the final-state
bottom quarks and it was demonstrated in~\cite{Melnikov:1993np}
that they vanish at NLO. In their calculation, the authors of
\cite{Melnikov:1993np} first consider the subgraph $I_\mu$,
which corresponds to the cut to the right of the gluon of the third diagram in
Figure~\ref{fig:cancellation}. Their result for $I_\mu$ scales as
$\sqrt{\alpha_s}\,\Gamma_0/|\mathbf{k}|$, where $\mathbf{k}$ is
the gluon three-momentum, which is either potential or ultrasoft in their
case. They then show by explicit computation
that all NLO corrections that involve the subgraph $I_\mu$ and/or
its CP conjugate $J_\mu$ vanish. In our approach, where loop integrals
are strictly expanded according to the scaling of the
momentum regions, the only non-vanishing contribution to the subgraph
$I_\mu$ comes from the region of hard loop momentum, where no
inverse powers of $|\mathbf{k}|$ can appear because the external
momenta are expanded out. Therefore, the absence of any `jet-jet'
interactions at NLO is a matter of simple power counting, which
implies that corrections can first appear at the relative order
$\alpha_{\rm EW}$, i.e. at NNLO. We have already proven that
there are also no `jet-jet' interactions at NNLO by demonstrating
that electroweak corrections to the QCD vertex in
Figure~\ref{fig:cancellation} vanish.

\subsubsection{\boldmath Absorptive part from field renormalization,
$\sigma_{C^{(k)}_{\text{Abs,}Z_t}}$
\label{sec:resCZt}}

The hard matching coefficients $C^{(k)}$ become complex at NNLO
due to $bW^+$ loop corrections. The
imaginary part contributes to the finite-width divergence of the resonant
cross section $\sigma_\text{res, rest}$ in (\ref{eq:resrest}) and, thus,
it has to be determined in $d$ dimensions in accordance with the scheme
used to evaluate the other components of part~(I). The bare absorptive
contribution to $C^{(k)}$ on the other hand, is contained in part
(II) and therefore has to be computed in a different scheme.
Since the two parts are treated using different conventions, we find it
convenient to separate them in notation. The matching coefficients
up to NNLO are expanded as
\begin{equation}
C^{(k)}=C_0^{(k)}\left[1+c_v^{(1)}\left(\frac{\alpha_s}{4\pi}\right)
 +c_v^{(2)}\left(\frac{\alpha_s}{4\pi}\right)^2+\frac{y_t^2}{2}c_{vH}^{(2)}+\dots\right]
 +\left(C_{\text{EW}}^{(k)}+iC_{\text{Abs}}^{(k)}\right)
\frac{\alpha}{4\pi}+\dots,
 \label{eq:Ckfull}
\end{equation}
\begin{equation}
 C_{\text{Abs}}^{(k)}=C_{\text{Abs}, Z_t}^{(k)}+C_{\text{Abs,bare}}^{(k)}.
 \label{eq:splitCabs}
\end{equation}
The real part of the electroweak corrections, $C_{\text{EW}}^{(k)}$, does not
yield a finite-width divergence at NNLO. Thus, it is not necessary to split
it as well. The absorptive part~\eqref{eq:splitCabs} of the hard matching
coefficient is available in four dimensions~\cite{Hoang:2004tg}.
We have repeated the calculation of the individual contributions in the
schemes described above. In four dimensions we reproduce the
result of~\cite{Hoang:2004tg},
\begin{equation}
 C_{\text{Abs}}^{(v,a)} = \frac{4m_t^2}{\alpha^2}
\left(C_\text{V,A}^\text{bW,abs}\right)_\text{from \cite{Hoang:2004tg}}
+ \mathcal{O}(\epsilon),
\end{equation}
where the normalization factor is necessary because the definition of the
hard matching coefficients in~\cite{Hoang:2004tg} differs
from~\eqref{eq:Ckfull}.

The bare part $C_{\text{Abs,bare}}^{(k)}$ will be given in
Section~\ref{sec:resAbs}. From the field renormalization, we obtain
\begin{eqnarray}
 C_{\text{Abs}, Z_t}^{(v)} & = & \frac{\pi\Gamma_0^{(d)}}
{m_t\alpha s_w^2 (4 c_w^2-x_W)(1 - x_W) (1 + 2 x_W (1 - \epsilon))}
\nonumber \\
 & &\times\Big[(1+4e_e s_w^2)
(2-\epsilon+x_W(2-5\epsilon+2\epsilon^2)+2x_W^2(1-\epsilon)^2)
\nonumber\\
 & & -2s_w^2e_t(1+e_e(4-x_W))(3-2\epsilon)(1+x_W(1-2\epsilon)+2x_W^2(1-\epsilon))
\Big],
\label{eq:CVabs}
\\
 C_{\text{Abs}, Z_t}^{(a)} & = & \frac{-\pi\Gamma_0^{(d)}}
{m_t\alpha s_w^2 (4 c_w^2-x_W)(1 - x_W) (1 + 2 x_W (1 - \epsilon))}
\,\Big[2-\epsilon+x_W(2-5\epsilon+2\epsilon^2) \nonumber\\
 & & +2x_W^2(1-\epsilon)^2-2 e_t s_w^2(3-2\epsilon)(1+x_W(1-2\epsilon)
+2x_W^2(1-\epsilon))\Big].
 \label{eq:CAabs}
\end{eqnarray}
The contribution to the NNLO cross section is given by
\begin{equation}
 \sigma_{C^{(k)}_{\text{Abs,}Z_t}}=\sigma_0\,\frac{12\alpha N_c}{s}
\,\Big[C_0^{(v)}C_{\text{Abs}, Z_t}^{(v)}+
C_0^{(a)}C_{\text{Abs}, Z_t}^{(a)}\Big]\,
\text{Re}\left[G_0^{(w)}(E + i\Gamma)\right],
\label{eq:sigmaCabsZt}
\end{equation}
where the finite terms from the multiplication of the $1/\epsilon$
divergence in the real part of the Green function~\eqref{eq:G0} with
the $\mathcal{O}(\epsilon)$ parts of \eqref{eq:CVabs} and
\eqref{eq:CAabs} are included.

\subsubsection{\boldmath
Electroweak contributions to the hard matching coefficient,
$\sigma_{C_{\text{EW}}^{(k)}}$
\label{sec:resCV}}

The real part of the electroweak contributions to the NNLO matching
coefficients $C^{(k)}$ has been computed
in~\cite{Grzadkowski:1986pm,Guth:1991ab,Hoang:2006pd}. Pure QED corrections
have been neglected there. Therefore, we split
\begin{equation}
 C_{\text{EW}}^{(k)} = C_{\text{QED}}^{(k)} + C_{\text{WZ}}^{(k)}.
 \label{eq:CEW}
\end{equation}
The hard QED vertex correction to the $\gamma e^+e^-$ and $Ze^+e^-$
vertices contains divergences that cancel among initial-state
radiation (ISR) contributions (see Section~\ref{sec:resRad}).
We therefore assign it to $\sigma_\text{IS}^{\rm conv}$ to render both,
$\sigma_{C_{\text{EW}}^{(k)}}$ and $\sigma_\text{IS}^{\rm conv}$, separately
finite.
There is no contribution from the box diagram involving two photons,
since only its interference with the production of the top pair
through the vector component of the $s$-channel $\gamma$ or $Z$ boson
is of NNLO and the correlator of three vector currents vanishes~\cite{
Furry:1937zz}. The box diagram with a photon and $Z$ boson is considered
to be a non-QED correction to the photon-exchange contribution and is
therefore already part of $C_{\text{WZ}}^{(k)}$.
Thus, the only pure QED effects are the hard photon vertex correction
to the $\gamma t\bar{t}$ and $Z t\bar{t}$ currents and the photon
self energy, which yield
\begin{eqnarray}
 C_{\text{QED}}^{(v)} & = & -8e_t^2C_0^{(v)}
- \frac{4\pi\,e_ee_t}{\alpha\,s}\,\Pi_\text{R}^\text{AA}(s,\mu_\alpha^2),
\nonumber\\
 C_{\text{QED}}^{(a)} & = & -8e_t^2C_0^{(a)}.
 \label{eq:CQED}
\end{eqnarray}
As in~\cite{Beneke:2014pta}, the renormalized photon self-energy
$\Pi_\text{R}^\text{AA}(s,\mu_\alpha^2)$ is defined in the scheme
of~\cite{Jegerlehner:2011mw}, and will be discussed below.
The non-QED contributions are
\begin{eqnarray}
 C_{\text{WZ}}^{(v)} & = &
\frac{4m_t^2}{\alpha_0^2}C_{V}^{\text{ew}}(\nu=1)
-C_0^{(v)}\,\frac{4\pi}{\alpha}\,\frac{y_t^2}{2}\,c_{vH}^{(2)}
+ \frac{\pi\,e_ee_t}{\alpha_0\,m_t^2}\,\Pi_\text{R}^\text{AA}(4m_t^2,0),
\nonumber\\
 C_{\text{WZ}}^{(a)} & = &
\frac{4m_t^2}{\alpha_0^2}C_{A}^{\text{ew}}(\nu=1)
-C_0^{(a)}\,\frac{4\pi}{\alpha}\,\frac{y_t^2}{2}\,c_{vH}^{(2)},
\label{eq:CWZ}
\end{eqnarray}
where $C_{V,A}^{\text{ew}}(\nu=1)$ is given in~\cite{Hoang:2006pd},
$\alpha_0$ is the fine-structure constant and
$\Pi_\text{R}^\text{AA}(4m_t^2,0)$ coincides with the expression for
$\Pi_\text{R}^\text{AA}$ from~\cite{Hoang:2006pd}.
The subtraction terms are present because Higgs effects which only
involve the top Yukawa coupling have already been included separately
as part of $\sigma_H$ in~\cite{Beneke:2015lwa} and the photon self-energy
is contained in the QED contribution~\eqref{eq:CQED}. Corrections that
involve Higgs couplings to gauge bosons or Goldstone bosons remain
in~\eqref{eq:CWZ}.

We note that the photon self-energy terms in~\eqref{eq:CQED}
and~\eqref{eq:CWZ} differ, because we use a renormalization scheme which is
different from~\cite{Grzadkowski:1986pm,Guth:1991ab,Hoang:2006pd}.
The matching coefficients given
in~\cite{Grzadkowski:1986pm,Guth:1991ab,Hoang:2006pd} are expressed in terms
of the fine-structure constant $\alpha_0$. This scheme suffers from a
large spurious dependence on the light fermion masses, that cancels
explicitly with the self-energy corrections to the matching coefficients
when the fine-structure constant is expressed in terms of a less
 infrared-dependent definition of the electroweak coupling constant.
Therefore, we write the cross section in terms of the running on-shell
coupling $\alpha\equiv\alpha(\mu_\alpha)$ from~\cite{Jegerlehner:2011mw}.
In this scheme the renormalized photon self-energy takes the form
\begin{equation}
\Pi_\text{R}^\text{AA}(s,\mu_\alpha^2) = \left.\Pi^\text{AA}(s)
- \frac{s}{\mu_\alpha^2}\Pi^\text{AA}(\mu_\alpha^2)\right|_{
\alpha_0\to\alpha},
\end{equation}
where the bare self energy $\Pi^\text{AA}$ is taken
from~\cite{Grzadkowski:1986pm}. The explicit factor $1/\mu_\alpha^2$ appears,
because \cite{Grzadkowski:1986pm} defines the photon vacuum polarization
$\Pi^{\text{AA}}(s)$ as a dimensionful quantity and does not imply a
power-dependence of the cross section on the scale $\mu_\alpha$. In the
limits $\mu_\alpha\to0$ and $s\to4m_t^2$ the scheme
of~\cite{Jegerlehner:2011mw} converges to the scheme
of~\cite{Grzadkowski:1986pm,Guth:1991ab,Hoang:2006pd}, i.e.
$\alpha\to\alpha_0$ and $\Pi^\text{AA}(\mu_\alpha^2)/\mu_\alpha^2\to
\Pi^{\prime,\text{AA}}(0)$, and the self-energy terms in~\eqref{eq:CQED}
and~\eqref{eq:CWZ} coincide with each
other and with the respective expression in~\cite{Hoang:2006pd}.

The cross section receives the contribution
\begin{equation}
\sigma_{C_{\text{EW}}^{(k)}}=\sigma_0\,\frac{12\alpha N_c}{s}\,\Big[
C_0^{(v)}C_{\text{EW}}^{(v)}+C_0^{(a)}C_{\text{EW}}^{(a)}\Big]\,
\text{Im}\left[G_0(E + i\Gamma)\right],
\label{eq:CkEW}
\end{equation}
from the electroweak corrections to the hard matching coefficients
of the production operators.

\subsubsection{\boldmath Initial-state radiation,
$\sigma_{\text{IS}}^\text{conv}$
\label{sec:resRad}}

In this section we take into account effects from QED initial-state
radiation. As such we count all corrections that involve an additional
photon attached only to the external $e^\pm$ states relative to the
LO cross section.
ISR was already considered in the 1980s~\cite{Fadin:1987wz} at the
leading logarithmic order, i.e. summing corrections of the form
$(\alpha\ln(s/m_e^2))^k$ to all orders, whereas later works
concentrated on the `partonic' $t\bar t$ cross section.
We extend the treatment of ISR to NNLO$+$LL accuracy below. The
non-resonant part is only affected by QED radiation at NNNLO
and will not be considered. With the exception of the effects from the
hard momentum region, all  contributions are universal and our treatment
closely follows the one for $W$ pair production near threshold
in~\cite{Beneke:2007zg,Actis:2008rb}. In fact, the equations in this
section can often be obtained directly from those in~\cite{Actis:2008rb}
by substituting $c_{p,LR}^{(1,\rm fin)} \to -4+\pi^2/12$, and by adapting
the different tree-level process.\footnote{
Compared to our results the expressions in~\cite{Beneke:2007zg,Actis:2008rb}
contain an extra factor $(1-\epsilon)$ from the $d$-dimensional spin sum
over the initial state which we treat in $d=4$ dimensions as described
at the beginning of Section~\ref{sec:PartI}.}

When the electron mass is neglected the ISR contribution
involves the hard, $k\sim m_t$, and ultrasoft, $k\sim m_t v^2$, momentum
regions, and in addition two hard-collinear momentum regions,
$\bar{n}_i \cdot k\sim m_t, n_i\cdot k \sim m_t v^2, k_{i\perp} \sim m_t v$
($i=1,2$) familiar from SCET, where $n_i,\bar{n}_i$ are pairs of
light-like vectors with $n_i\cdot \bar{n}_i=2$
defined by the electron ($i=1$) and positron ($i=2$) momentum.
Real collinear emission is kinematically forbidden in the resonant
part, because it carries away a hard momentum fraction, which pushes
the top pair off-shell. Virtual collinear corrections are scaleless.
Hence, the hard-collinear regions vanish~\cite{Beneke:2007zg}, and we are
left with the hard and ultrasoft contributions. We evaluate these separately.
A hard photon cannot be
exchanged between the incoming and outgoing electrons, since this
would also push the top pair off-shell. Thus the only correction from
the hard region is the QED $\gamma/Zee$ vertex correction which
contributes to the hard matching coefficients $C^{(v,a)}$. We find
\begin{eqnarray}
C_{\gamma/Zee}^{(v,a)} &=& \text{Re}\left[C_0^{(v,a)}
\frac{\alpha}{4\pi}\left(\frac{\mu^2}{-4m_t^2-i0}\right)^\epsilon
 \left(-\frac{2}{\epsilon^2}-\frac{3}{\epsilon}-8+\frac{\pi^2}{6}
\right)\right] \\
&&\hspace*{-1cm}
= \,-C_0^{(v,a)}\frac{\alpha}{4\pi} \left(\frac{2}{\epsilon^2}
+\frac{1}{\epsilon}\left(3+2\ln\frac{\mu^2}{4m_t^2}\right)
+\ln^2\frac{\mu^2}{4m_t^2}+3\ln\frac{\mu^2}{4m_t^2}
+8-\frac{7\pi^2}{6}\right).\quad\nonumber
\end{eqnarray}
As it should be, this agrees with the QCD analogue of the hard matching 
coefficient of the vector current to SCET, first obtained in this context 
in~\cite{Manohar:2003vb} in DIS kinematics.
We only kept the real part, because the imaginary part comes from cuts that
do not correspond to the final state $b\bar{b}W^+W^-$. The correction to
the cross section from hard ISR is
\begin{equation}
\sigma_{\text{IS}}^{(\text{H})}=\sigma_0\,
\frac{48\pi N_c}{s}\,[C_0^{(v)}C_{\gamma/Zee}^{(v)}
+C_0^{(a)}C_{\gamma/Zee}^{(a)}]\,
\text{Im}\left[G_0(E + i\Gamma)\right].
\label{eq:sigmaISh}
\end{equation}

\begin{figure}
 \begin{center}
    \includegraphics[width=0.42\textwidth]{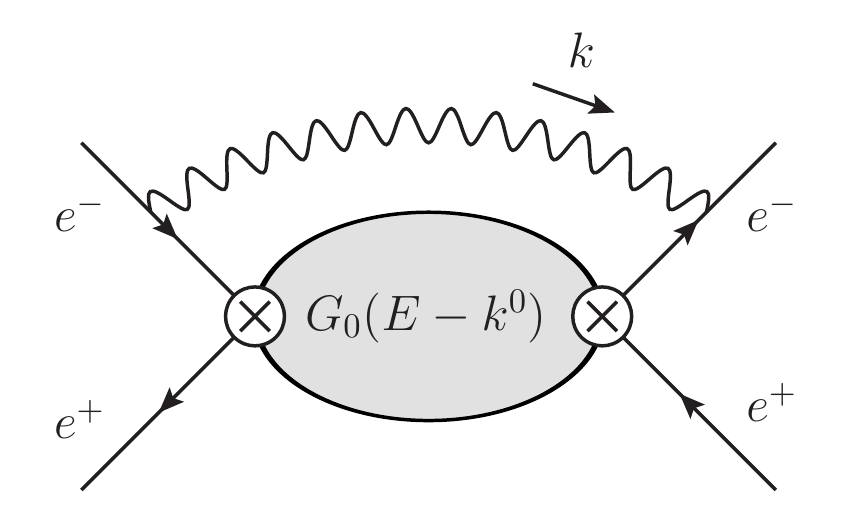}
    \includegraphics[width=0.42\textwidth]{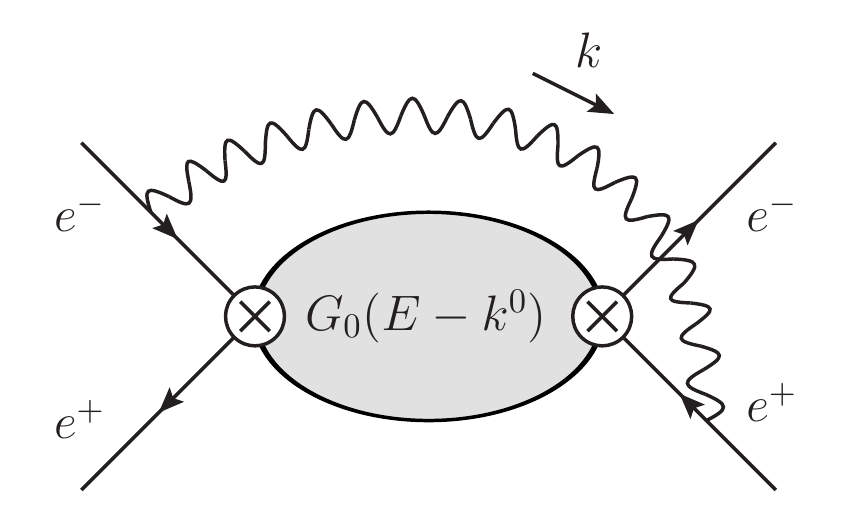}
  \caption{\label{fig:Res_usoft}Ultrasoft photon corrections to the
  resonant cross section. The symmetric diagrams, obtained by the
  interchange of electrons and positrons, are not shown.}
 \end{center}
\end{figure}

The contributions from the ultrasoft momentum region are shown in
Figure~\ref{fig:Res_usoft}. Virtual ultrasoft corrections are scaleless.
The diagram with the photon attached to incoming and outgoing electron
vanishes, because it is proportional to the square of the light-like
direction $n_1$ of the electron beam.
No ultrasoft corrections that couple to the collinear and
non-relativistic sector occur at NNLO, because the leading ultrasoft
photon coupling to the final state vanishes, as discussed in
Section~\ref{sec:resGF}. Thus, the contribution to the cross section
from the ultrasoft region is due to the right diagram in
Figure~\ref{fig:Res_usoft} and reads
\begin{equation}
\sigma_{\text{IS}}^{(\text{US})}=\sigma_0\,\frac{24\pi N_c}{s}\,
\Big[C_0^{(v)^2}+C_0^{(a)^2}\Big]\,
\frac{\alpha}{4\pi}\frac{-8\sqrt{\pi}}{\epsilon\Gamma(1/2-\epsilon)}
\left(\mu^2e^{\gamma_E}\right)^\epsilon\,
 \text{Im}\left[\,\int\limits_0^\infty dk\frac{G_0(E + i\Gamma-k)}
{k^{1+2\epsilon}}\right].
 \label{eq:sigmaISus}
\end{equation}
When the small electron mass is neglected, the photon radiation
corrections are given by the sum of \eqref{eq:sigmaISh} and
\eqref{eq:sigmaISus}.
We observe that the $1/\epsilon^2$ pole cancels, but a collinear
divergence remains, because the cross section is not infrared safe
for $m_e=0$. Subtracting this divergence defines a scheme-dependent
`partonic' cross section.

The divergence is regularized by the non-zero electron
mass, which in turn yields large logarithms $\ln(s/m_e^2)$. They can
be resummed into an electron distribution function $\Gamma_{ee}^\text{LL}(x)$,
which describes the probability of finding an electron with momentum $xp$
in the ``parent electron'' with momentum $p$. The cross section with
resummed ISR from the electron and positron is given by
\begin{equation}
\sigma_\text{ISR}(s)=\int\limits_0^1dx_1\int\limits_0^1dx_2
\,\Gamma_{ee}^\text{LL}(x_1)\Gamma_{ee}^\text{LL}(x_2)
\,\sigma^\text{conv}(x_1x_2s).
\label{eq:convolutionstructurefunctions}
\end{equation}
Expressions for the structure function at leading-logarithmic (LL) accuracy
can be found in~\cite{Kuraev:1985hb,Skrzypek:1992vk,Beenakker:1994vn,Beenakker:1996kt}, where LL implies that all terms of the form
$\alpha^n\ln^n(s/m_e^2)$ are summed to all orders. The resummation of the
next-to-leading logarithms (NLL) $\alpha^{n+1}\ln^n(s/m_e^2)$
is crucial for the precision program at a future lepton collider, but the
structure functions are presently unknown at this order.

At LO, the `partonic' cross section $\sigma^\text{conv}(s)$ is
given by~\eqref{eq:sigmaLO}. At higher orders it depends
on the scheme used to regularize and subtract the collinear divergence.
The scheme dependence cancels in the convolution with the structure functions.
This implies that we have to adapt the results \eqref{eq:sigmaISh} and
\eqref{eq:sigmaISus}, which correspond to a minimal subtraction scheme,
to the conventional scheme in which the structure functions
$\Gamma_{ee}^\text{LL}(x)$ are defined. This procedure has been described in
detail in~\cite{Beneke:2007zg,Actis:2008rb}. First, we need to convert the
dimensional regulator of the collinear divergences to a finite electron
mass regulator.
Then, the $\mathcal{O}(\alpha)$ terms that appear in the convolution of
the structure functions with the LO cross section have to be subtracted
from the fixed order NNLO partonic cross section to avoid double counting.

The first point is accomplished by noting that the presence of the
additional scale $m_e\ll m_t v^2$ introduces the
additional momentum regions
\begin{equation}
\begin{array}{llll}
\text{hard-collinear:}&\hspace{0.5cm}\bar{n}_i\cdot k \sim m_t,
&\hspace{0.5cm} \displaystyle
n_i\cdot k \sim \frac{m_e^2}{m_t},
&\hspace{0.5cm} k_{i\perp} \sim m_e,
\\[0.4cm]
\text{soft-collinear:}&\hspace{0.5cm}\bar{n}_i\cdot k \sim m_tv^2,
&\hspace{0.5cm} \displaystyle
n_i\cdot k \sim \frac{m_e^2 v^2}{m_t},
&\hspace{0.5cm} k_{i\perp} \sim m_e v^2,
\end{array}
\end{equation}
with $k^2\sim m_e^2$ and $k^2\sim m_e^2 v^4$, respectively.
The soft-collinear region contributes in the diagrams shown in
Figure~\ref{fig:Res_usoft}. As before, the left diagram vanishes,
and one finds
\begin{equation}
\sigma_{\text{IS}}^{(\text{SC})}=\sigma_0\,\frac{24\pi N_c}{s}\,\Big[
C_0^{(v)^2}+C_0^{(a)^2}\Big]\,\frac{\alpha}{4\pi}8\Gamma(\epsilon)\,
\left(\frac{m_t^2}{m_e^2}\right)^\epsilon
\left(\mu^2e^{\gamma_E}\right)^\epsilon\,
\text{Im}\left[\,\int\limits_0^\infty dk
\frac{G_0(E + i\Gamma-k)}{k^{1+2\epsilon}}\right].
\label{eq:sigmaISsc}
\end{equation}
The hard-collinear contribution comes from $\gamma/Zee$ vertex correction
diagram and gives
\begin{eqnarray}
\sigma_{\text{IS}}^{(\text{HC})} & = & \sigma_0\,
\frac{24\pi N_c}{s}\,\Big[C_0^{(v)^2}+C_0^{(a)^2}\Big]\,
\text{Im}\left[G_0(E + i\Gamma)\right] \nonumber\\
& & \times\frac{\alpha}{4\pi}\left[\frac{4}{\epsilon^2}
+\frac{1}{\epsilon}\left(6+4\ln\frac{\mu^2}{m_e^2}\right)
+2\ln^2\frac{\mu^2}{m_e^2}+6\ln\frac{\mu^2}{m_e^2}+\frac{\pi^2}{3}+12\right].
\label{eq:sigmaIShc}
\end{eqnarray}
The collinear $1/\epsilon$ poles cancel in the sum of the
hard and hard-collinear, and ultrasoft and soft-collinear
contributions, separately. The collinear sensitivity is
instead expressed through the large logarithms $\ln(4m_t^2/m_e^2)$.
The remaining singularities cancel in the sum over all regions.
To make the cancellation explicit, one can expand the factor
$1/k^{1+2\epsilon}$ in the distribution sense:
\begin{equation}
\frac{\mu^{2\epsilon}}{k^{1+2\epsilon}}=
-\frac{(a/\mu)^{-2\epsilon}}{2\epsilon}\delta(k)
+\frac{1}{\left[k\right]_{a+}}+\mathcal{O}(\epsilon),
\end{equation}
where $a>0$ is arbitrary and we have introduced the modified
plus-distribution
\begin{equation}
\int\limits_0^\infty dk\frac{f(k)}{\left[k\right]_{a_+}}
=\int\limits_0^\infty dk\frac{f(k)-f(0)\theta(a-k)}{k}.
\end{equation}
We obtain
\begin{eqnarray}
\sigma_{\text{IS}} & = & \sigma_{\text{IS}}^{(\text{H})}
 +\sigma_{\text{IS}}^{(\text{HC})}+\sigma_{\text{IS}}^{(\text{US})}
 +\sigma_{\text{IS}}^{(\text{SC})} \\
 & = & \sigma_0\,\frac{24\pi N_c}{s}\,\Big[C_0^{(v)^2}+C_0^{(a)^2}\Big]\,
\frac{\alpha}{4\pi}\,\Bigg\{8\ln\left(\frac{4 m_t^2}{m_e^2}\right)
\int\limits_0^\infty dk\frac{\text{Im}\left[G_0(E + i\Gamma-k)\right]}
{\left[k\right]_{a+}} \nonumber \\
 & & +\left[\frac{4\pi^2}{3}-4+6\ln\left(\frac{4 m_t^2}{m_e^2}\right)
+4\ln\left(\frac{a^2}{m_t^2}\right)\ln\left(\frac{4 m_t^2}{m_e^2}\right)
\right]\text{Im}\left[G_0(E + i\Gamma)\right]\Bigg\}, \nonumber
 \label{eq:sigmaIS}
\end{eqnarray}
which is finite, such that the four-dimensional expression~\eqref{eq:G0}
for the LO Green function can be used. The $a$-dependence cancels.

We determine the subtraction terms by expanding the convolution
of the LO cross section with the structure function in the coupling
constant. We take the expression for the electron structure function
from~\cite{Beenakker:1996kt} with
$\beta\equiv\beta_\text{exp}=\beta_\text{S}=\beta_\text{H}=
2(\alpha/\pi)(\ln(s/m_e^2)-1)$, given by
\begin{eqnarray}
 \Gamma_{ee}^{\text{LL}}(x) & = &
\frac{\exp\left((\frac{3}{8}-\frac{1}{2}\,\gamma_E)\,\beta\right)}
         {\Gamma(1+\frac{1}{2}\,\beta)}\,\frac{\beta}{2}\,
    (1-x)^{\beta/2 - 1} - \frac{1}{4}\,\beta\,(1+x) \nonumber \\
                                                & & %\hspace*{-2cm}
    -\, \frac{1}{4^2\,2!}\,\beta^2\,\left[ \frac{1+3x^2}{1-x}\,\ln(x)
    + 4(1+x)\,\ln(1-x) + 5 + x \right] \nonumber \\
                                                & & %\hspace*{-2cm}
    -\, \frac{1}{4^3\,3!}\,\beta^3\,\left\{ \vphantom{\frac{1}{4}}
    (1+x)\, \left[ 6\, \text{Li}_2 (x) + 12\,\ln^2(1-x) - 3\pi^2 \right]
    \right. \nonumber \\
                                                & & %\hspace*{-2cm}
    + \left. \frac{1}{1-x}\,\left[\frac{3}{2}\,(1+8 x + 3 x^2)\,\ln(x)
    + 6\,(x+5)\,(1-x)\,\ln(1-x)
    \right.\right. \nonumber \\
                                                & & %\hspace*{-2cm}
    + 12\, (1+x^2)\,\ln(x)\,\ln(1-x)
    - \frac{1}{2}\,(1+7 x^2)\,\ln^2(x) \nonumber \\
                                                & & %\hspace*{-2cm}
    + \left.\left. \frac{1}{4}\,(39 - 24 x - 15 x^2)
    \right] \right\}.
\label{eq:structurefn}
\end{eqnarray}
The perturbative expansion of the structure function can be written as
\begin{equation}
\Gamma_{ee}^\text{LL}(x)=\delta(1-x)+\Gamma_{ee}^{\text{LL}(1)}(x)
+\mathcal{O}(\alpha^2).
\end{equation}
For the determination of the subtraction term only the limit $x\rightarrow1$
is important,
\begin{equation}
\Gamma_{ee}^{\text{LL}(1)}(x)\mathop{\longrightarrow}^{x\rightarrow1}
\frac{\alpha}{4\pi}\,\Big[\ln\left(\frac{s}{m_e^2}\right)-1\Big]
\left\{\frac{4}{[1-x]_+}+3\delta(1-x)\right\}.
\end{equation}
The $\mathcal{O}(\alpha)$ term in the convolution of the leading
order partonic cross section with the structure functions is
\begin{eqnarray}
 2\int\limits_0^1dx\,\Gamma_{ee}^{\text{LL}(1)}(x)\sigma^\text{LO}(xs) & = &
\sigma_0\,\frac{24\pi N_c}{s}\,\Big[C_0^{(v)^2}+C_0^{(a)^2}\Big]\,
\frac{\alpha}{4\pi}
\,\Big[\ln\left(\frac{4 m_t^2}{m_e^2}\right)-1\Big]\nonumber \\
 & & \hspace*{-3cm}
\times\,\left\{6\,\text{Im}\left[G_0(E + i\Gamma)\right]
+8\int\limits_0^{m_t} dk\,\frac{\text{Im}\left[G_0(E + i\Gamma-k)\right]}
{\left[k\right]_+}\right\},
 \label{eq:ISsubtraction}
\end{eqnarray}
where in $\sigma^\text{LO}(xs)$
the non-relativistic Green function was evaluated at
$\sqrt{xs}-2m_t=E-m_t(1-x)+\dots$ and we have substituted $k=m_t(1-x)$.
We also set $a=m_t$ and neglected the difference between $s$ and
$4 m_t^2$ in the argument of the logarithm. The initial-state QED correction
to the partonic cross section in the conventional scheme for the
electron structure function is given by \eqref{eq:sigmaIS}
with \eqref{eq:ISsubtraction} subtracted, resulting in
\begin{eqnarray}
\sigma^\text{conv}_{\rm IS}(s) & = & \sigma_0\,\frac{24\pi N_c}{s}\,
\Big[C_0^{(v)^2}+C_0^{(a)^2}\Big]\,\frac{\alpha}{4\pi} \\
 & & \times\left\{8\int\limits_0^{m_t} dk\,
\frac{\text{Im}\left[G_0(E + i\Gamma-k)\right]}{\left[k\right]_+}+\left(\frac{4\pi^2}{3}+2\right)\text{Im}\left[G_0(E + i\Gamma)\right]\right\},\nonumber
 \label{eq:sigmaISconv}
\end{eqnarray}
The imaginary part of the Green function is neglected
for $E<-m_t$ outside the non-relativistic regime. The
photon radiation contribution~\eqref{eq:sigmaISconv} to the cross section
in this scheme is finite and free of large logarithms of the
electron mass.

%%%%%%%%%%%%%%%%%%%%%%%%%%%%%%%%%%%%%%%%%%%%%%%%%%%%%%%%%%%%%%%%%%%%%%
%%%%%%%%%%%%%%%%%%%   Squared Contribution      %%%%%%%%%%%%%%%%%%%%%%
%%%%%%%%%%%%%%%%%%%%%%%%%%%%%%%%%%%%%%%%%%%%%%%%%%%%%%%%%%%%%%%%%%%%%%

\subsection{The squared contribution\label{sec:squared}}

In this section we discuss the calculation of the squared contribution
$\sigma_\text{sq}$ in (\ref{eq:splitfull}), which is given by the diagrams
in Figure~\ref{fig:squared}. The computer programs
Package-X~\cite{Patel:2015tea},
FeynCalc~\cite{Mertig:1990an,Shtabovenko:2016sxi} and
LoopTools~\cite{Hahn:1998yk} have been employed for certain steps of the
computation. The result for the scalar four-point integral in the diagram
$h_{1b}$ was taken from~\cite{Beenakker:2002nc}.
The individual diagram contributions to the hadronic tensor $H$ are evaluated
in $d$ dimensions and written in the form~\eqref{eq:EPsubtraction}. The
numerical $t$ (or $t^*$) integral contains all terms with positive integer or
half-integer powers of $(1-y)$. With the exception of $h_{1b}$, the
subtracted integrands were all obtained in analytical form. The integrand for
$h_{1b}$ contains an additional numerical angular integral.
The expressions for the integrands are rather lengthy and will not be given
explicitly. The numerical integrals are plagued by integrable singularities
involving $1/\sqrt{1-t}$ and $\ln(1-t)$ terms, that cause numerical
instabilities in the evaluation of some diagrams. As a remedy, we computed
additional terms in the expansion in $(1-t)$
analytically and used them as further subtractions.

The contributions corresponding to the second term on the right-hand side
of~\eqref{eq:EPsubtraction} are given by the sum of the respective
expressions in~\cite{Jantzen:2013gpa} and terms from the
$\mathcal{O}(\epsilon)$ contributions to $\hat{g}_{ix}^{(1,b)}$. The latter
encapsulate the dependence of the squared contribution on the computational
scheme and are therefore specified below. In the notation
of~\cite{Jantzen:2013gpa}, we obtain\footnote{Note that in the
expressions from~\cite{Jantzen:2013gpa} quoted below and in
Section~\ref{sec:interference}, $\mu$
in~\cite{Jantzen:2013gpa} must be interpreted as $\mu_w$ and 
$x$ in~\cite{Jantzen:2013gpa} must be identified with $x_W$.}
\begin{eqnarray}
 H_{1a} & = & H_{1a}|_{\text{from~\cite{Jantzen:2013gpa}}} +
N_H\Bigg[\frac{192(2+2 x_W+5 x_W^2)\ln\frac{1-x_W}{2}-623-239x_W-
1154x_W^2}{144 (1-x_W) (1+2 x_W)}\,v_t^Lv_t^R
\nonumber\\
 & & -\frac{2+3 x_W-(1+2 x_W)\ln\frac{1-x_W}{2}}{2+4 x_W}\,v_t^La_t^R
+\frac{11+16 x_W-6(1+2 x_W)\ln\frac{1-x_W}{2}}{18+36 x_W}\,a_t^La_t^R\Bigg]
\nonumber\\
 & & +H_{1a}^{\rm (EP\,fin)},
\nonumber\\
 H_{1b} & = & H_{1b}|_{\text{from~\cite{Jantzen:2013gpa}}} +
N_H\Bigg[\frac{1-2x_W+15 x_W^2-3 \left(1+x_W+2 x_W^2\right)
\ln \frac{1-x_W}{2}}{2 (1-x_W) (1+2 x_W)}\,v_t^Lv_t^R
\nonumber\\
 & & -\frac{17-28 x_W-6 (1-2 x_W) \ln \frac{1-x_W}{2}}{18(1+2x_W)}\,
v_t^La_t^R +\frac{5-4 x_W-6(1-2 x_W) \ln \frac{1-x_W}{2}}{18(1+2x_W)}\,
a_t^L v_t^R\Bigg]\nonumber\\
 & & +H_{1b}^{\rm (EP\,fin)},
\end{eqnarray}
where $H_{1x}^{\rm (EP\,fin)}$ is the contribution from the first term on the
right-hand side of~\eqref{eq:EPsubtraction}. The prefactor is defined as
\begin{equation}
N_H=
m_t\Gamma_0 N_c\frac{\alpha_sC_F}{4\pi}.
\end{equation}
The other diagrams in Figure~\ref{fig:squared} do not contain
$1/\epsilon$ poles from the endpoint divergence and, therefore, no terms
of this type are present, i.e.~$H_{1x}=
H_{1x}|_{\text{from~\cite{Jantzen:2013gpa}}}
+H_{1x}^{\rm (EP\,fin)}$.

The contribution of an individual diagram $h_{ix}$, $g_i$ to the
non-resonant part is
\begin{equation}
\sigma_{ix}=-8\pi^2\alpha^2n_s\sum_{L,R=\gamma,Z}
\frac{v_e^Lv_e^R+a_e^La_e^R}{(4 m_t^2-m_L^2)(4 m_t^2-m_R^2)}\,
\text{Re}\left(H_{ix}\right),
 \label{eq:sigmaix}
\end{equation}
where $n_s$ is a symmetry factor, that is either two for diagrams
which are symmetric with respect to the cut, or four for diagrams
which are not symmetric with respect to the cut. The photon
couplings are $v_f^\gamma=-e_f$ and $a_f^\gamma=0$, where $e_f$
is the fermion charge measured in units of the positron charge,
and couplings of the fermions to $Z$ bosons are given
by~\eqref{eq:Zffcouplings}. The photon mass obviously vanishes, $m_\gamma=0$.
In~\eqref{eq:sigmaix}, $\mathcal{O}(\epsilon)$ terms in the leptonic
tensor have been discarded, as discussed at the beginning of this section.
We have checked explicitly that IR and UV divergences
cancel in the sum over the diagrams in the squared contribution.

%%%%%%%%%%%%%%%%%%%%%%%%%%%%%%%%%%%%%%%%%%%%%%%%%%%%%%%%%%%%%%%%%%%%%%
%%%%%%%%%%%%%%%%%%%%%%%%%%%%%    Part II      %%%%%%%%%%%%%%%%%%%%%%%%
%%%%%%%%%%%%%%%%%%%%%%%%%%%%%%%%%%%%%%%%%%%%%%%%%%%%%%%%%%%%%%%%%%%%%%

\section{Part (II)\label{sec:PartII}}

It would be a natural choice to use the same scheme for part~(II), given
by (see (\ref{eq:splitfull}))
\begin{equation}
\sigma_\text{int}^{(\text{EP div})}+
\sigma_{C^{(k)}_{\text{Abs,bare}}},
\end{equation}
as for part~(I). We can however simplify the computation of this part by
performing the Dirac algebra and one of the loop integrations in four
dimensions. The details of this scheme and the computation of
$\sigma_{C^{(k)}_{\text{Abs,bare}}}$ and $\sigma_\text{int}^{(\text{EP div})}$
are shown in Section~\ref{sec:resAbs} and~\ref{sec:interference}, respectively.

%%%%%%%%%%%%%%%%%%%%%%%%%%%%%%%%%%%%%%%%%%%%%%%%%%%%%%%%%%%%%%%%
%%%%%%%%%%%%%%%%%%     Absorptive part      %%%%%%%%%%%%%%%%%%%%
%%%%%%%%%%%%%%%%%%%%%%%%%%%%%%%%%%%%%%%%%%%%%%%%%%%%%%%%%%%%%%%%

\subsection{Absorptive contribution to the matching coefficient
\label{sec:resAbs}}

The bare absorptive part of the matching coefficients
$C_{\text{Abs,bare}}^{(k)}$ is given by the diagrams shown in the second
row of Figure~\ref{fig:Cabs}. We define the scheme as follows:
The coefficients $C_{\text{Abs,bare}}^{(k)}$ are calculated in four
dimensions, but the loop integrations in the third row of
Figure~\ref{fig:Cabs}, i.e. the ones related to the non-relativistic Green
function, are performed in $d$ dimensions.
The Dirac algebra is completely treated in four dimensions. We describe in
Section~\ref{sec:interference} how the interference contribution must be
treated to achieve consistency with this scheme.

Our results for $C_{\text{Abs,bare}}^{(k)}$ in four dimensions are
\begin{eqnarray}
 C_{\text{Abs,bare}}^{(v)} & = & -\frac{\pi}{24 s_w^4 x_W \left(1-x_W^2\right)
(4 c_w^2-x_W)}\Big[(1-x_W)\left(5+44 x_W+28 x_W^2-4 x_W^3-x_W^4\right)
\nonumber \\
 & & -(1-x_W)s_w^2\big[e_e(1-x_W)^2\left(e_t (4 - 21 x_W - 3 x_W^2 + 2 x_W^3)
-4 + 4 x_W - 4 x_W^2\right) \nonumber \\
 & & +e_t (1-x_W)^2 (1 - 5 x_W - 2 x_W^2)+4 + 48 x_W + 36 x_W^2 + 8 x_W^3\big]
\nonumber \\
 & & -12 x_W (1+x_W) (4 c_w^2-x_W) \,\text{arctanh}(1-x_W)\Big],
\end{eqnarray}
\begin{eqnarray}
 C_{\text{Abs,bare}}^{(a)} & = & \frac{\pi}{24 s_w^4 x_W \left(1-x_W^2\right)
(4 c_w^2-x_W)}\Big[(1-x_W)\left(5 + 44 x_W + 28 x_W^2 - 4 x_W^3 - x_W^4\right)
\nonumber \\
 & & -(1-x_W)s_w^2\big[e_t(1-x_W)^2(1 - 5 x_W - 2 x_W^2)+4 + 48 x_W
+ 36 x_W^2 + 8 x_W^3\big] \nonumber\\
 & & -12 x_W (1 + x_W) (4 c_w^2-x_W) \,\text{arctanh}(1-x_W)\Big].
\end{eqnarray}
The contribution to the cross section is given by
\begin{equation}
 \sigma_{C^{(k)}_{\text{Abs,bare}}}=\sigma_0\,\frac{12\alpha N_c}{s}\,
\Big[C_0^{(v)}C_{\text{Abs,bare}}^{(v)}+C_0^{(a)}C_{\text{Abs,bare}}^{(a)}
\Big]\,
\text{Re}\left[\frac{3}{3-2\epsilon}G_0^{(w)}(E+i\Gamma)\right].
 \label{eq:sigmaCAbs}
\end{equation}
We recall that at LO the Dirac structure of the top anti-top pair becomes
trivial in the non-relativistic regime and only yields a prefactor
$3-2\epsilon$. By introducing the factor $3/(3-2\epsilon)$ in front of the
Green function in~\eqref{eq:sigmaCAbs}, we adapted the expression to the
scheme described above, which involves four-dimensional Dirac
algebra. The contribution~\eqref{eq:sigmaCAbs} is not affected by loose cuts.

%%%%%%%%%%%%%%%%%%%%%%%%%%%%%%%%%%%%%%%%%%%%%%%%%%%%%%%%%%%%%%%%%%%%%%%
%%%%%%%%%%%%%%%    Interference part (EP div)      %%%%%%%%%%%%%%%%%%%%
%%%%%%%%%%%%%%%%%%%%%%%%%%%%%%%%%%%%%%%%%%%%%%%%%%%%%%%%%%%%%%%%%%%%%%%

\subsection{Endpoint divergence of the interference contribution
\label{sec:interference}}

The endpoint-divergent part of the interference contribution has been computed
in~\cite{Jantzen:2013gpa}. The expression~\eqref{eq:sigmaintEPdiv} also
contains an endpoint-finite term from the $\mathcal{O}(\epsilon)$ terms
in $\hat{g}_{ia}^{(1,2)}$ multiplying the $1/\epsilon$ pole. This term
carries the dependence on the computational scheme and must, therefore, be
treated in the same scheme as the contribution~\eqref{eq:sigmaCAbs}. We
evaluate it using the expansion by regions approach described
in~\cite{Jantzen:2013gpa}. For each of the diagrams in
Figure~\ref{fig:interference}, we treat the loop contained in the
corresponding diagram in Figure~\ref{fig:Cabs}, i.e. the right loop in
$h_{2a}$ and the left loop in $h_{3a}$ and $h_{4a}$, in four dimensions. The
Dirac algebra is also done in four dimensions, but the remaining loop
integrations are performed in $d$ dimensions. In the notation
of~\cite{Jantzen:2013gpa} we obtain
\begin{eqnarray}
 H_{2a}^{(\text{EP div})} & = & H_{2a}|_{\text{from~\cite{Jantzen:2013gpa}}}
+ N_H\,\frac{1-5x_W-2 x_W^2}{36 (1+x_W) (1+2 x_W)}
\left(8-3 \ln\frac{\mu_w^2}{4m_t^2}\right)v_t^L(v_b^R+a_b^R),\quad
\\
 H_{3a}^{(\text{EP div})} & = & H_{3a}|_{\text{from~\cite{Jantzen:2013gpa}}}
+ N_H\,\frac{2+5 x_W-2 x_W^2}{36 x_W (1+2 x_W)}
\left(-8+3 \ln\frac{\mu_w^2}{4m_t^2}\right)I_\text{WW}^Lv_t^R
\end{eqnarray}
with $I_\text{WW}^L=1$ for diagram $h_{3a}$ with a photon attached to
the $WW$ vertex, and $-c_w/s_w$ for the $WWZ$ vertex.
The endpoint divergent contributions of $h_{2a}$ and $h_{3a}$ follow from
equation~\eqref{eq:sigmaix}
with $n_s=4$. The contribution of $h_{4a}$ is given by
\begin{eqnarray}
\sigma_{4a}^{(\text{EP div})} & = &
n_s \,\Delta\sigma_{4a}|_{\text{from~\cite{Jantzen:2013gpa}}}
+ n_s N_H\frac{\pi^2\alpha^2}{s_w^2}\frac{1}{s}
\left(\frac{e_te_e}{s}+\frac{v_t(v_e+a_e)}{s-M_Z^2}\right)
\left(-2+\ln\frac{\mu_w^2}{4m_t^2}\right) \nonumber\\
 & & \times\frac{(1-x_W) \left(1-2x_W-23 x_W^2\right)+12 x_W^2
\ln \left(\frac{2}{x_W}-1\right)}{3x_W (1-x_W)^3 (1+2 x_W)}.
 \label{eq:h4a}
\end{eqnarray}
with symmetry factor $n_s=4$.

To verify that the treatment of the scheme is consistent we computed the
finite sum of the contributions from the diagrams $h_{2a}$ and $h_{3a}$ and
the contributions from the corresponding diagrams in
$C_{\text{Abs,bare}}^{(k)}$ also in the scheme of part (I) and found
perfect agreement with the results presented above. Applying the scheme
of part (II) simplifies the computation, especially for $h_{4a}$,
since it avoids the more complicated integration of the left loop in
$h_{4a}$ in $d$ dimensions.

%%%%%%%%%%%%%%%%%%%%%%%%%%%%%%%%%%%%%%%%%%%%%%%%%%%%%%%%%%%%%%%%%%
%%%%%%%%%%%%%%%%%%%%%%     Part III      %%%%%%%%%%%%%%%%%%%%%%%%%
%%%%%%%%%%%%%%%%%%%%%%%%%%%%%%%%%%%%%%%%%%%%%%%%%%%%%%%%%%%%%%%%%%

\section{Part (III)\label{sec:PartIII}}

The part (III) $\sigma_\text{int}^{(\text{EP fin})} +
\sigma_\text{aut}$ contains the automated part $\sigma_\text{aut}$, which is
evaluated with \texttt{MadGraph}. The automated part is UV divergent and
therefore scheme dependent. The divergence and the corresponding scheme
dependence cancel with the endpoint-finite part
$\sigma_\text{int}^{(\text{EP fin})}$
of the interference contribution. We first describe the implementation
of the automated part in \texttt{MadGraph} in Section~\ref{sec:automated}.
This fixes the scheme in which $\sigma_\text{int}^{(\text{EP fin})}$ is
computed in Section~\ref{sec:interferenceEPfin}.

%%%%%%%%%%%%%%%%%%%%%%%%%%%%%%%%%%%%%%%%%%%%%%%%%%%%%%%%%%%%%%%%%%%%%
%%%%%%%%%%%%%%%%%%%%%%     Automated      %%%%%%%%%%%%%%%%%%%%%%%%%%%
%%%%%%%%%%%%%%%%%%%%%%%%%%%%%%%%%%%%%%%%%%%%%%%%%%%%%%%%%%%%%%%%%%%%%

\subsection{The automated part\label{sec:automated}}

We first recall some aspects of \texttt{MadGraph}, which are relevant to our
definition of the computational scheme.
\begin{enumerate}
\item \label{en:FKS}
The subtraction of IR singularities is performed automatically using the 
FKS method~\cite{Frixione:1995ms,Frederix:2009yq}. The IR singularities
in the real corrections are subtracted before the phase-space integration
and the subtraction terms are then integrated over the phase space of the
real emission and added to the virtual corrections, where they cancel the IR
singularities that arise in the loop integrals. The phase-space
integration is then always done in four dimensions.
\item In the virtual corrections, \texttt{MadGraph} uses rational $R_2$
terms~\cite{Ossola:2008xq} to absorb the $(-2\epsilon)$-dimensional parts of
the numerators. For a given diagram with amplitude $\mathcal{C}$ the
decomposition takes the form
\begin{equation}
\mathcal{C}\equiv\int d^d\bar{l}\,\frac{\bar{N}(\bar{l})}{\prod_i \bar{D}_i}
=\int d^d\bar{l}\,\frac{N(l)}{\prod_i \bar{D}_i}+R_2,
\label{eq:R2terms}
\end{equation}
where $D_i=(\bar{l}+p_i)^2-m_i^2$, quantities with a bar are
 $(4-2\epsilon)$-dimensional and quantities without are 4-dimensional.
The non-$R_2$ term can be written as a sum over 4-dimensional coefficients
multiplying $d$-dimensional tensor integrals. The $(-2\epsilon)$-dimensional
parts related to the implementation of the 't Hooft-Veltman
scheme~\cite{'tHooft:1972fi} in \texttt{MadGraph} are all contained in the
$R_2$ terms. \label{en:R2}
\item  The amplitudes for the non-$R_2$ terms, the $R_2$ terms, the UV
counterterms and the FKS subtraction terms are written as separate lists,
each of them containing the coefficient of the $1/\epsilon^2$ pole,
the $1/\epsilon$ pole and the finite part. Afterwards, only four-dimensional
 operations are performed, i.e. the multiplication with the conjugated
four-dimensional born amplitude and the four-dimensional phase-space
integration.
\label{en:Asquared}
\end{enumerate}
Given the way that $\sigma_\text{aut}$ is defined, we never have to modify
the construction of an amplitude $\mathcal{A}_i$, but we have to remove
certain contributions $\mathcal{A}_i\mathcal{A}_j^*$ in the squared
amplitude $|\mathcal{A}|^2=\sum_{i,j}\mathcal{A}_i\mathcal{A}_j^*$.
All the contributions associated with the diagrams in
Figure~\ref{fig:squared} have to be removed, i.e. also the $R_2$ parts,
the UV counterterms and the FKS subtraction terms. There is however an
ambiguity in the subtraction of the contributions in
Figure~\ref{fig:interference}, which determines the scheme in which
$\sigma_\text{int}^{(\text{EP fin})}$ must be computed. We choose to only
subtract the non-$R_2$ terms of $h_{ia}$ with $i=2,3,4$. Following the
discussion of the items~\ref{en:FKS} and~\ref{en:R2} above, this implies
that $\sigma_\text{int}^{(\text{EP fin})}$ has to be computed by using
dimensional regularization for the tensor integrals. All other steps in the
computation of $\sigma_\text{int}^{(\text{EP fin})}$ are then performed in
four dimensions.

In the following, we describe the steps we performed in \texttt{MadGraph}
to obtain the automated part in the scheme defined above. It is obvious that
this cannot be achieved by modifying the process generation, because
the automated part does not correspond to a squared amplitude.
We therefore first generate the full process $e^+e^-\rightarrow \bar{t}W^+b$
including QCD corrections. By not invoking the complex mass scheme, we make
sure that the self-energy insertions are treated perturbatively. Hence, the
cross section diverges rapidly for center-of-mass energies
approaching $\sqrt{s}=2m_t$ from below. We remove the contribution from the
endpoint divergent born diagram $h_1$, the diagrams shown in
Figures~\ref{fig:squared} and the non-$R_2$ terms from
Figure~\ref{fig:interference} by editing the code generated by
\texttt{MadGraph}.

Finally, we have to deactivate some checks inside the code, that are
invalidated by the modifications. \texttt{MadGraph} checks if the
$1/\epsilon^2$ and $1/\epsilon$ poles vanish for a number of phase space
points. Here, this is not the case because the automated part is UV
divergent. We have addressed this issue with in two ways -- by deactivating
the check or by performing a minimal subtraction of the UV divergence --
and found agreement of both approaches. The minimal subtraction was also
used to verify the cancellation of the UV divergence with the endpoint-finite
part of the interference contribution $\sigma_\text{int}^{(\text{EP fin})}$.
Furthermore, due to the subtractions, the tree-level cross section and the
real corrections are no longer the squared absolute value of an amplitude
and, thus, no longer positive for all phase-space points. The positivity of
these expressions is not necessary to make the code run properly, but is
only used as an internal check~\cite{Frederix}. Therefore, we can safely
switch it off. The code can now be evaluated directly at the threshold
$\sqrt{s}=2m_t$. The contribution $\sigma_\text{aut}$ is given by the
difference of fixed-order runs at NLO and LO, multiplied by a factor two to
account for the $t\bar{b}W^-$ contributions. Further details on the
implementation and modifications in \texttt{MadGraph} are provided in
Appendix~\ref{app:MADGRAPH}. The evaluation of the automated part in the code
\texttt{QQbar\_Threshold} relies on a precomputed grid as described in
Appendix~\ref{sec:impl_gridgen}. Since the contribution $\sigma_\text{aut}$
is rather small, we do not aim for more precision than about 10\% in the
automated part. The resulting error of the cross section is less than
one per mille. To reach this target precision we set up
\texttt{MadGraph} to generate an integration grid from four iterations
with 15000 points per integration channel and perform the actual
integration using six iterations with 100000 points for each point of
the \texttt{QQbar\_Threshold} grid. More precise results are possible at
the cost of a considerably increased computing time for the generation
of the grid.

%%%%%%%%%%%%%%%%%%%%%%%%%%%%%%%%%%%%%%%%%%%%%%%%%%%%%%%%%%%%%%%%%%%%%%%%
%%%%%%%%%%%%%%%%%%%%  Interference part (EP fin)      %%%%%%%%%%%%%%%%%%
%%%%%%%%%%%%%%%%%%%%%%%%%%%%%%%%%%%%%%%%%%%%%%%%%%%%%%%%%%%%%%%%%%%%%%%%

\subsection{Endpoint-finite part of the interference contribution
\label{sec:interferenceEPfin}}

We recall that the endpoint-finite part of the interference contribution has
the form~\eqref{eq:sigmaintEPfin}. As detailed in Section~\ref{sec:automated}
it must be evaluated by taking only the tensor integrals of the virtual loop
in $d$ dimensions and then performing all other steps in the computation
strictly in four dimensions. Within this scheme, we have determined the
endpoint subtracted integrands for the diagrams $h_{2a},h_{3a}$ analytically
and for $h_{4a}$ as a one-dimensional angular integral. We refrain from
giving the lengthy results for the integrands. As described in
Section~\ref{sec:squared} we use additional terms in the expansion of the
amplitudes in $1-t$ as subtractions to deal with integrable divergences that
appear in the limit $t\rightarrow1$ of the $t$-integration. The result for
$\sigma_\text{int}^{(\text{EP fin})}$ is given by applying the same
prefactors and symmetry factors as for the endpoint divergent part of the
interference contribution $\sigma_\text{int}^{(\text{EP div})}$ in
Section~\ref{sec:interference}.

%%%%%%%%%%%%%%%%%%%%%%%%%%%%%%%%%%%%%%%%%%%%%%%%%%%%%%%%%%%%%%%%%
%%%%%%%%%%%%%%%%%    Implementation      %%%%%%%%%%%%%%%%%%%%%%%%
%%%%%%%%%%%%%%%%%%%%%%%%%%%%%%%%%%%%%%%%%%%%%%%%%%%%%%%%%%%%%%%%%

\section{Checks and implementation
\label{sec:checks_and_imp}}

\subsection{Consistency checks\label{sec:checks}}

Having performed the computation of the non-resonant part in the presence
of the invariant mass cut~\eqref{eq:definitioncut}, denoted by
$c_{\Delta M_t}(p_i)$, allows us to perform a very powerful numerical
consistency check. The non-resonant cross section
$\sigma_\text{non-res}(\bar{c}_{\Delta M_t})$ in the presence of the
complementary cut $\bar{c}_{\Delta M_t}(p_i)=1-c_{\Delta M_t}(p_i)$ is finite.
Therefore, we can evaluate it using unedited \texttt{MadGraph} code. On the
other hand, it can be obtained from our result by taking the difference
$\sigma_\text{non-res}-\sigma_\text{non-res}(c_{\Delta M_t})$.
The comparison for various values of the cut $\Delta M_t$ numerically tests
the whole non-resonant result in (\ref{eq:splitnonres}),
with the exception of the contributions from
the $\mathcal{O}(\epsilon)$ parts of the $\hat{g}_{ix}^{(1,b)}$
terms in~\eqref{eq:EPsubtraction},
which originate from the $t^{(*)}\rightarrow1$
region and are independent of the value of the cut, i.e. are not
present in $\sigma_\text{non-res}(\bar{c}_{\Delta M_t})$.

%%%%%%%%%%%%%%%%%%%%%%%%%%%%%%%%%%%%%%%%%%%%%%%%%%%%%%%%%%%%%%%%%%%%%%%%
\begin{figure}[t]
\begin{center}
\includegraphics[width=0.8\textwidth]{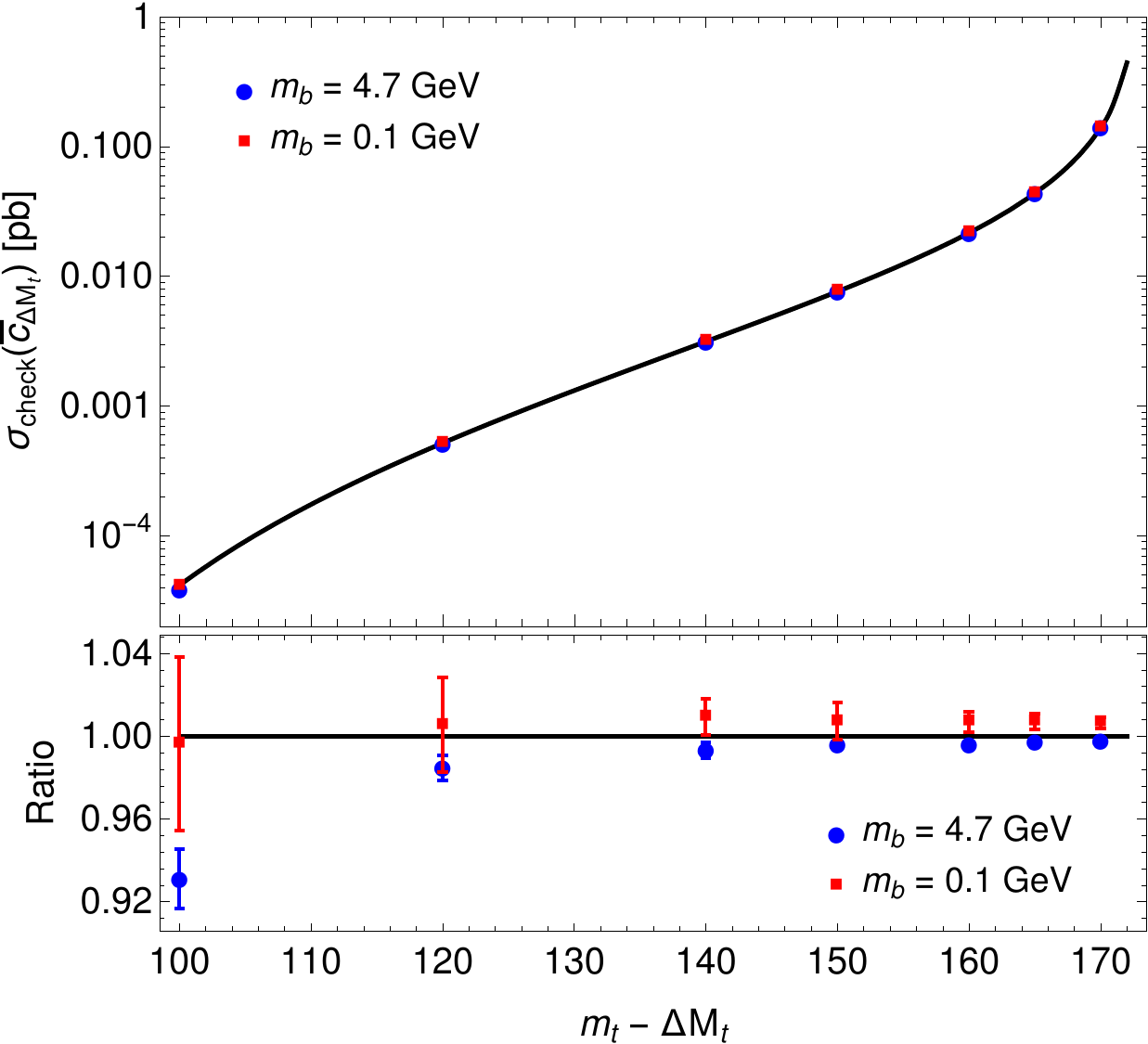}
\caption{\label{fig:ConsistencyCheck}Consistency check for various values for
the complementary cut $p_t^2\leq(m_t-\Delta M_t)^2$. The line in the upper
panel is our semi-analytical result~\eqref{eq:cc_semianalytical_part}
for $\sigma_\text{check}(\bar{c}_{\Delta M_t})$ in pb, given by the sum of
the contributions from the tree-level diagram $h_1$, the squared and the
interference contributions. The points give the same
quantity~\eqref{eq:cc_MadGraph_part} obtained from the difference of
\texttt{MadGraph} runs with the unedited and edited code. The lower panel
shows the same results normalized to~\eqref{eq:cc_semianalytical_part}.
The \texttt{MadGraph} results have been obtained for the default value of the
bottom-quark mass $m_b=4.7\,$GeV and a negligible value $m_b=0.1\,$GeV. The
error bars in the lower plot are obtained by adding the standard deviation
of ten runs of the unedited \texttt{MadGraph} code and the standard deviation
of ten runs of the edited \texttt{MadGraph} code in quadrature, while the
estimated uncertainty of an individual run is ignored. The increase of the
relative uncertainty for large values of $\Delta M_t$ is related to large
cancellations between the results of the edited and unedited code. For this
check, we have used the default values of \texttt{MadGraph}, $m_t=173\,$GeV,
$\mu=m_Z$, $\alpha_s(m_Z)=0.118$ and $\alpha=1/132.507$.}
\end{center}
\end{figure}
%%%%%%%%%%%%%%%%%%%%%%%%%%%%%%%%%%%%%%%%%%%%%%%%%%%%%%%%%%%%%%%%%%%%%%%%

The result of our check is shown in Figure~\ref{fig:ConsistencyCheck}.
Here, we have rearranged the contributions as follows,
\begin{eqnarray}
\label{eq:cc_semianalytical_part}
 \sigma_\text{check}(\bar{c}_{\Delta M_t}) & \equiv &
\sigma_{h_1}(\bar{c}_{\Delta M_t}) +
\sigma_\text{sq}(\bar{c}_{\Delta M_t}) +
\sigma_\text{int}(\bar{c}_{\Delta M_t})
\\
&    =   & \sigma_\text{non-res}(\bar{c}_{\Delta M_t})
- \left[ \sigma_\text{aut}(\bar{c}_{\Delta M_t})
- \sigma_{h_1}(\bar{c}_{\Delta M_t})\right],
\label{eq:cc_MadGraph_part}
\end{eqnarray}
where $\sigma_{h_1}(\bar{c}_{\Delta M_t})$ is the contribution to the
non-resonant part from the diagram $h_1$ at NLO (Figure~\ref{fig:NonresNLO})
in the presence of the complementary cut. The line in
Figure~\ref{fig:ConsistencyCheck} shows our semi-analytical result
for $\sigma_\text{check}(\bar{c}_{\Delta M_t})$ obtained by means
of~\eqref{eq:cc_semianalytical_part}. The points show the same quantity
determined by evaluating~\eqref{eq:cc_MadGraph_part} using \texttt{MadGraph}.
The contribution from diagram $h_1$ is included in
$\sigma_\text{check}(\bar{c}_{\Delta M_t})$, because our edited
\texttt{MadGraph} code, described in Section~\ref{sec:automated},
corresponds to the combination $\sigma_\text{aut}(\bar{c}_{\Delta M_t})
- \sigma_{h_1}(\bar{c}_{\Delta M_t})$
that appears in~\eqref{eq:cc_MadGraph_part}.
We performed the same check for the individual contributions from the
diagrams $h_{ia}$ with $i=2,3,4$.
In particular, this provides very welcome reassurance that the scheme
dependence within part (III) has been treated consistently. Within estimated
numerical errors we find good agreement, if the bottom-quark mass
$m_b$ is neglected, as is done in our calculation.

\subsection{Implementation in \texttt{QQbar\_Threshold}
\label{sec:implementation}}

All of the aforementioned NNLO corrections have been implemented in the
new version 2 of the public code
\texttt{QQbar\_threshold}~\cite{Beneke:2016kkb}.  A summary of
the code changes and some code examples for the new
functions are provided in Appendix~\ref{app:implementation}.
\texttt{QQbar\_threshold} can be downloaded from
\url{https://www.hepforge.org/downloads/qqbarthreshold/}. An updated
online manual is available under
\url{https://qqbarthreshold.hepforge.org/}.

%%%%%%%%%%%%%%%%%%%%%%%%%%%%%%%%%%%%%%%%%%%%%%%%%%%%%%%%%%%%%%%%%%%%%
%%%%%%%%%%%%%%%%%%%%%%%   Comparison      %%%%%%%%%%%%%%%%%%%%%%%%%%%
%%%%%%%%%%%%%%%%%%%%%%%%%%%%%%%%%%%%%%%%%%%%%%%%%%%%%%%%%%%%%%%%%%%%%

\subsection{Comparison to other approaches
\label{sec:comparison}}

While a complete calculation of NNLO electroweak and non-resonant
contributions to the top-pair threshold as reported here has never been
done before, NNLO non-resonant corrections have been evaluated in
certain approximations in~\cite{Hoang:2010gu} and \cite{Penin:2011gg}.
We briefly comment on these approximations and their limitations here.

\subsubsection{Comparison to \cite{Hoang:2010gu}}

The leading NNLO non-resonant contributions for the case of ``not-too-loose''
cuts satisfying  $\Gamma_t\ll \Delta M_t \ll m_t$ were determined
in~\cite{Hoang:2010gu} within the so-called phase-space matching (PSM)
approach. This result captures the first terms in the expansion
in $\Lambda/m_t$ ($\Lambda^2\equiv 2m_t\Delta M_t-\Delta M_t^2$) of the full
non-resonant result, namely the terms of order
$m_t^2/\Lambda^2$, $m_t/\Lambda$ and $(m_t/\Lambda)^0 \log \Lambda$. The latter
correspond to the endpoint-divergent terms computed in~\cite{Jantzen:2013gpa},
which give the approximate result labelled ``aNNLO'' in
Figure~\ref{fig:SigmaNonresCutPlot} below. Because of the ``not-too-loose''
cut condition, the PSM approach does not allow the calculation of
the $b\bar b W^+ W^-X$ total cross section near the top anti-top threshold.

The agreement between the PSM result and the full non-resonant computation
of the non-analytic terms in the expansion in the invariant-mass
cut parameter $\Lambda/m_t$ can be understood as a consequence of the
cancellation of singularities between
adjacent regions of loop momentum~\cite{Jantzen:2013gpa}.
The $\Lambda/m_t$ non-resonant terms are obtained in the PSM approach
by computing the ultraviolet behaviour of the  resonant amplitude where the
cut on the invariant mass of the top and anti-top quark has been implemented.
Therefore $\Lambda$ effectively acts as a regulator of the ultraviolet
singular behaviour of the resonant part of the amplitude, that is obtained
assuming on-shell top quarks, when the latter is further taken into the
off-shell limit, {\it i.e.} for $|\mathbf{p_t}| \gg \Gamma_t$. On the other
hand, the endpoint-divergent terms are obtained from the non-resonant part
of the amplitude, that assumes off-shell tops with $|p_t^2-m_t^2| \gg
\Gamma_t$, upon going to the (infrared) on-shell limit within a distance
regulated by $\Lambda$. The fact that both expansions provide the same
divergent terms in $\Lambda/m_t$ is thus a consequence of the cancellation
of the dependence on the cut-off $\Lambda$ that separates the resonant and
non-resonant regions. For the limitations on the PSM result to provide
higher-order terms in  the $\Lambda/m_t$ expansion we refer the reader
to~\cite{Jantzen:2013gpa}.

\subsubsection{Comparison to \cite{Penin:2011gg}}

Another approach, introduced in \cite{Penin:2011gg}, aims at the computation
of the non-resonant contribution to the total cross section in an expansion
in $\rho^{1/2}$, where $\rho=1-m_W/m_t$ is treated as a small parameter.
Even though $\rho$ is not small in reality, one may hope that with
sufficiently many terms in the expansion, a good approximation might be
obtained. Indeed, the exact NLO non-resonant result
from \cite{Beneke:2010mp} was reproduced by combining a deep expansion
with Pad\'{e} approximants.

Our concern here is the computation of the first two terms in the
$\rho^{1/2}$ expansion of the NNLO non-resonant contribution. In the
present notation, the first term in the expansion given in
\cite{Penin:2011gg} reads explicitly
\begin{equation}
\sigma_{\!\mbox{\footnotesize\cite{Penin:2011gg}}}^{(1), \rm nr} =
\sigma_0\,\frac{24\pi N_c}{s}\,
\Big[C_0^{(v)^2}+C_0^{(a)^2}\Big]\,\frac{m_t\Gamma_0}{\rho} \,
\frac{\alpha_s C_F}{4\pi}
\left\{2\ln\frac{|E+i\Gamma|}{m_t\rho}+4\ln 2+1 +
{\cal O}\left(\sqrt{\rho}\,\right)
\right\}.
\label{eq:ppnonres}
\end{equation}
It is immediately clear from this expression that the meaning of
``non-resonant'' is different from ours, in which case the non-resonant
contribution is analytic in energy and has a $1/\epsilon$ pole.
It appears that \cite{Penin:2011gg} does not distinguish between what
we call non-resonant and absorptive matching coefficient contribution
to the resonant part and directly constructs the expansion of
the diagram in  $\rho^{1/2}$, such that \eqref{eq:ppnonres} gives the
sum of all contributions at order ${\cal O}(\alpha_s/\rho)$.

It is instructive to construct the ${\cal O}(\alpha_s/\rho)$ terms
from the results in \cite{Ruiz-Femenia:2014ava} and in the present paper.
We find that they arise only from
\begin{equation}
\sigma_{\rm sq}+\sigma_{C^{(k)}_{\text{Abs,}Z_t}}
\label{eq:LOrhoxsec}
\end{equation}
in part (I) and specifically from diagrams $h_{1a}$ and $h_{1b}$ in
$\sigma_{\rm sq}$. Each of
the two terms contains a $1/\epsilon$ pole, which cancels in the sum.
This holds separately for the two diagrams $h_{1a}$ and $h_{1b}$ plus
their corresponding resonant counterparts{\footnote{The resonant
counterparts of $h_{1a}$ and $h_{1b}$ correspond to the same diagrams
but taking the loop momenta in the top anti-top loops in the potential
region, and keeping only the NNLO term of the self-energy insertion in
$h_{1a}$.}
that contribute to $\sigma_{C^{(k)}_{\text{Abs,}Z_t}}$. We note that the
leading term \eqref{eq:ppnonres} from \cite{Penin:2011gg} originates only
from diagram $h_{1a}$. Our result for this diagram including its resonant
counterpart indeed agrees with the above except for the constant term $+1$
(see \eqref{eq:h1ares}, \eqref{eq:h1anonres} in
Appendix~\ref{app:furthercomparison}). However, as was already mentioned
in \cite{Jantzen:2013gpa,Ruiz-Femenia:2014ava}, contrary to the statement
made in \cite{Penin:2011gg} there is a non-vanishing contribution from
$h_{1b}$ at the same order. We computed the ${\cal O}(\alpha_s/\rho)$
from this diagram explicitly, and find that the complete
${\cal O}(\alpha_s/\rho)$ contribution to the total cross section reads
\begin{equation}
\sigma^{{\cal O}(\alpha_s/\rho)}%^{\text{NNLO,}\,\rho}
= \sigma_0 \frac{24\pi N_c}{s}\,
\Big[C_0^{(v)^2}+C_0^{(a)^2}\Big]\,\frac{m_t\Gamma_0}{\rho}
\,\frac{\alpha_sC_F}{4\pi}\left\{ \ln\frac{|E+i\Gamma|}{m_t}+4\ln 2 \right\}.
\label{eq:as/rhoxsec}
\end{equation}
Note the absence of a logarithmic dependence on $\rho$ in the sum
of all contributions (see \eqref{eq:h1ares}--\eqref{eq:h1bnonres}}
for the individual results).
This can be traced to the cancellation of $1/\epsilon$ divergences and
the scaling of the leading momentum regions that contribute to the
$1/\rho$ enhanced term. Furthermore, the coefficient of $\ln|E+i\Gamma|$ 
differs by a factor of two, which is related to the contribution of the 
diagram $h_{1b}$ as described in Appendix~\ref{app:furthercomparison}.
We therefore disagree with the NNLO non-resonant
result given in \cite{Penin:2011gg} already from the leading term
in the $\rho^{1/2}$ expansion.

The authors of \cite{Penin:2011gg} did not actually attempt the calculation
of diagram $h_{1b}$ but referred to \cite{Melnikov:1993np} to claim that
it must not contribute. However, as already discussed in
Section~\ref{sec:resPotnew}, the purported vanishing of $h_{1b}$,
called ``jet-jet'' contribution in  \cite{Melnikov:1993np}, refers to a
different order in
the non-relativistic expansion, namely NLO, and is reflected in the
present framework as the non-renormalization of the coupling of the
top quark to a potential gluon and the Coulomb potential by electroweak
effects. Moreover, when the $\rho^{1/2}$ expansion is constructed from momentum
regions, the leading $1/\rho$ term arises from a momentum region that
was missed in \cite{Penin:2011gg}. Further details on the comparison
and diagram $h_{1b}$ can be found in Appendix~\ref{app:furthercomparison}.

%%%%%%%%%%%%%%%%%%%%%%%%%%%%%%%%%%%%%%%%%%%%%%%%%%%%%%%%%%%%%%%%%%%%%
%%%%%%%%%%%%%%%%%%%%%     Phenomenology      %%%%%%%%%%%%%%%%%%%%%%%%
%%%%%%%%%%%%%%%%%%%%%%%%%%%%%%%%%%%%%%%%%%%%%%%%%%%%%%%%%%%%%%%%%%%%%

\section{Discussion of results \label{sec:results}}

Recent experimental studies~\cite{Seidel:2013sqa,Horiguchi:2013wra}
concluded that the statistical uncertainties of the top threshold scan at
a future $e^+ e^-$  collider can be very small in
realistic running scenarios. Thus, when discussing the impact of the
electroweak and non-resonant corrections in this section, we focus
on the theoretical uncertainties. An experimental analysis based on the
theory prediction available in
\texttt{QQbar\_threshold}~\cite{Beneke:2016kkb} is in
progress~\cite{Simon:2016pwp} and will combine statistical and
systematic experimental errors with theory uncertainties.

To avoid the IR renormalon ambiguities, we exclusively employ the PS
shift (PSS) mass scheme defined
in~\cite{Beneke:2013PartII,Beneke:2014pta,Beneke:2016kkb}.
For the numerical evaluation we adopt the input values
\begin{equation}
\begin{array}{lll}
m_t^\text{PS} = 171.5\,\text{GeV},\hspace{1cm} &
\alpha_s(m_Z) = 0.1184,\hspace{1cm} & \alpha(m_Z) = 1/128.944\\
m_H = 125\,\text{GeV},\hspace{1cm} & m_Z = 91.1876\,\text{GeV},
\hspace{1cm} & m_W = 80.385\,\text{GeV},\\
\Gamma_t = 1.33\,\text{GeV},\hspace{1cm} &
\mu_r = 80\,\text{GeV},\hspace{1cm} & \mu_w = 350\,\text{GeV},
\end{array}
\label{eq:pheno_inputs}
\end{equation}
where $m_t^\text{PS}$ is the top-quark PS mass~\cite{Beneke:1998rk} and the
running electroweak coupling is taken from~\cite{Jegerlehner:2011mw},
see the discussion in Section~\ref{sec:resCV}.

%%%%%%%%%%%%%%%%%%%%%%%%%%%%%%%%%%%%%%%%%%%%%%%%%%%%%%%%%%%%%%%%%%%%%%%%
%%%%%%%%%%%%%%%%%%%%%%%%%%%    Corrections      %%%%%%%%%%%%%%%%%%%%%%%%
%%%%%%%%%%%%%%%%%%%%%%%%%%%%%%%%%%%%%%%%%%%%%%%%%%%%%%%%%%%%%%%%%%%%%%%%

\subsection{Size of the electroweak effects\label{sec:pheno}}

%%%%%%%%%%%%%%%%%%%%%%%%%%%%%%%%%%%%%%%%%%%%%%%%%%%%%%%%%%%%%%%%%%%%%%
\begin{figure}[t]
  \centering
  \includegraphics[width=0.7\textwidth]{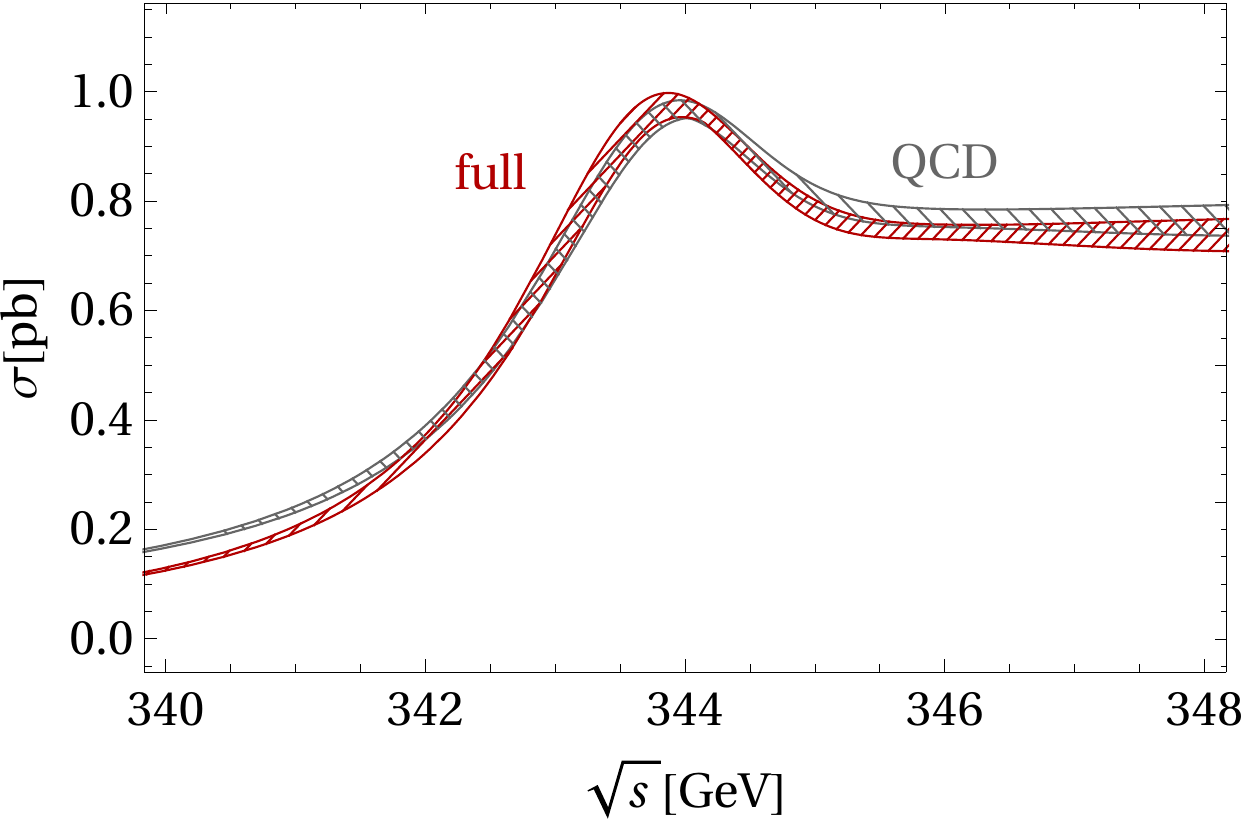}\\[1.0cm]
  \includegraphics[width=0.7\textwidth]{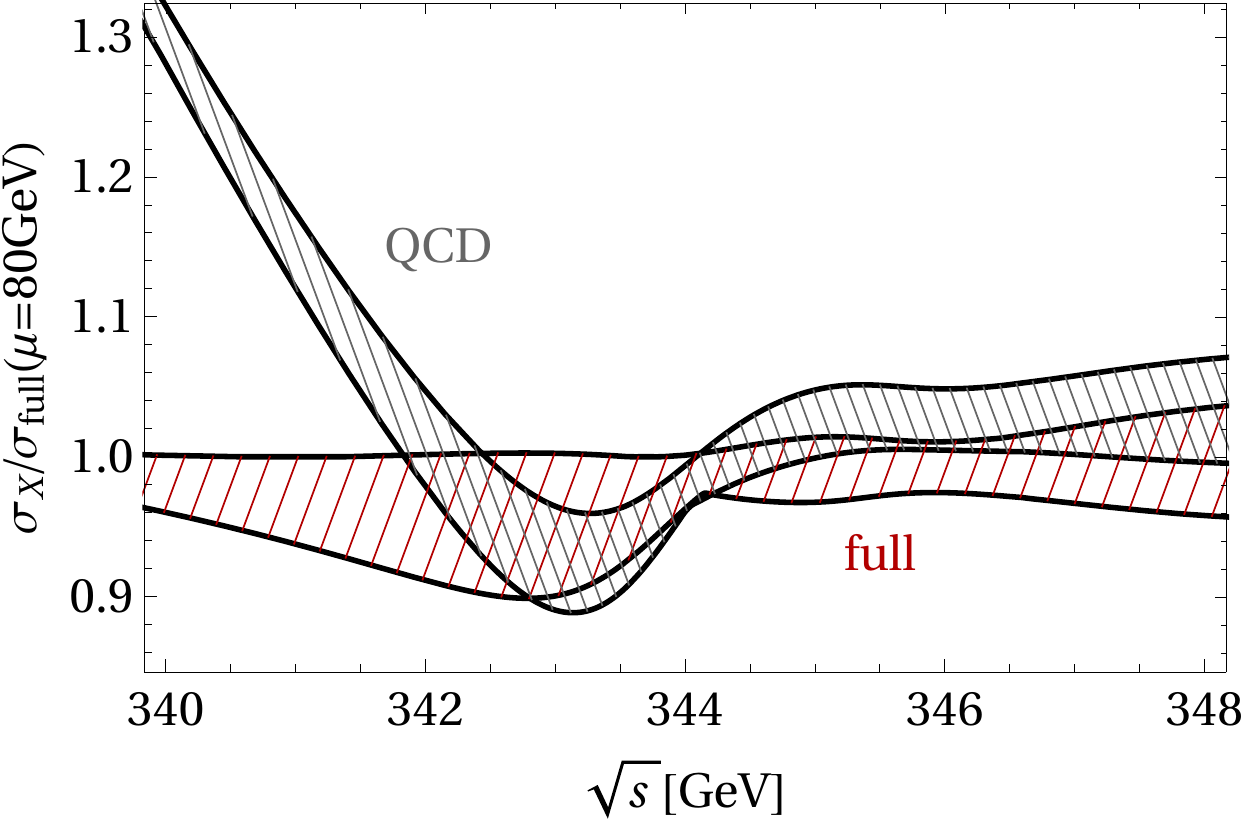}
  \caption{The cross section in pure QCD (grey hatched band) and including
  the electroweak and non-resonant corrections (red hatched band). The bands
  represent the uncertainty from scale variation. The upper panel shows the
  cross section in pb and the lower panel shows the results normalized to the
  full one for the central scale $\mu_r=80\,$GeV.\label{fig:AllPlots}}
\end{figure}
%%%%%%%%%%%%%%%%%%%%%%%%%%%%%%%%%%%%%%%%%%%%%%%%%%%%%%%%%%%%%%%%%%%%%%%%%

We define a reference QCD prediction by adding the small P-wave
contribution~\cite{Beneke:2013kia} to the S-wave result
of~\cite{Beneke:2015kwa}. The result is shown by the grey hatched band
labelled ``QCD'' in Figure~\ref{fig:AllPlots}, which is spanned by variation
of the renormalization scale $\mu_r$ between 50\,GeV and
350\,GeV. Figure~\ref{fig:AllPlots} also shows the net effect of all the
corrections discussed above, excluding ISR, which will be considered below.
These non-QCD effects slightly increase the height of the peak and move it
towards smaller center-of-mass energies. Above the peak the cross section
is slightly decreased by about $3.0-3.6\%$. Overall, the effect of the
non-QCD corrections is to make the resonance more pronounced.
The largest effect is observed below the peak, where the absorptive parts
of the matching coefficients and the non-resonant contribution dominate the
non-QCD corrections. Here, the bands cease to overlap at around
$\sqrt{s}=341.8\,$GeV. The size of the uncertainty band is somewhat increased
and now reaches up to $\pm5.2\%$ directly below the peak, where the
uncertainty estimate for the QCD result is $\pm3.8\%$. In the remaining
regions it is about $\pm3\%$. The increase of the scale uncertainty is
mainly due to the Higgs potential insertion as was already
observed in~\cite{Beneke:2015lwa}.

%%%%%%%%%%%%%%%%%%%%%%%%%%%%%%%%%%%%%%%%%%%%%%%%%%%%%%%%%%%%%%%%%%%%%%
\begin{figure}
  \centering
  \includegraphics[width=0.6\textwidth]{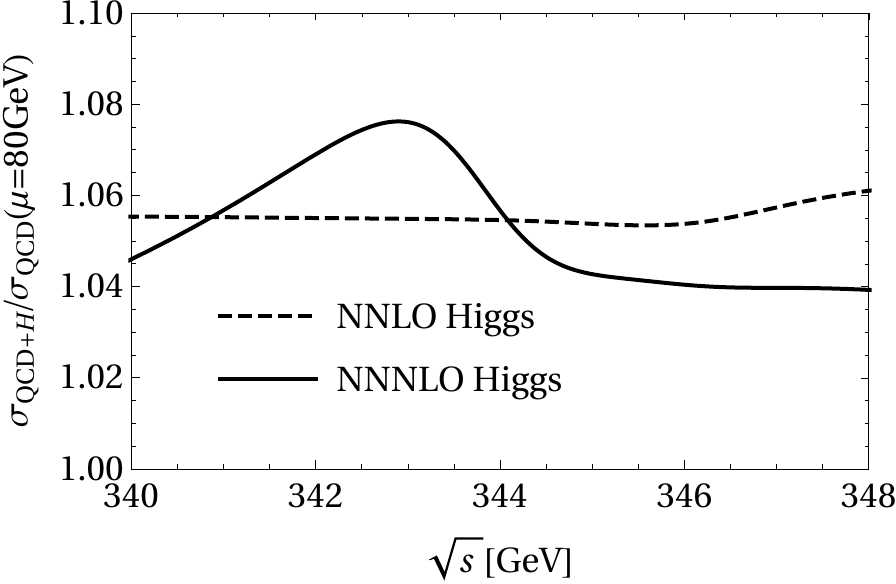}\\[0.2cm]
  \includegraphics[width=0.6\textwidth]{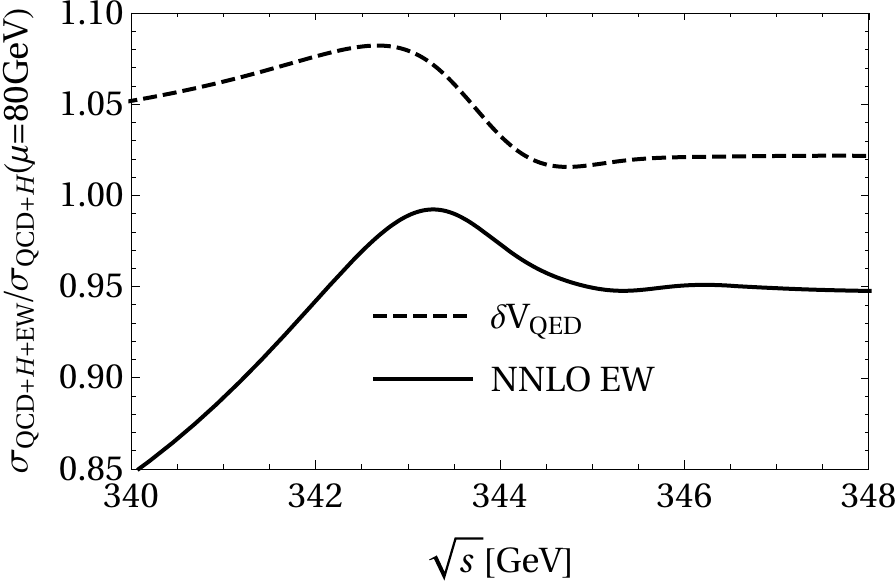}\\[0.2cm]
  \includegraphics[width=0.6\textwidth]{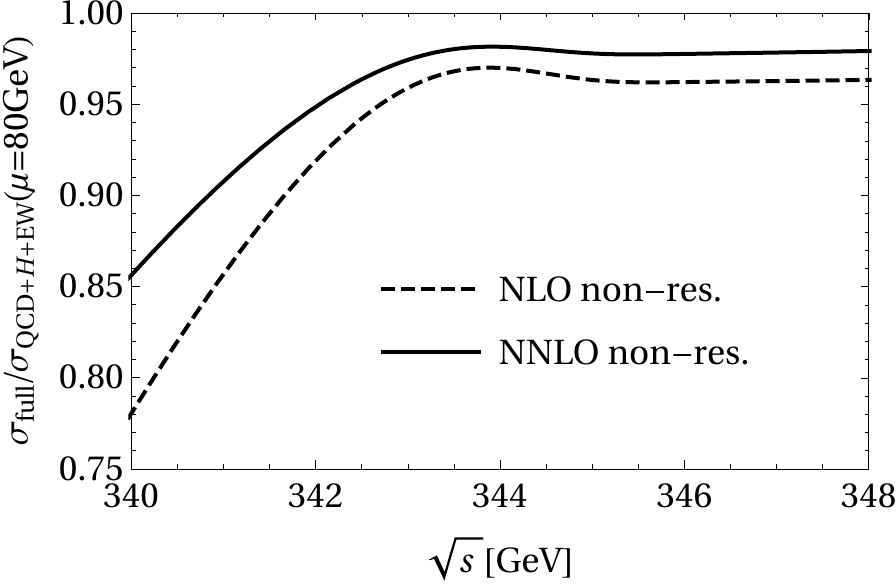}
  \caption{Relative corrections to the cross section by adding Higgs (top),
  electroweak (middle) and non-resonant (bottom) effects cumulatively.
  \label{fig:NonQCDPlots}}
\end{figure}
%%%%%%%%%%%%%%%%%%%%%%%%%%%%%%%%%%%%%%%%%%%%%%%%%%%%%%%%%%%%%%%%%%%%%%

The size of the individual contributions is shown in
Figure~\ref{fig:NonQCDPlots}.
We have already discussed the Higgs, QED Coulomb and NLO non-resonant
corrections in~\cite{Beneke:2015lwa}, but briefly recapitulate the results
here to give a complete overview over the non-QCD correction up to NNLO.
In the top-panel we show the relative effect of the Higgs contribution
$\sigma_{H}$ at NNLO and NNNLO. At NNLO there is an almost constant
relative shift, because only the hard-matching coefficient $c_{vH}^{(2)}$
is present. At NNNLO, there is also a contribution from the local Higgs
potential, which modifies the position of the peak. Due to the attractive
nature of the potential, the binding energy is increased and the peak is
shifted to the left. At the same time the Higgs corrections increase the
cross section by $3-8\%$, depending on the value of $\sqrt{s}$, and make
the peak more pronounced. The comparison of the dashed and solid curves
demonstrates that the inclusion of the NNNLO corrections is important for
correctly capturing the energy dependence of the Higgs effects, which is
crucial for a reliable measurement of the top Yukawa coupling.

The remaining electroweak contributions to the `partonic' resonant cross
section, $\sigma_{\delta V_\text{QED}}$, $\sigma_{\Gamma}$,
$\sigma_{C_{\text{EW}}^{(k)}}$, $\sigma_{C^{(k)}_{\text{Abs,bare}+Z_t}}$,
are shown in the middle panel. The dashed line corresponds to the correction
from the QED Coulomb potential only. It is attractive and therefore leads to
an increase of the cross section by $2-8\%$ and a shift of the peak towards
smaller center-of-mass energy. The solid line shows the full correction.
The width contribution $\sigma_\Gamma$ decreases the cross section by
$0-1.5\%$ depending on the energy.
Including the real part of the electroweak matching coefficient leads to
an almost constant relative shift of about $-3.3\%$.
The absorptive part of the matching coefficient multiplies the real part of
the non-relativistic Green function, which has a broad peak, roughly centered
around the point where the imaginary part has its maximal slope, on top of
a smooth background. Thus, the absolute contribution has only a mild energy
dependence and is of the order of $-3\%$ near and above the peak. However,
it becomes even more important below the peak, where the cross section is
small and modified by up to $-15\%$.

The lower panel illustrates the behaviour of the non-resonant contribution to
the total cross section. Its absolute size is nearly energy-independent. Thus,
the shape of the curves is given by the ``inverse'' of the resonant cross
section. At NLO, the effect is of the order $-(3-4)\%$ near and above the
peak and reaches up to $-22\%$ for low center-of-mass energies, where the
resonant cross section becomes small. The NNLO corrections compensate about
$40\%$ of the NLO result. This is in contrast to the findings
of~\cite{Jantzen:2013gpa,Ruiz-Femenia:2014ava}, where an enhancement of the
negative non-resonant correction from an approximate NNLO result was
observed. The apparent discrepancy is entirely explained by the very
different choice made in \cite{Jantzen:2013gpa,Ruiz-Femenia:2014ava} for
the finite-width scale ($\mu_w=30\,$GeV) compared to the present
($\mu_w=350\,$GeV). The dependence
of the full result on $\mu_w$
is very mild as discussed below and, thus, mainly the size of the individual
contributions is affected -- most notably the non-resonant correction and the
one from the absorptive part of the hard matching coefficients.

%%%%%%%%%%%%%%%%%%%%%%%%%%%%%%%%%%%%%%%%%%%%%%%%%%%%%%%%%%%%%%%%%%%%%%
\begin{figure}
  \centering
  \includegraphics[width=0.51\textwidth]{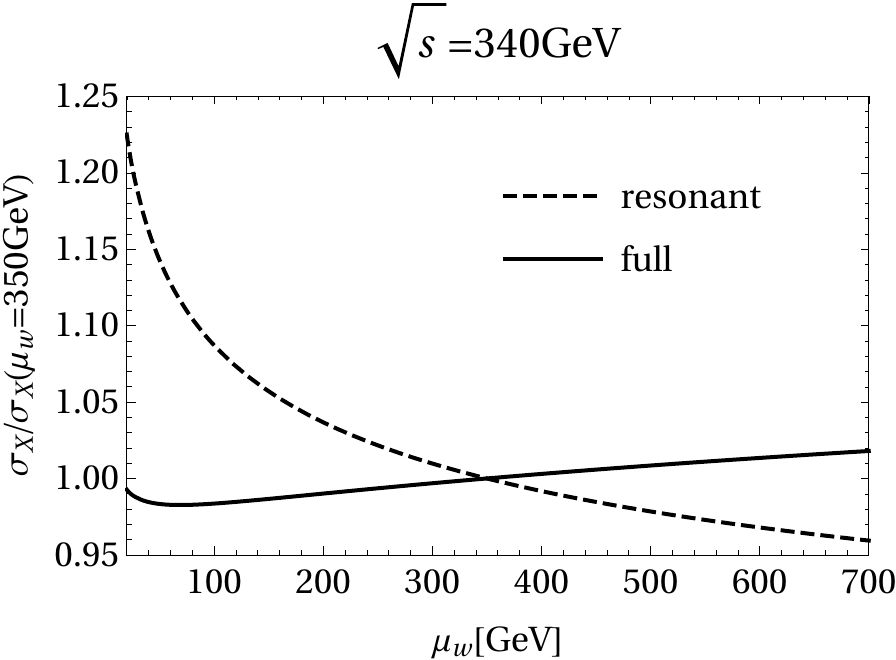}\\[0.3cm]
  \includegraphics[width=0.51\textwidth]{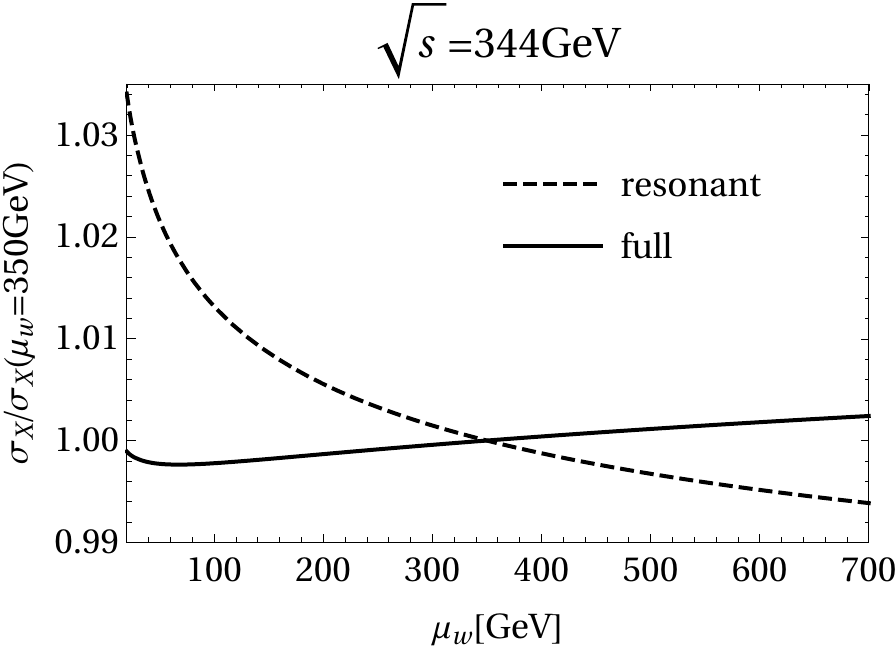}\\[0.3cm]
  \includegraphics[width=0.51\textwidth]{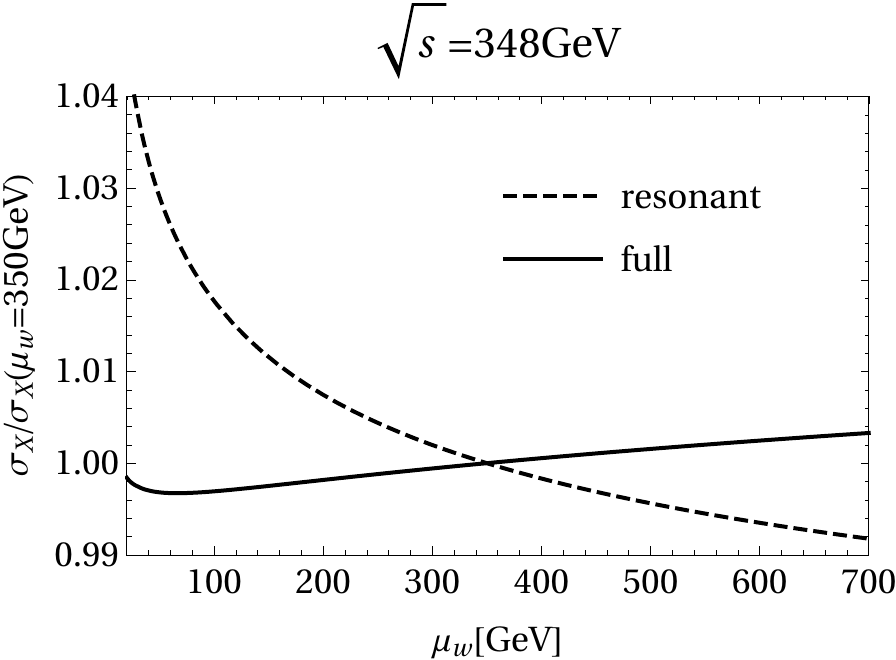}
  \caption{Dependence of the resonant-only and full cross section on the
  scale $\mu_w$ normalized to the one at $\mu_w=350\,$GeV for
  $\sqrt{s}=340\,$GeV (top panel), $\sqrt{s}=344\,$GeV (middle panel)
  and $\sqrt{s}=348\,$GeV (bottom panel).\label{fig:MuwPlots}}
\end{figure}
%%%%%%%%%%%%%%%%%%%%%%%%%%%%%%%%%%%%%%%%%%%%%%%%%%%%%%%%%%%%%%%%%%%%%%

We recall that the bands in Figure~\ref{fig:AllPlots} only include the
variation of the renormalization scale between 50\,GeV and 350\,GeV, while
the scale $\mu_w=350\,$GeV is kept fixed. The dependence on the scale
$\mu_w$ cancels exactly between all contributions of a given order. We show
the $\mu_w$ dependence of the resonant cross section and the full cross
section in Figure~\ref{fig:MuwPlots}. For the resonant-only cross section,
it is mild near and above the peak, but is significantly larger than the
renormalization scale dependence below the peak. The sensitivity to $\mu_w$
is greatly reduced for the full cross section, where the variation between
20 and 700 GeV considered in the plots only yields a $\pm(0.2-0.3)\%$ effect
near and above the peak and only a mild $\pm1.8\%$ below the peak. The
remaining $\mu_w$ dependence is of NNNLO, where the full QCD corrections,
but only a few electroweak effects are included and therefore no full
cancellation is achieved.

The central value  $\mu_w=350\text{ GeV}$ for the finite-width scale is
chosen near the hard scale to make the corresponding logarithms in the
non-resonant part small. The logarithms of $\mu_w$ are introduced by the
separation into different momentum regions and are therefore spurious in
nature. Explicitly, some of the `large' logarithms $\ln v$ contained in
the full cross section are split as follows
\begin{equation}
 \sigma_\text{full} \supset \ln v = \underbrace{\ln\frac{\mu_w}{m_t}}_{\subset\,\sigma_\text{non-res}}
 + \underbrace{\ln\frac{m_t v}{\mu_w}}_{\subset\,\sigma_\text{res}},
\end{equation}
where the first logarithm to the right of the equality sign
is part of the non-resonant
contribution and the second one of the resonant. Choosing
$\mu_w\sim m_t$ captures the `large' logarithms present at NNNLO in the
resonant part and renders the logarithms contained in the non-resonant
part small. While the NNNLO resonant contributions are already partially
known, the NNNLO non-resonant corrections are beyond the present
computational limits. Thus our scale choice minimizes the uncertainty from
the missing NNNLO contributions.\footnote{The same argument motivated the
different choice made in \cite{Jantzen:2013gpa,Ruiz-Femenia:2014ava}, since
in these papers the NNLO resonant electroweak contribution
was not available.}

\begin{figure}
  \centering
  \includegraphics[width=0.7\textwidth]{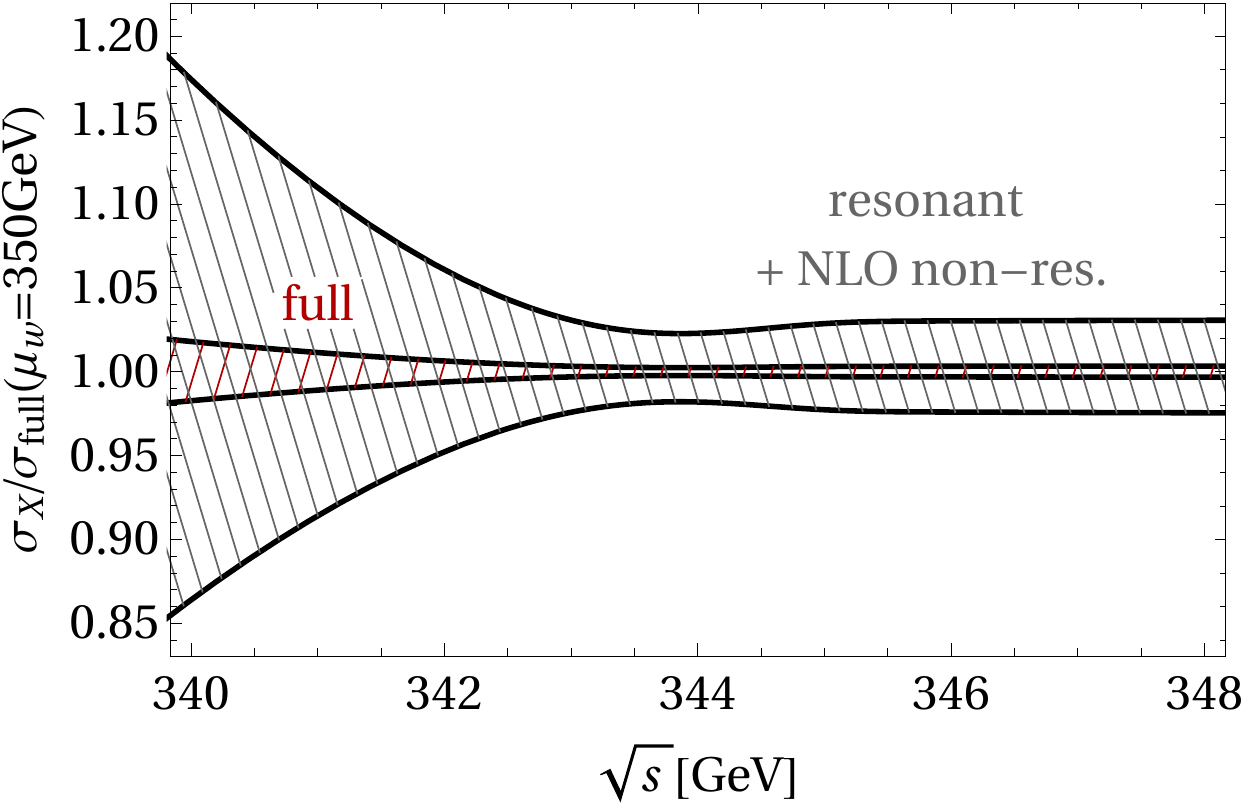}
  \caption{The resonant plus NLO non-resonant (grey hatched band) and
  full cross section (red hatched band) are shown for $\mu_r=80\,$GeV.
  The bands are the envelope of the values obtained by varying $\mu_w$
  between 20 and 700 GeV. The results have been normalized to the full
  cross section for the central scale $\mu_w=350\,$GeV.
  \label{fig:AllMuWPlots}}
\end{figure}

Variation of the scale $\mu_w$ can be used to estimate the size of the
missing NNNLO non-resonant corrections. The corresponding bands for the
resonant plus NLO non-resonant and full cross section are shown in
Figure~\ref{fig:AllMuWPlots}, where we have varied $\mu_w$ between 20
and 700 GeV. We observe that the inner (red) band is entirely contained in the
outer (grey) one and much narrower. Thus, the chosen range of the finite-width
scale variation provides a reasonable estimate of the NNLO
non-resonant correction. However, an estimate of the missing
NNNLO non-resonant correction based on the width of the ``full'' (red) band in
Figure~\ref{fig:AllMuWPlots} is potentially less reliable, because the
leading NNNLO terms might not cause any $\mu_w$ dependence. This would be
similar to the situation at NLO, where the leading non-resonant effect
arises, yet there is no $\mu_w$ dependence of the resonant contribution
at this order at all, since the divergence from factorizing resonant
and non-resonant contributions is purely linear.

%%%%%%%%%%%%%%%%%%%%%%%%%%%%%%%%%%%%%%%%%%%%%%%%%%%%%%%%%%%%%%%%%%%%%%
%%%%%%%%%%%%%%%%%%%%%%%%%%%%%%%%%%%%%%%%%%%%%%%%%%%%%%%%%%%%%%%%%%%%%%

%%%%%%%%%%%%%%%%%%%%%%%%%%%%%%%%%%%%%%%%%%%%%%%%%%%%%%%%%%%%%%%%%%%%%%
\begin{figure}
  \centering
  \includegraphics[width=0.65\textwidth]{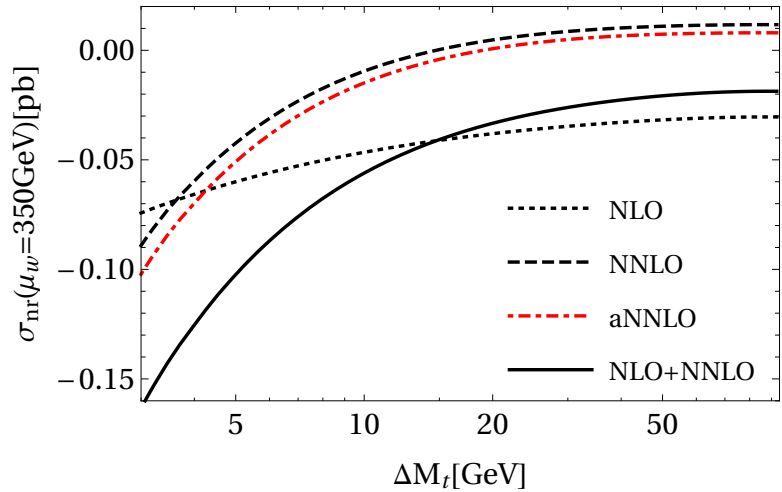}
  \caption{The dependence of the non-resonant contribution to the cross
  section on the invariant mass cut~\eqref{eq:definitioncut}. The dotted
  line shows the result at NLO and the dashed line the NNLO correction
  (without the NLO terms). The sum of both is drawn as a solid line.
  The red dot-dashed line denotes the approximate NNLO result (without
  the NLO terms) from~\cite{Jantzen:2013gpa}. The full cross section
  corresponds to $\Delta M_t=m_t-m_W$.\label{fig:SigmaNonresCutPlot}}
\end{figure}
%%%%%%%%%%%%%%%%%%%%%%%%%%%%%%%%%%%%%%%%%%%%%%%%%%%%%%%%%%%%%%%%%%%%%%

We discussed the possibility of imposing loose cuts, which affect only
the non-resonant part of the cross section, in Section~\ref{sec:nonres}.
The dependence on the cut defined in~\eqref{eq:definitioncut} is shown in
Figure~\ref{fig:SigmaNonresCutPlot}, where the dotted and solid lines denote
the NLO and NNLO non-resonant contribution. Very loose cuts with
$\Delta M_t\geq30\,$GeV have only a mild influence on the cross section.
Tighter cuts $\Delta M_t = (30,20,10,5)\,$GeV reduce the cross section
by $(0.007,0.014,0.037,0.084)\,$pb. We observe that for $\Delta M_t$ around
4\,GeV the NNLO non-resonant contribution becomes as large as the NLO one.
Here, the assumption that the cut is loose is no longer appropriate and
our description breaks down.
The dashed line in Figure~\ref{fig:SigmaNonresCutPlot} shows the approximate
NNLO result~\cite{Jantzen:2013gpa}, which includes only the
endpoint-divergent terms as $\Delta M_t\to 0$, for comparison. It describes
the dependence on the cut very well, since the endpoint-divergent terms are
most sensitive to it, but it is shifted by -0.004\,pb for the full cross
section and up to -0.013\,pb including invariant mass cuts. In the absence
of any cuts the exact result corresponds to a $46\%$ correction with respect
to the approximate NNLO result. We note, however, that for the scale choice
of~\cite{Jantzen:2013gpa} and in the range of loose, but not too loose
cuts $\Gamma_t\ll\Delta M_t\ll m_t$ the approximation is much better.

%%%%%%%%%%%%%%%%%%%%%%%%%%%%%%%%%%%%%%%%%%%%%%%%%%%%%%%%%%%%%%%%%%%%%%
\begin{figure}
\centering
\includegraphics[width=0.65\textwidth]{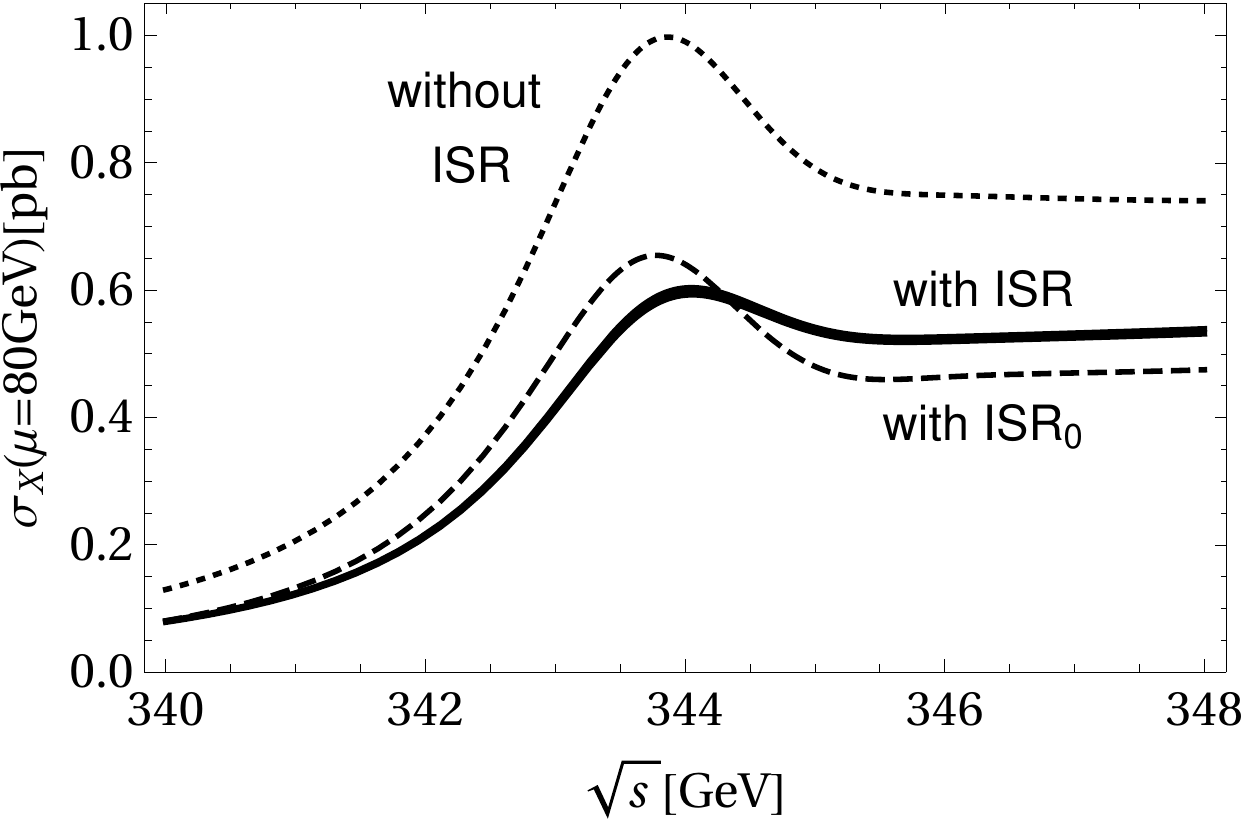}
\caption{The effect of initial-state QED radiation on the cross section.
The dotted curve shows the full result without ISR. The solid band
(with ISR) is the envelope of results obtained by convoluting the full
`partonic' cross section with the structure functions with different
systematics (see text). The dashed line (with ISR$_0$) is obtained by
convoluting only the leading order `partonic' cross section with the
structure functions and adding the full `partonic' corrections on
top.\label{fig:ISRPlot}}
\end{figure}
%%%%%%%%%%%%%%%%%%%%%%%%%%%%%%%%%%%%%%%%%%%%%%%%%%%%%%%%%%%%%%%%%%%%%%

We finally discuss the effects of initial-state QED radiation, which have so
far only been taken into account in the experimental studies.
Figure~\ref{fig:ISRPlot} shows the partonic cross section
$\sigma^\text{conv}$ and its convolution with the electron structure
functions. The QED contribution $\sigma^\text{conv}_{\rm IS}$ to the partonic
cross section from (\ref{eq:sigmaISconv}) is a small effect of the order
$-(0.6-1.3)\%$. The convolution, however, reduces the cross section by
$28-44\%$. The black band is spanned by four different implementations of
the convolution (\ref{eq:convolutionstructurefunctions}) of the full NNNLO QCD
plus NNLO EW cross section with the structure functions. This involves an
extrapolation of the cross section for energy values outside of the range
of the grids available in \texttt{QQbar\_Threshold}\cite{Beneke:2016kkb}.
We either use the shape of the LO cross section below $\sqrt{s}=328$\,GeV,
rescaled to match the full result at $\sqrt{s}=328$\,GeV, or an alternative
implementation that interpolates linearly between
$\sigma(\sqrt{s}=320\,\text{GeV})=0\,$pb and our result at
$\sqrt{s}=328$\,GeV. Numerically, we find a small difference of $0.1\%$
near and above the peak, which goes up to $0.8\%$ at
$\sqrt{s}=340\,$GeV.\footnote{While the grid could technically be extended
to smaller values of $\sqrt{s}$, the PNREFT and unstable-particle EFT
breaks down far below the threshold. Improving the accuracy in this
region would require matching the EFT description to the fixed-order
calculation of the full non-resonant process as discussed for a
single-particle resonance in \cite{Beneke:2003xh,Beneke:2004km}.}
For both extrapolations we consider the
convolution~\eqref{eq:convolutionstructurefunctions} with the structure
functions as defined in (\ref{eq:structurefn}) and a purely LL approximation
where we set $\beta = (2\alpha/\pi)\ln(s/m_e^2)$ in the structure function
and accordingly modify the non-logarithmic ISR
contribution~\eqref{eq:sigmaISconv} for the different subtraction
term (\ref{eq:ISsubtraction}). The difference is formally a NLL effect
and provides a rough estimate of the overall size of NLL ISR corrections.
It amounts to about $1.4$\% above the peak and reaches up to $2.1\%$ in
the region where the slope is large.

For comparison we furthermore show as the dashed line the expression
\begin{equation}
 \sigma_{\text{ISR}_0}(s)=\sigma^\text{conv}(s)-\sigma^\text{LO}(s)
+\int\limits_0^1dx_1\int\limits_0^1dx_2 \,\Gamma_{ee}^\text{LL}(x_1)
\Gamma_{ee}^\text{LL}(x_2)\sigma^\text{LO}(x_1x_2s),
\end{equation}
where the ISR resummation is only applied to the LO cross section.
Since the LL resummation modifies a N$^k$LO correction by order one,
the difference between ISR and ISR$_0$ is formally a NLO effect.
This emphasizes that it is mandatory to perform the convolution
with the full partonic result.

We see, as it is of course expected, that ISR is a huge effect,
reducing the cross section by $28-44\%$. It also leads to a significant
modification of the shape. The peak is shifted by almost
200\,MeV to the right and smeared out considerably. Its height is reduced
by about 40\%. This emphasizes the need for a full NLL treatment of ISR
and a proper analysis of the convergence and remaining uncertainty, which is
of universal importance for high-energy $e^+e^-$ collider processes,
but beyond the
scope of this work. We further note that at the level of NNLO electroweak
accuracy the partonic cross section depends on the scheme employed for
the electron structure function, and a phenomenological convolution
as often applied in experimental studies in an unspecified scheme is
no longer adequate.

%%%%%%%%%%%%%%%%%%%%%%%%%%%%%%%%%%%%%%%%%%%%%%%%%%%%%%%%%%%%%%%%%%%
%%%%%%%%%%%%%%%%%%%%%%    Sensitivity      %%%%%%%%%%%%%%%%%%%%%%%%
%%%%%%%%%%%%%%%%%%%%%%%%%%%%%%%%%%%%%%%%%%%%%%%%%%%%%%%%%%%%%%%%%%%

\subsection{Sensitivity to Standard Model parameters
\label{sec:parameters}}

%%%%%%%%%%%%%%%%%%%%%%%%%%%%%%%%%%%%%%%%%%%%%%%%%%%%%%%%%%%%%%%%%%%
\begin{figure}[t]
  \centering
  \includegraphics[width=0.65\textwidth]{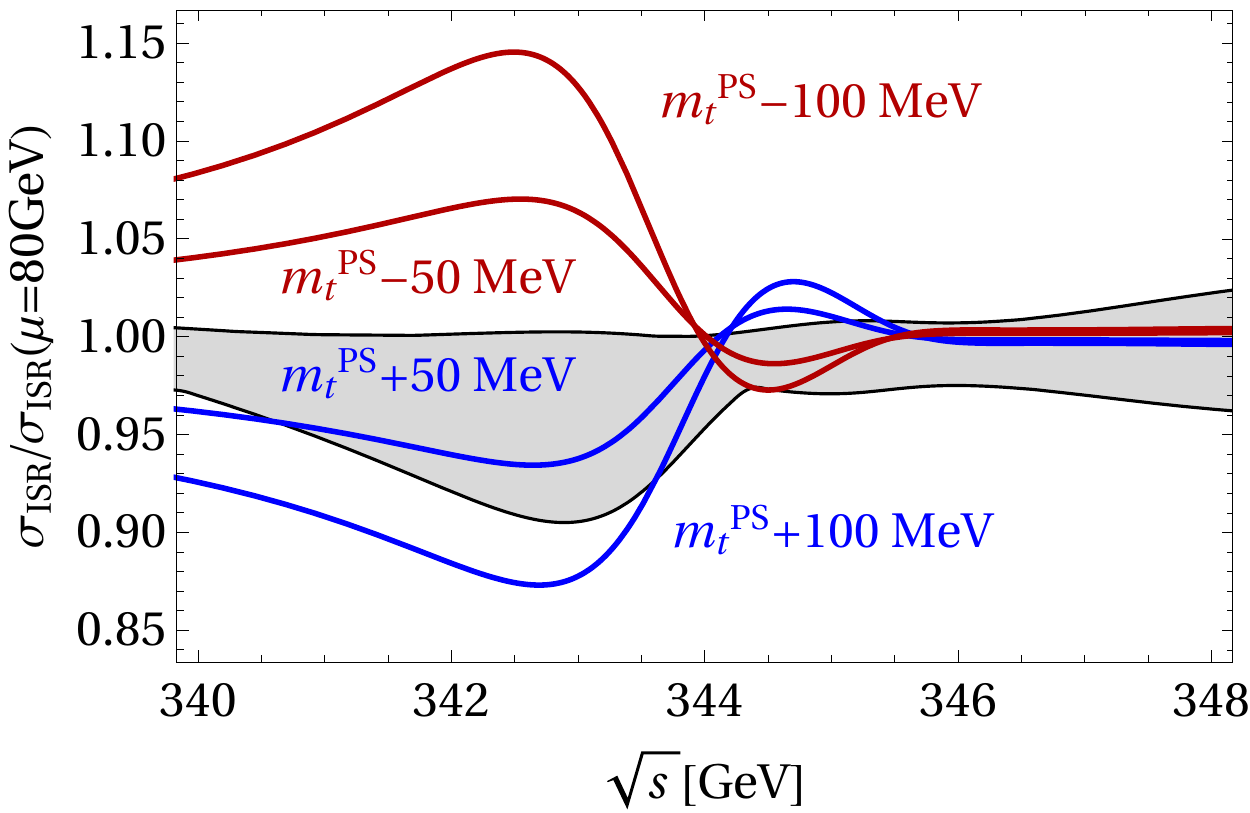}\\[1.0cm]
  \includegraphics[width=0.65\textwidth]{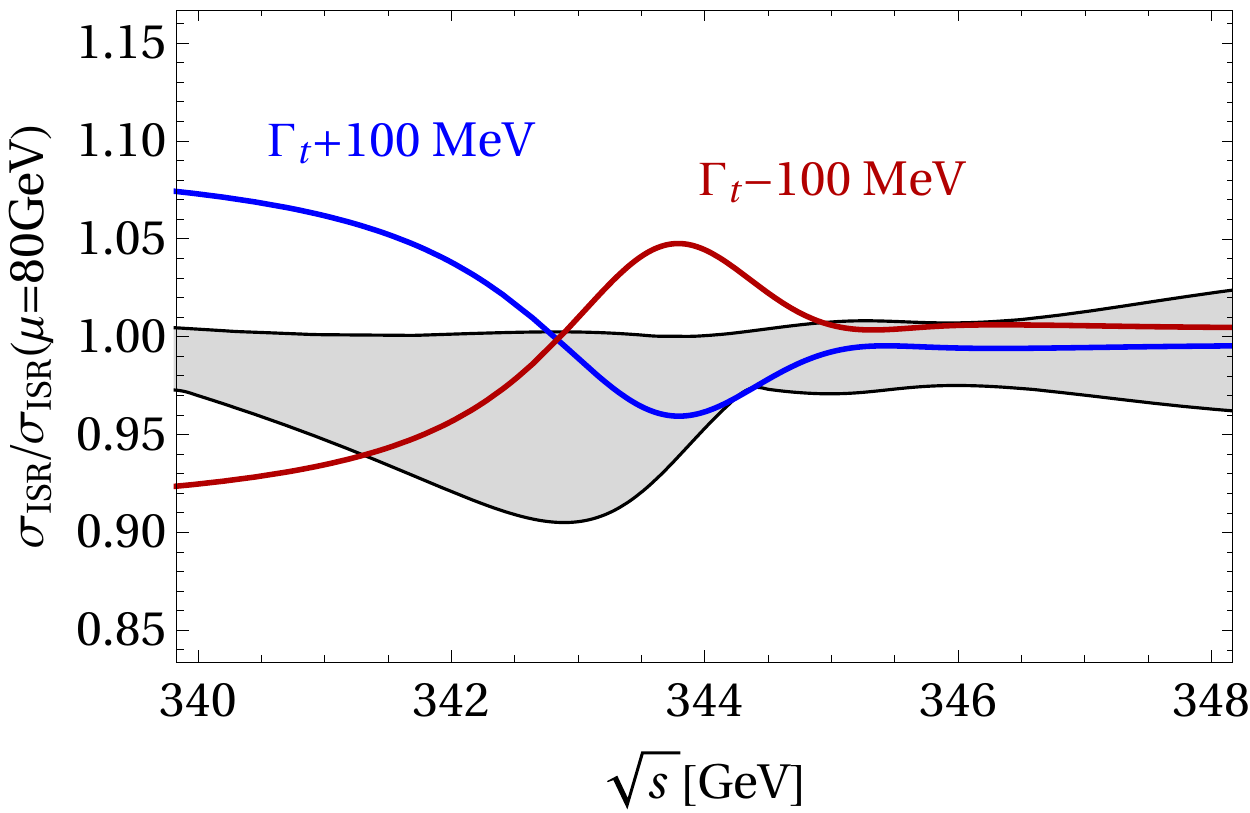}
  \caption{The cross section for the input values~\eqref{eq:pheno_inputs}
  up to variations of the top mass (top panel) and top width (bottom
  panel) is shown in comparison the uncertainty band from scale variation
  (cf. Figure~\ref{fig:AllPlots}). The prediction is normalized to the
  full cross section.}\label{fig:MassWidthPlot}
\end{figure}
%%%%%%%%%%%%%%%%%%%%%%%%%%%%%%%%%%%%%%%%%%%%%%%%%%%%%%%%%%%%%%%%%%%

\begin{figure}[t]
  \centering
  \includegraphics[width=0.65\textwidth]{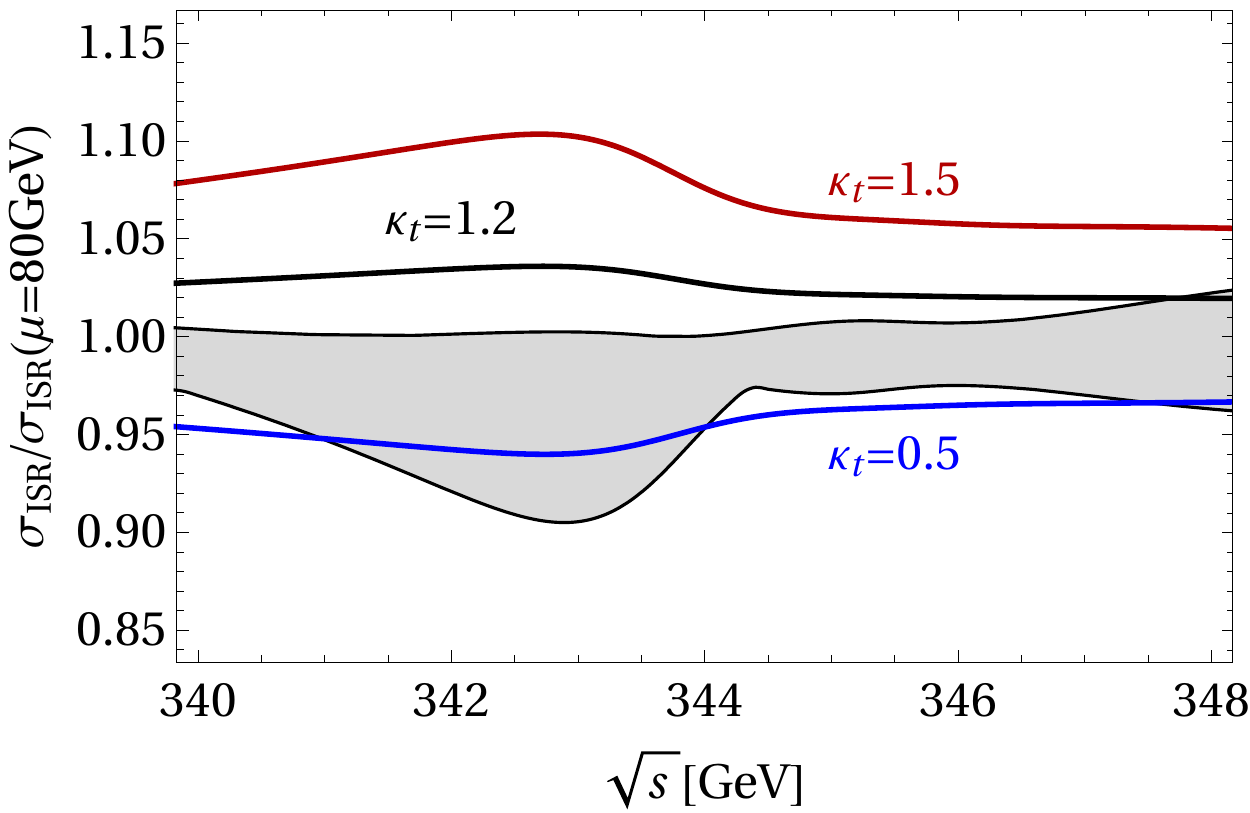}\\[1.0cm]
  \includegraphics[width=0.65\textwidth]{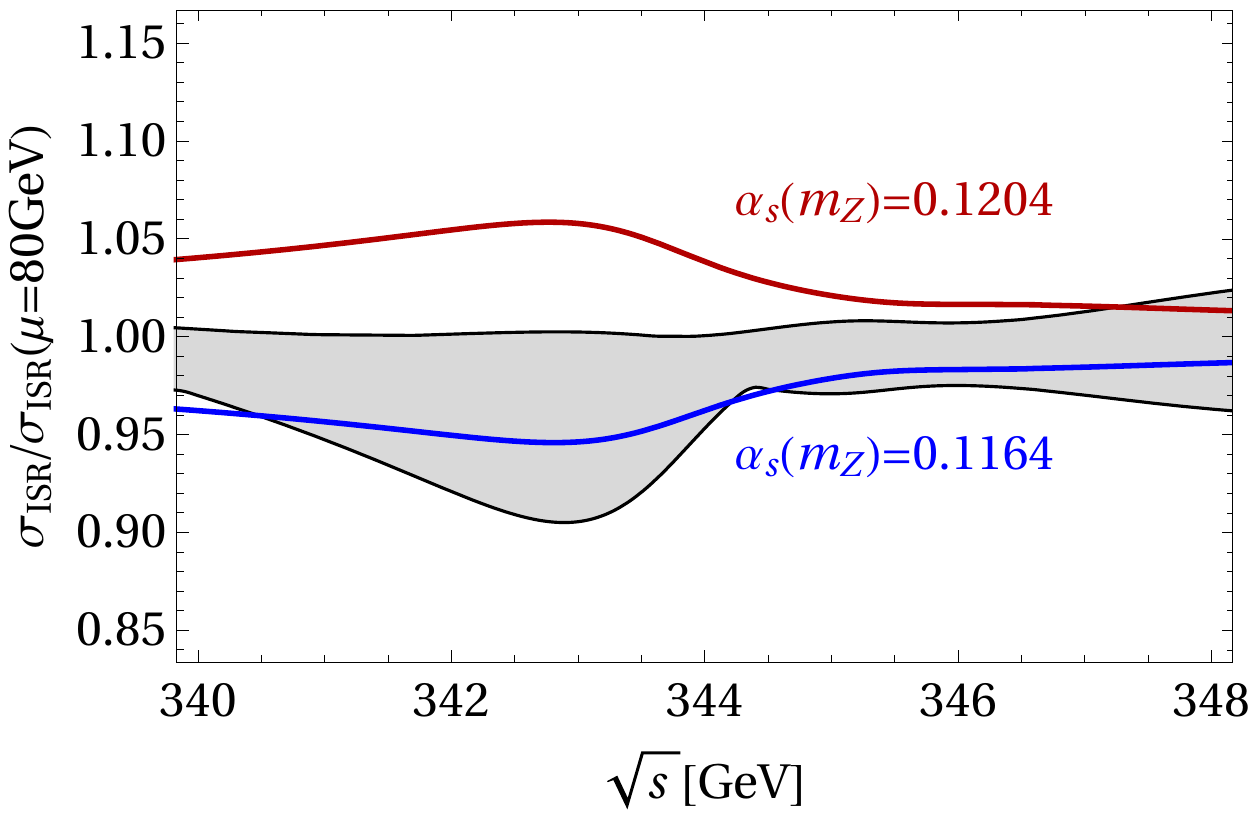}
  \caption{The upper (lower) panel show the effects of variations of the
  Yukawa coupling (strong coupling) on the cross section. The Yukawa coupling
  is parametrized as $y_t = \kappa_t\,y_t^\text{SM}$, where $y_t^\text{SM} =
  \sqrt{2}\,m_t/v$ is the Standard Model value. The predictions are normalized
  to the full cross section and the uncertainty band is the same as in
  Figure~\ref{fig:MassWidthPlot}.}
\label{fig:yukasvar}
\end{figure}

\begin{figure}
\centering
\includegraphics[width=0.65\textwidth]{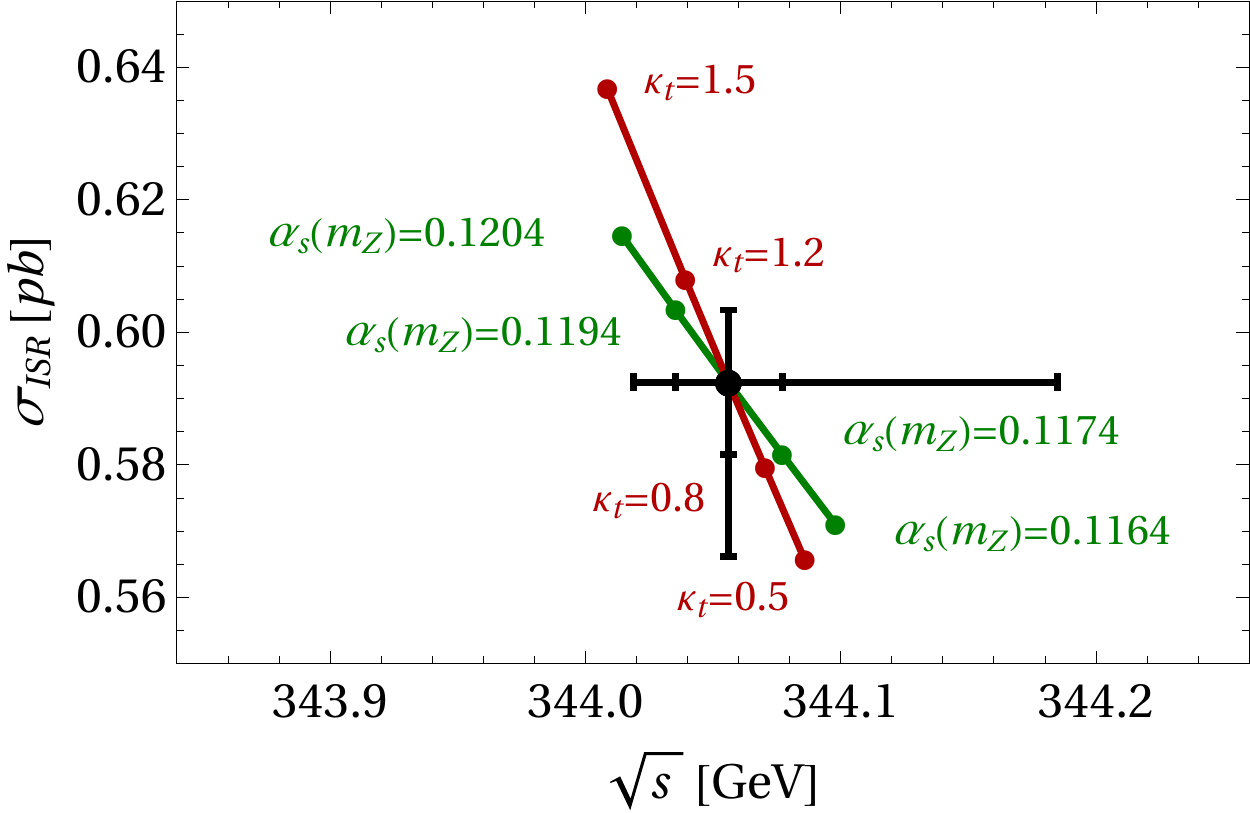}
\caption{The effect of a variation of the Yukawa coupling (strong coupling)
on the position and height of the peak is indicated by the red (green)
line and points. The black cross represents the theoretical uncertainty
with the default input parameters. The outer error bar is obtained by
adding the uncertainties from the renormalization scale and variation of
$\alpha_s(m_Z)$ by $\pm0.001$ in quadrature, while the inner bar shows
only the latter contribution.}
\label{fig:yukaspeak}
\end{figure}

Since the non-QCD effects computed in this paper cause substantial
corrections to the cross section we provide an update of the discussion
in~\cite{Beneke:2015kwa,Beneke:2015lwa} of the sensitivity of the
top threshold scan to Standard Model parameters.
Figures~\ref{fig:MassWidthPlot}, \ref{fig:yukasvar} and \ref{fig:yukaspeak}
estimate the sensitivity by comparing
the effects of parameter variations to the scale uncertainty in terms
of the relative variation to a reference cross section.

%%%%%%%%%%%%%%%%%%%%%%%%%%%%%%%%%%%%%%%%%%%%%%%%%%%%%%%%%%%%%%%%%%%
%%%%%%%%%%%%%%%%%%%%%%%%%%%%%%%%%%%%%%%%%%%%%%%%%%%%%%%%%%%%%%%%%%%

All electroweak and non-resonant effects discussed in this paper are
included in the figures, in particular also the ISR corrections. There
are small quantitative differences with respect
to~\cite{Beneke:2015kwa,Beneke:2015lwa}, such as a small reduction of the
height of the peaks present in the top-mass
variation curves near 344.5~GeV, but the essence of the results and
the associated conclusions remain unchanged. It is especially noteworthy that
the huge ISR correction discussed above does not degrade the sensitivity.
Since the bulk of the ISR correction is produced by the convolution
with the luminosity function, we expect that the additional convolution
of the cross section with the collider-specific beam function will
not dilute the sensitivity to the parameters, either.

From Figures~\ref{fig:MassWidthPlot} and \ref{fig:yukasvar} we expect
the threshold scan to be sensitive to variations of about $\pm40$ MeV
 for the top-quark PS mass, $\pm60$ MeV for the top-quark width,
$_{-25}^{+20}\,\%$ for the top-quark Yukawa coupling and $\pm0.0015$ for the
strong coupling constant $\alpha_s(m_Z)$, when only a single parameter
is varied at a time. These numbers are obtained from comparing the width
of the band for the parameter variation with the one from the theoretical
uncertainty, and requiring that the former is larger than the latter
for a sufficient range in energy. This leaves open the question of how
well the corrections from the simultaneous variation of several parameters
can be disentangled from their energy dependence, which particularly
concerns the Yukawa and the strong coupling, where variations lead to similar
effects as seen in Figure~\ref{fig:yukasvar} for the energy dependence
and in Figure~\ref{fig:yukaspeak} for the position and height of the peak
in the cross section. This needs to be addressed within realistic
simulations, which include experimental uncertainties as well.

%%%%%%%%%%%%%%%%%%%%%%%%%%%%%%%%%%%%%%%%%%%%%%%%%%%%%%%%%%%%%%%%%%%%%%%%%%
%%%%%%%%%%%%%%%%%%%%%%%%%%%%%%%%%%%%%%%%%%%%%%%%%%%%%%%%%%%%%%%%%%%%%%%%%%

%%%%%%%%%%%%%%%%%%%%%%%%%%%%%%%%%%%%%%%%%%%%%%%%%%%%%%%%%%%%%%%%%
%%%%%%%%%%%%%%%%%%%     Conclusions      %%%%%%%%%%%%%%%%%%%%%%%%
%%%%%%%%%%%%%%%%%%%%%%%%%%%%%%%%%%%%%%%%%%%%%%%%%%%%%%%%%%%%%%%%%

\section{Conclusions\label{sec:conclusions}}

The recent advance in the QCD calculation of the top
anti-top threshold~\cite{Beneke:2015kwa}
has motivated the consideration of non-QCD effects of potentially
similar size to the third-order QCD correction. While Higgs/top-Yukawa
coupling effects up to the third order were already obtained
in \cite{Beneke:2015lwa}, the present work completed the calculation
of NNLO electroweak corrections and in particular the NNLO non-resonant
contribution to the $e^+ e^-\to b\bar{b}W^+W^- X$ process near
the top-pair production threshold. This elevates the theoretical prediction
to NNNLO QCD plus NNLO electroweak accuracy, including for the first
time initial-state radiation in a scheme consistent with one-loop
QED corrections. The new effects are indeed non-negligible compared
to the $\pm 3\%$ accuracy estimated for the pure QCD calculation and
are therefore essential for accurate top and Standard Model parameter
determinations from the threshold. They have been implemented in the
new version 2 of the public code
\texttt{QQbar\_threshold}~\cite{Beneke:2016kkb}.

Despite the level of sophistication already achieved, further
improvement could be considered or might be necessary, such as the
combination of the NNLL summation of logarithms of $E/m_t$ in the
QCD part \cite{Hoang:2013uda}
with the NNNLO fixed-order calculation~\cite{Beneke:2015kwa}, the inclusion
of already known NNNLO electroweak corrections
(see \cite{Eiras:2006xm,Kiyo:2008mh}) or the one-loop correction
to the Higgs potential (a N4LO effect) together with the terms
required to make these additions factorization-scheme independent.
To cancel the finite-width $\mu_w$ scale dependence of the
NNNLO QCD result completely, the non-resonant part is needed
to the same accuracy, which appears prohibitive at present. Finally,
a consistent implementation of QED initial-state radiation with
next-to-leading logarithmic accuracy seems to be a general prerequisite
for accurate predictions of scattering at a future high-energy
$e^+ e^-$ collider.

%%%%%%%%%%%%%%%%%%%%%%%%%%%%%%%%%%%%%%%%%%%%%%%%%%%%%%%%%%%%%%%%%%%%%%%%
%%%%%%%%%%%%%%%%%%%%    Acknowledgements      %%%%%%%%%%%%%%%%%%%%%%%%%%
%%%%%%%%%%%%%%%%%%%%%%%%%%%%%%%%%%%%%%%%%%%%%%%%%%%%%%%%%%%%%%%%%%%%%%%%

\noindent
\subsubsection*{Acknowledgements}

We are grateful to B.~Chokouf\'e Nejad, C.~Degrande, R.~Frederix,
O.~Mattelaer, J.~Piclum, V.~Shtabovenko, F.~Simon, and
J.F.~von Soden-Fraunhofen for helpful discussions. We made use of
Axodraw~\cite{Vermaseren:1994je} and JaxoDraw~\cite{Binosi:2008ig}
to draw Feynman diagrams. AM was supported by a
European Union COFUND/Durham Junior Research Fellowship under EU grant
agreement number 267209. PRF acknowledges
financial support from the ``Spanish Agencia Estatal de Investigacion"
(AEI), the EU ``Fondo Europeo de Desarrollo Regional" (FEDER) through
the project FPA2016-78645-P, and from the grant ``IFT Centro de Excelencia
Severo Ochoa SEV-2016-0597". MB and TR have been supported by the
Gottfried Wilhelm Leibniz programme of the Deutsche Forschungsgemeinschaft
(DFG), and the DFG cluster of excellence ``Origin and Structure of the
Universe.''

%%%%%%%%%%%%%%%%%%%%%%%%%%%%%%%%%%%%%%%%%%%%%%%%%%%%%%%%%%%%%%%%%%%%%%%%%%%

\appendix

%%%%%%%%%%%%%%%%%%%%%%%%%%%%%%%%%%%%%%%%%%%%%%%%%%%%%%%%%%%%%%%%%%%%%%%%%%%
\section{Implementation in \texttt{QQbar\_threshold}}
\label{app:implementation}

The new NNLO corrections have been implemented in the
new version 2 of the public code \texttt{QQbar\_threshold}.
In the following, we summarize the changes and give code examples for new
functions. \texttt{QQbar\_threshold} can be downloaded from\\[-0.3cm]

\centerline{\url{https://www.hepforge.org/downloads/qqbarthreshold/}}

\vskip0.3cm \noindent An updated
online manual is available under
\url{https://qqbarthreshold.hepforge.org/}.

\subsection{Non-resonant corrections}
\label{sec:impl_nonres}

By default, the \lstinline[language=C++]!ttbar_xsection! function
now includes the NNLO non-resonant contribution to the cross
section. The NLO and NNLO corrections can be controlled individually
with the \lstinline[language=C++]!contributions! option. For example,
\begin{lstlisting}[language=C++,caption={}]
const double a_NLO = 1.0;
const double a_NNLO = 0.0;
options opt;
opt.contributions.nonresonant = {{a_NLO, a_NNLO}};
ttbar_xsection(sqrt_s, {mu, mu_w}, {mt, width}, order, opt);
\end{lstlisting}
will calculate the cross section with the NLO non-resonant correction
multiplied by \lstinline[language=C++]!a_NLO! and the NNLO correction
multiplied by \lstinline[language=C++]!a_NNLO!. The equivalent
\texttt{Mathematica} code is
\begin{lstlisting}[language=Mathematica,caption={}]
aNLO = 1.0;
aNNLO = 0.0;
TTbarXSection[
   sqrts, {mu, muw}, {mt, width}, order,
   Contributions -> ExceptContributions[nonresonant -> {aNLO, aNNLO}]
]
\end{lstlisting}
As before, the complete nonresonant contribution can be disabled by
setting the option \lstinline[language=C++]!resonant_only! to
\lstinline[language=C++]!true! (
\lstinline[language=C++]!ResonantOnly -> True! in \texttt{Mathematica}).

\subsection{Initial-state radiation}
\label{sec:impl_ISR}

Initial-state radiation requires the computationally expensive
convolution with structure functions. Therefore, this correction is not
included automatically.

After defining the luminosity function
\begin{equation}
\label{eq:lumi_def}
{\cal L}(x) = \int_y^1 \frac{dy}{y} \,
\Gamma_{ee}^\text{LL}(y)\Gamma_{ee}^\text{LL}(x/y)
\end{equation}
with the electron structure functions $\Gamma_{ee}$
from~\eqref{eq:convolutionstructurefunctions} the cross section after
initial-state radiation is given by
\begin{equation}
  \label{eq:ISR_lumi}
  \sigma_{\rm ISR}(s) = \int_0^1dx \,{\cal L}(x)\,\sigma^{\text{conv}}(xs)\,.
\end{equation}
Here, $\sigma^{\text{conv}}$ is the partonic cross
section including the non-logarithmic initial-state radiation correction
$\sigma^{\text{conv}}_{\text{IS}}$ (see~\eqref{eq:sigmaISconv}).
The non-logarithmic correction can be included with the option setting
\begin{lstlisting}[language=C++,caption={}]
options opt;
opt.ISR_const = true;
\end{lstlisting}
in C++ and \lstinline[language=Mathematica]!ISRConst -> True! in
Mathematica. The default setting for this option is
\lstinline[language=C++]!false!. It should be set to
\lstinline[language=C++]!true! if (and only if) the logarithmically
enhanced component of the initial-state radiation is also included via
convolution with the luminosity function.

In principle, the convolution integral in~(\ref{eq:ISR_lumi}) covers
the whole energy range from zero to the nominal center-of-mass
energy. However, our prediction for the cross section is only valid in
the vicinity of the threshold. Sufficiently below the threshold the
actual cross section becomes negligible. This implies that we can
introduce a lower cut-off $x_\text{min}$ in the integral
(cf. Section~\ref{sec:pheno}). In the following we choose $x_\text{min}
= (330\,\text{GeV})^2/s$.

A further, purely numerical problem arises from the integrable
divergence of the luminosity function for $x \to 1$. In order to
eliminate this divergence, we can change the integration variable to $t
= (1-x)^\beta$ and write the cross section as
\begin{equation}
  \label{eq:ISR_lumi_mod}
    \sigma_{\rm ISR}(s) = \int_0^{t_\text{max}}\!\!dt \,\bar{{\cal
      L}}(t)\,\sigma^{\text{conv}}\big(x(t)\, s\big)\,,\qquad \bar{{\cal
      L}}(t) = \frac{(1-x)^{1-\beta}{\cal L}(x)}{\beta}\,,
     \qquad x(t) = 1 - t^\frac{1}{\beta}\,,
\end{equation}
with the modified luminosity function $\bar{{\cal L}}(t)$ and a
cut-off $t_\text{max} = (1 - x_\text{min})^\beta$. The function $\beta =
-2\alpha(\mu_\alpha)/\pi[\log(m_e^2/s) + 1]$ is available in
\texttt{QQbar\_threshold} as
\mbox{\lstinline[language=C++]!ISR_log(sqrt_s, alpha)!} in C++ and
\lstinline[language=Mathematica]!ISRLog! in Mathematica.

Finally, version 2 of \texttt{QQbar\_threshold} provides an
\lstinline!integrate! function in the header \texttt{integrate.hpp},
which can be used to compute the convolution integral as shown below.

The following C++ code prints the cross section $\sigma = 0.591736\,$pb
after initial state radiation for a center-of-mass energy of $\sqrt{s} =
344\,$GeV, including all known perturbative corrections:
\begin{lstlisting}[language=C++]
#include <iostream>
#include <cmath>

#include "QQbar_threshold/load_grid.hpp"
#include "QQbar_threshold/xsection.hpp"
#include "QQbar_threshold/structure_function.hpp"
#include "QQbar_threshold/integrate.hpp"
#include "QQbar_threshold/constants.hpp"

int main(){
  namespace QQt = QQbar_threshold;
  QQt::load_grid(QQt::grid_directory() + "ttbar_grid.tsv");
  constexpr double sqrt_s = 344.;
  constexpr double mu = 80.;
  constexpr double mu_width = 350.;
  constexpr double mt_PS = 171.5;
  constexpr double width = 1.33;
  QQt::options opt = QQt::top_options();
  opt.ISR_const = true;

  const double beta = QQt::ISR_log(sqrt_s, QQt::alpha_mZ);
  const auto integrand = [=](double t){
    const double x = 1 - std::pow(t, 1/beta);
    const double L = QQt::modified_luminosity_function(t, beta);
    const double sigma = QQt::ttbar_xsection(
        std::sqrt(x)*sqrt_s, {mu, mu_width}, {mt_PS, width}, QQt::N3LO,
        opt
    );
    return L*sigma;
  };

  constexpr double x_min = 330.*330./(sqrt_s*sqrt_s);
  const double t_max = std::pow(1 - x_min, beta);

  std::cout << QQt::integrate(integrand, 0, t_max) << '\n';
}
\end{lstlisting}
The corresponding Mathematica code is
\begin{lstlisting}[language=Mathematica]
Needs["QQbarThreshold`"];

LoadGrid[GridDirectory <> "ttbar_grid.tsv"];
sqrts = 344.;
mu = 80.;
muWidth = 350.;
mtPS = 171.5;
width = 1.33;
order = "N3LO";
beta = ISRLog[sqrts, alphamZ];

xmin = (330./sqrts)^2;
tmax = (1-xmin)^beta;

Print[
   NIntegrate[
      ModifiedLuminosityFunction[t, beta]*
      TTbarXSection[
         Sqrt[1-t^(1/beta)] * sqrts, {mu, muWidth}, {mtPS, width}, order,
         ISRConst -> True
      ],
      {t, 0, tmax}
   ]
];
\end{lstlisting}
Numerically, the structure function under the replacement $\beta \to
2\beta$ is very close to the luminosity function. The same holds for the
modified versions of both functions. Indeed, substituting
\mbox{\lstinline[language=C++]!modified_luminosity_function(t, beta)!}
with \mbox{\lstinline[language=C++]!modified_structure_function(t,
2*beta)!}  in the example changes the result for the cross section to
$\sigma = 0.59169\,$pb, i.e. by less than $10^{-4}$. This
observation can be used to somewhat accelerate the computation of the
convolution at the cost of accuracy.

\subsection{Width corrections}
\label{sec:impl_width}

Among the resonant NNLO electroweak corrections listed in~\eqref{eq:resrest},
only $\sigma_{\text{IS}}^\text{conv}$ and
$\sigma_\Gamma$ are not already available in version 1 of
\texttt{QQbar\_threshold}. In version 2, the correction $\sigma_\Gamma$
proportional to the top-quark width is included by default in the
prediction for the cross section. Its components (cf.~\eqref{eq:delta2GammaG},~\eqref{eq:resEWwidth}) can be controlled
individually with the new \lstinline[language=C++]!contributions!
options \lstinline[language=C++]!v_width_kinetic! (Eq.~\eqref{eq:delta_XD2_G}), \lstinline[language=C++]!v_width2! (Eq.~\eqref{eq:delta_X_G}), and
\lstinline[language=C++]!width_ep! (Eq.~\eqref{eq:Oepswidthterms}). The
respective Mathematica contribution names are
\lstinline[language=C++]!vwidthkinetic!,
\lstinline[language=C++]!vwidth2!, and
\lstinline[language=C++]!widthep!.

To incorporate the width corrections to the quarkonium energy levels
from~\eqref{eq:delGammaE2} the function
\mbox{\lstinline[language=C++]!ttbar_energy_level(n, mu, \{m, width\}, order, opts)!} can now take both the mass and width as arguments. Similarly, the
\lstinline[language=C++]!ttbar_residue! function can now take both arguments.
In this way, this function
includes the width corrections to the wave functions
from~\eqref{eq:delGammaF2}. The corresponding Mathematica functions
\lstinline[language=C++]!TTbarEnergyLevel! and
\lstinline[language=C++]!TTbarResidue!
are similarly extended. Finally, the toponium width including the
corrections in~\eqref{eq:delGamma2} can be computed with the new
function
\mbox{\lstinline[language=C++]!ttbar_width(n, mu, \{m, width\}, order,
  opts)!} (\lstinline[language=C++]!TTbarWidth! in
Mathematica).\footnote{This new function should not be confused with the
  older \lstinline[language=C++]!top_width! function, which calculates
  the width of the top quark itself as opposed to the width of a toponium
bound state.}

\subsection{Note on backwards compatibility}
\label{sec:backwards_compat}

Disabling the new corrections in version 2 via the
\lstinline[language=C++]!contributions! option will produce results that
are similar, but not identical to version 1 of the code. There are two
causes for the difference.

First, as detailed in Sections~\ref{sec:resCZt} and~\ref{sec:resAbs},
the calculational scheme for the NNLO electroweak corrections to the
resonant cross section has been changed. Consequently, predictions for
the cross section at or beyond NNLO that include electroweak corrections
will differ between the two versions, even if all new corrections are
disabled. The numerical differences are typically less than 1\%, but can
amount to almost 10\% for a small
renormalization scale and far below the threshold, where the cross section
is already very small and the (new and old) NNLO electroweak corrections
are sizeable.

Second, in contrast to the original code, version 2 now captures the
full dependence on the scale $\mu_w$. The numerical effect of this
change is of the order of a few per mille for low energies and
significantly less than one per mille in the peak region.

\subsection{Calculation of the non-resonant correction}
\label{sec:impl_gridgen}

The dynamic numeric evaluation of the NNLO non-resonant corrections
is computationally prohibitively expensive. Hence,
\texttt{QQbar\_threshold} internally uses interpolation of a precomputed
grid. For reference purposes, a copy
\texttt{NNLO\_nonresonant\_grid.tsv} of this internal grid is provided
in the directory given by the function
\mbox{\lstinline[language=C++]!grid_directory!} in C++ and the variable
\lstinline[language=Mathematica]!GridDirectory! in Mathematica. The
coordinates of the grids are given by $x_W = m_W^2/m_t^2$, accounting for
variations of the top-quark mass, and $y_w = (1-y)/(1-x_W)$, which covers
changes in the invariant mass cut discussed in
Section~\ref{sec:nonres}. The remaining two grid entries
$\Sigma_\text{automated}(x_W,y_w)$ and $\Sigma_\text{manual}(x_W,y_w)$
parametrize the
automated and the manual part of the non-resonant cross section for
$\mu_w = m_t$. To obtain their contribution to the cross section, these
entries have to be multiplied by a factor of
$\alpha_s(\mu_r) \sigma_0\Gamma_t$. The complete NNLO correction to the
non-resonant cross section for arbitrary $\mu_w$ is then given by
\begin{equation}
  \label{eq:sigma_non-res_grid} \sigma_{\text{non-res}}^{\text{NNLO}} =
\alpha_s(\mu_r)\sigma_0\Gamma_t\bigg( \Sigma_\text{automated} +
\Sigma_\text{manual} + \Sigma_{\log}\log \frac{m_t^2}{\mu_w^2}\bigg) -
\frac{\delta \Gamma_1}{\Gamma_0}\sigma_{\text{non-res}}^{\text{NLO}}\,,
\end{equation}
where the coefficient of the logarithm reads
\begin{equation}
  \label{eq:muw_dep}
  \Sigma_{\log} = \frac{3 N_c C_F m_t}{s}\bigg[{C_0^{(v)}}^2 + {C_0^{(a)}}^2
  + {C^{(v)^2}_{0,\text{P-wave}}} + {C^{(a)^2}_{\text{0,P-wave}}} +
  \frac{3\alpha(\mu_\alpha)m_t}{4\pi\Gamma_t}
  (C_0^{(v)}C^{(v)}_\text{Abs} + C_0^{(a)}C^{(a)}_\text{Abs})\bigg]\,.
\end{equation}
As mentioned before, the dependence on $\mu_w$ has to cancel exactly
against the dependence in the resonant cross section. Like in the
resonant part, we therefore do not expand out the energy dependence
of the $s$-channel propagators in $\Sigma_{\log}$.
The last term in~\eqref{eq:sigma_non-res_grid} is required because
we have expressed the non-resonant cross section in terms of the
all-order width $\Gamma_t$. The NLO non-resonant part is proportional
to $\Gamma_t$ and therefore implicitly contains the NNLO correction
$(\delta \Gamma_1/\Gamma_0)\sigma_{\text{non-res}}^{\text{NLO}}$,
where $\delta \Gamma_1$ is the NLO QCD correction to the top-quark width.
The same contribution appears in the NNLO calculation of the non-resonant
part and we must include the last term in~\eqref{eq:sigma_non-res_grid}
to subtract this double counting.

Note that highly unphysical top-quark masses, i.e. $x_W < 0.15$ or $x_W >
0.3$ and extremely tight invariant mass cuts $y_w < 0.01$ are not
supported. Furthermore, the default values are assumed for the remaining
Standard Model parameters, such as the values of $m_W$ and $m_Z$.

Our code for producing the NNLO non-resonant grid depends on a number of
software packages, including
\texttt{MadGraph5\_aMC@NLO}~\cite{Alwall:2014hca},
\texttt{FastJet}~\cite{Cacciari:2011ma}, and
\texttt{Cuba}~\cite{Hahn:2004fe}. In order to facilitate reproducing our
results without having to install all dependencies, we provide an
image that can be executed using the \texttt{Docker} virtualisation
software. After installing \texttt{Docker} and downloading the image
\texttt{nnlo\_nonres\_grid\_entry.tar.gz} from
\url{https://www.hepforge.org/downloads/qqbarthreshold/}, it can be
imported with
\begin{lstlisting}[language=bash,caption={}]
docker load -i nnlo_nonres_grid_entry.tar.gz
\end{lstlisting}
and run with
\begin{lstlisting}[language=bash,caption={}]
docker run -it amaier/nnlo_nonres_grid_entry <xw> <yw>
\end{lstlisting}
with \lstinline!xw! and \lstinline!yw! replaced by the respective values
for the parameters introduced above. The last line of the output
corresponds to a grid entry in the same format as in the reference
grid.

For convenience we provide a Mathematica interface to the calculation of
the NNLO non-resonant grid entries. After importing the \texttt{Docker}
image as described above and loading the \texttt{QQbarGridCalc}
Mathematica package distributed with \texttt{QQbar\_threshold}, a grid
entry can be computed with
\begin{lstlisting}[language=Mathematica,caption={}]
QQbarCalcNNLONonresonantGridEntry[xw, yw, Verbose -> True]
\end{lstlisting}
which returns a list $\{\Sigma_{\text{automated}},
\Sigma_{\text{manual}}\}$. Setting the
\lstinline[language=Mathematica]!Verbose! option to
\lstinline[language=Mathematica]!False! will suppress intermediate
output.

Especially for large values $y_w \sim 1$ the calculation of the
automated contribution can fail, in which case a slight change in the
input parameters may help. In practice, this is not a severe problem as
the automated contribution becomes essentially constant in this
region. The precision of the automated calculation is not very high, and
the values obtained can easily deviate from the ones in the reference
grid by around $10\%$. Since the NNLO non-resonant contribution itself
is not very large and typically dominated by the manual and logarithmic
contributions, this translates to an error of at most one per mille in
the final cross section. One way to reduce this error further would be
to calculate the grid entries several times and average over the results.

In principle it is possible to compute an entirely new grid with the
\texttt{Docker} container. In practice it is computationally much more
efficient to calculate $\Sigma_\text{automated}$ in the absence of an
invariant mass cut, i.e. for $y_w = 1$, and derive the entries for all other
values of $y_w$ exploiting complementary cuts as discussed in
Section~\ref{sec:nonres}. The entry for some $y_w$ with $0< y_w < 1$ is
then given by $\Sigma_{\text{automated}} =
\Sigma_{\text{automated}}\big|_{y_w = 1} -
\overline{\Sigma}_{\text{automated}}$, where the phase space integral in
$\overline{\Sigma}_{\text{automated}}$ is restricted to the
complementary region $0 \leq t \leq 1 - (1-x_W)\*y_w$. In this way, the
numerically problematic endpoint region $t\to 1$ is only computed once
for each value of $x_W$.

%%%%%%%%%%%%%%%%%%%%%%%%%%%%%%%%%%%%%%%%%%%%%%%%%%%%%%%%%%%%%%%%%%%%%%%%%%%

\section{Implementation of the subtractions in \texttt{MadGraph}}
\label{app:MADGRAPH}

We briefly describe the implementation of the subtractions of
the diagram $h_1$ in Figure~\ref{fig:NonresNLO} and all the
diagrams in Figures~\ref{fig:squared} and~\ref{fig:interference}
in \texttt{MadGraph5\_aMC@NLO 2.5.0.beta2}. The \texttt{MG5\_aMC}
code for the computation of the process $e^+e^-\to t\bar{b}W^-$
at NLO in QCD is created in the directory {\tt TBWsubtractions}
by entering the commands
\begin{lstlisting}[caption={}]
MG5_aMC>generate e+ e- > t b~ w- [QCD]
MG5_aMC>output TBWsubtractions
\end{lstlisting}
in the \texttt{MadGraph5} prompt.
First, we subtract the diagram $h_1$ from the code in the directory
\texttt{$\sim$/SubProcesses/{\allowbreak}P0\_epem\_wmtbx/}.\footnote{ Here and in the
following $\sim$/ refers to the code directory.} To this end, we
identify the corresponding diagram numbers in {\tt MadGraph} as 3 and 4
using {\tt born.ps}. The subtraction is achieved by removing the terms
proportional to \lstinline[language=fortran]!AMP(I)*DCONJG(AMP(J))! with
\lstinline[language=fortran]!I,J!$=3,4$ in the squared matrix
element. This affects the function \lstinline[language=fortran]!BORN! in
{\tt born.f} and \lstinline[language=fortran]!BORN_HEL! in {\tt
born\_hel.f}.  We note that one should avoid first adding and then
subtracting the terms to avoid numerical instabilities since these
contributions are divergent at threshold.

To subtract the real corrections $g_i$ in Figure~\ref{fig:squared} we
remove the respective terms in the squared real amplitude given by the
function \lstinline[language=fortran]!MATRIX_1! in {\tt matrix\_1.f}
where the corresponding set of \lstinline[language=fortran]!I,J! values
can be determined from {\tt matrix\_1.ps}. To maintain separate IR
finiteness of the real and virtual contributions we also have to edit
the FKS subtraction terms in the files \texttt{b\_sf\_001.f},
\texttt{b\_sf\_002.f}, and \texttt{b\_sf\_003.f} accordingly. This is
done by removing the terms containing the product of the tree-level
amplitudes 3 and 4 in the functions
\lstinline[language=fortran]!B_SF_00i!.

In the folder
{\tt $\sim$/SubProcesses/{\allowbreak}P0\_epem\_wmtbx/{\allowbreak}V0\_epem\_wmtbx}
for the virtual corrections we first apply the usual subtractions for
the squared tree-level amplitude to the function
\lstinline[language=fortran]!MATRIX! in {\tt born\_matrix.f}. The
interference of a given one-loop diagram with the tree-level diagrams is
evaluated by the function
\lstinline[language=fortran]!CREATE_LOOP_COEFS! in {\tt polynomial.f}.
We create a copy called
\lstinline[language=fortran]!CREATE_LOOP_COEFS_h1bcd! and which is
modified by removing the interference with the tree-level diagrams 3 and
4. This allows us to remove the diagrams $h_{1b},h_{1c},h_{1d}$ and
$h_{ia}$ with $i=1,\dots,4$ by modifying the calls to
\lstinline[language=fortran]!CREATE_LOOP_COEFS! in {\tt
coef\_construction\_1.f} for the loop diagrams corresponding to the
left-hand sides of the cuts. We either add the suffix
\lstinline[language=fortran]!h1bcd! or comment out the calls.  Again the
relevant diagram numbers in {\tt MadGraph} can be identified from the
graphical representation {\tt loop\_matrix.ps}. The same changes are
applied to the multiple precision version of the virtual corrections
given in {\tt mp\_compute\_loop\_coefs.f} and {\tt
mp\_coef\_construction\_1.f}.

Last but not least one needs to modify the counterterms
given in {\tt loop\_matrix.f}. The identification of the {\tt Madgraph}
IDs of the counterterm diagrams is more complicated since they are not
drawn but must be inferred from the code. The counterterm amplitudes
\lstinline[language=fortran]!AMPL(K,I)!, where the first index
\lstinline[language=fortran]!K!$=1,2,3$ denotes the finite part, the
$1/\epsilon$ pole and the $1/\epsilon^2$ pole, are defined in {\tt
helas\_calls\_ampb\_1.f} and {\tt helas\_calls\_uvct\_1.f}.  The first
file contains $R_2$-terms and mass renormalization counterterms which
are attributed to the loop diagrams in the same order in which they
appear in {\tt loop\_matrix.ps}. The second file contains the
multiplicative wave function renormalization counterterms for the
tree-level diagram.  To subtract the diagrams $h_{1e},h_{1f},h_{1g}$ and
the $R_2$ contributions of the remaining diagrams in the squared
contribution we remove the terms proportional to
\lstinline[language=fortran]!AMPL(K,I)*DCONJG(AMP(J))! with
\lstinline[language=fortran]!I!$=11,12,16\textup{--}23,28\textup{--}31$
and {\tt J}$=3,4$ in the function
\lstinline[language=fortran]!SLOOPMATRIX! in {\tt loop\_matrix.f}.

In addition we can implement the minimal subtraction of the UV
divergences in $h_{ia}$ with $i=2,\dots,4$ by modifying the interference
of the divergent part of the wave function renormalization of the
tree-level diagrams 3 and 4 with the other tree-level
diagrams. Explicitly, we multiply
\lstinline[language=fortran]!AMPL(2,I)!  with
\lstinline[language=fortran]!I!$=29,31$ in the subroutine
\lstinline[language=fortran]!HELAS_CALLS_UVCT_1! in
\texttt{helas\_calls\_uvct\_1.f} with a factor $2/3$.  As discussed in
Section~\ref{sec:automated} the $R_2$-terms and finite parts of the wave
function renormalization contributions for the interference contribution
are not modified. Alternatively, one can simply deactivate the check for
UV finiteness which yields the same results.

Following the discussion in Section~\ref{sec:automated} we have to
deactivate an internal {\tt MadGraph} consistency check for the
positivity of the squared real amplitude to make the modified code
run without producing error messages. This is done by removing the
code block
\begin{lstlisting}[language=fortran,caption={}]
      if(wgt.lt.0.d0)then
         ...
      endif
\end{lstlisting}
in {\tt $\sim$/SubProcesses/fks\_singular.f}.
Older {\tt MadGraph} versions also require modifications  in the file
{\tt $\sim$/SubProcesses/{\allowbreak}P0\_epem\_wmtbx/{\allowbreak}BinothLHA.f} to allow for negative values of the squared Born amplitude.

Our modified version of the {\tt MadGraph} code is shipped with the grid
generation routines in the new version of {\tt QQbar\_Threshold}. We
have checked our procedure by applying similar modifications to the
process $e^+e^-\to \bar{t}bW^+$ generated with the older MadGraph
version \texttt{2.4.3}. Furthermore, we have verified that an analogous
set of changes correctly removes the $Z$-boson exchange contribution to
the process $e^+e^-\to t\bar{t}$ at NLO by comparing the results to the
ones obtained by excluding the $Z$-boson exchange already in the process
generation.

%%% Local Variables:
%%% mode: latex
%%% TeX-master: "draft"
%%% End:

%%%%%%%%%%%%%%%%%%%%%%%%%%%%%%%%%%%%%%%%%%%%%%%%%%%%%%%%%%%%%%%%%%%%%%%%%%

\section{Further details on the comparison to \cite{Penin:2011gg}}
\label{app:furthercomparison}

We extend in this appendix the discussion about the discrepancy
with the result for the cross section at leading order in the $\rho^{1/2}$
expansion of \cite{Penin:2011gg} and its connection to diagram $h_{1b}$.
First we explain why the cancellation of finite-width and endpoint
divergences requires a non-vanishing contribution from diagram $h_{1b}$
at leading order. A similar argument
was already put forward in \cite{Jantzen:2013gpa,Ruiz-Femenia:2014ava}.
Then we show that the leading-order term in $h_{1b}$
comes from  a loop-momentum region which is not among those considered
to construct the unstable top EFT formulated in \cite{Penin:2011gg}.

%%%%%%%%%%%%%%%%%%%%%%%%%%%%%%%%%%%%%%%%%%%%%%%%%%%%%%%%%%%%%%%%%%%%%%%%%%%
\begin{figure}
  \centering
  \includegraphics[width=0.85\textwidth]{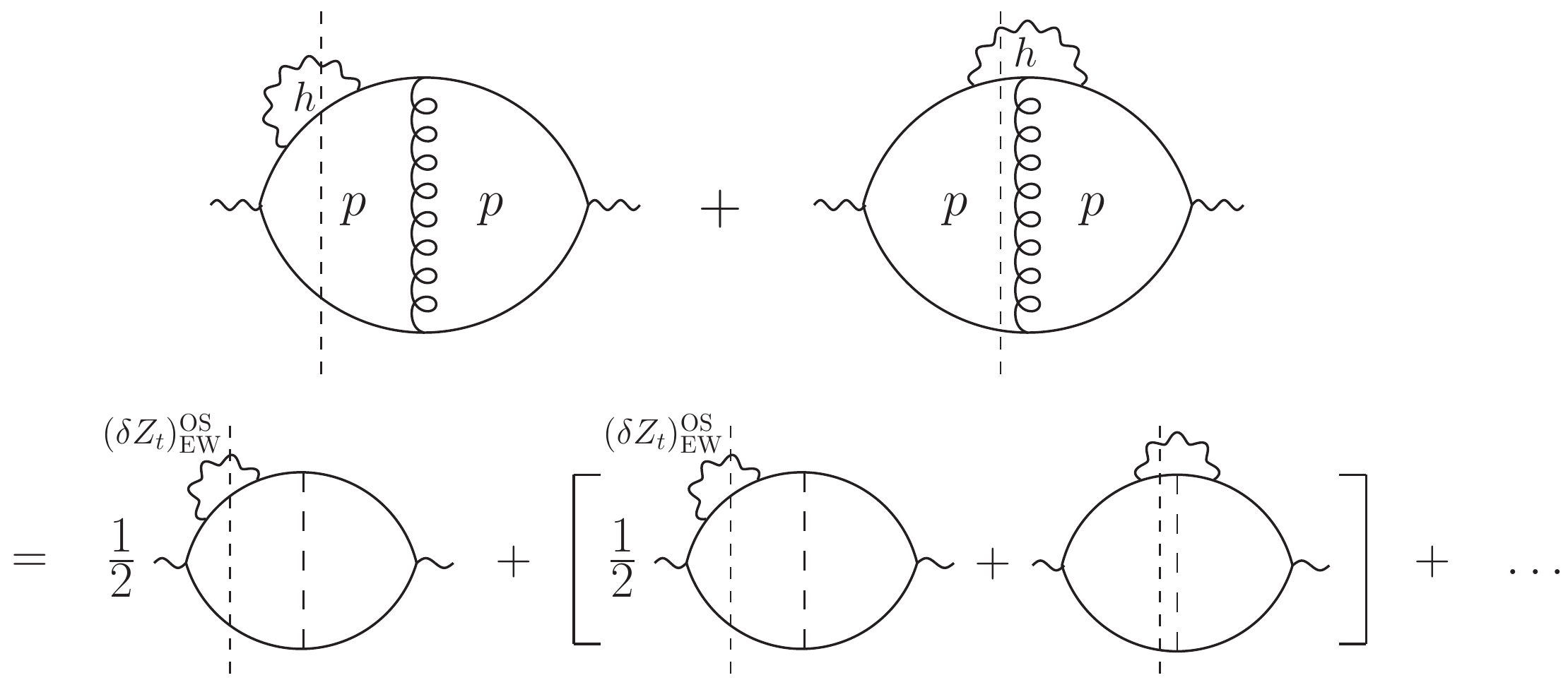}
  \caption{
Three-loop cut diagrams generating the NNLO resonant corrections related
to the top-quark instability. The top anti-top loops are potential, whereas
the $bW$ top-quark self-energy is hard. The symmetric cut diagrams are not
displayed. In the second line we display the terms that yield
$\sigma_{C^{(k)}_{\text{Abs,}Z_t}}$, and the dots stand for other resonant
contributions at NNLO and higher orders.\label{fig:res_h1ah1b}}
\end{figure}
%%%%%%%%%%%%%%%%%%%%%%%%%%%%%%%%%%%%%%%%%%%%%%%%%%%%%%%%%%%%%%%%%%%%%%%%%%%

The NNLO resonant corrections related to $h_{1a}$ and $h_{1b}$
are given by the diagrams displayed in Fig.~\ref{fig:res_h1ah1b}, where the
loop-momenta in the top anti-top loops are expanded according to the potential
($p$) scaling, while the loop momentum in the $bW$ loop is hard ($h$). It
should be understood that in the diagram with the self-energy insertion in
the propagator one has to consider only the NNLO piece (the cut self-energy
in the on-shell limit gives the top-quark width, which is a LO contribution
since $p^2-m_t^2\sim \Gamma$ in the potential region; such terms are already
accounted for by the replacement $E\to E+i\Gamma$ in the non-relativistic
propagator). The diagram with the cut self-energy contributes to
$\sigma_\Gamma$ \eqref{eq:resEWwidth} and $\sigma_{C^{(k)}_{\text{Abs,}Z_t}}$
\eqref{eq:sigmaCabsZt}, once the symmetric diagrams are considered.  In
particular, the latter arises from field renormalization due the absorptive
part of the electroweak one-loop self-energy. We have isolated that
contribution in the second line of Figure~\ref{fig:res_h1ah1b}, and split it
such that one half of it can be attributed to field renormalization
of the top quark leaving the production vertex, and the other half to
the renormalization of the top-quark field entering the $t\bar{t}g$ vertex.
The latter contribution is exactly cancelled by the electroweak correction to
the Coulomb potential (third diagram in the second line of
Figure~\ref{fig:res_h1ah1b}), because upon expanding out the external
(potential) momenta from the self-energy and vertex loops, these diagrams
are equivalent to the renormalized vertex in the on-shell scheme for zero
transfered momentum (see Figure~\ref{fig:cancellation}). Therefore, the
resonant counterpart of $h_{1b}$ is equal to minus one half of the diagram
with the field renormalization of the top quark leaving the production vertex,
which is proportional to
the coefficient $C^{(k)}_{\text{Abs,}Z_t}$ written in (\ref{eq:sigmaCabsZt}). It can be easily checked
that $C^{(k)}_{\text{Abs,}Z_t}$ behaves as $\Gamma_0/(1-x_W)\simeq \Gamma_0/\rho$, and that the
contribution to the cross section $\sigma_{C^{(k)}_{\text{Abs,}Z_t}}$ contains a $\alpha_s/\epsilon$
divergence from the real part of the Green function. Therefore both resonant
diagrams in the first line of Figure~\ref{fig:res_h1ah1b} contain $\alpha_s/\epsilon\times \Gamma_0/\rho$
divergences, that are cancelled with endpoint divergences from the corresponding non-resonant
diagrams $h_{1a}$ and $h_{1b}$ as was shown by explicit computation
in \cite{Ruiz-Femenia:2014ava}. In ~\cite{Penin:2011gg} only the non-resonant diagram analogue to $h_{1a}$ is considered,
while it is argued that the analogue to $h_{1b}$ must vanish quoting results from
\cite{Melnikov:1993np}. As already explained in Section~\ref{sec:comparison}, the results
from the latter
refer to the vanishing of resonant contributions related to the top-quark
instability at NLO, while the contribution under discussion here is of NNLO.

%%%%%%%%%%%%%%%%%%%%%%%%%%%%%%%%%%%%%%%%%%%%%%%%%%%%%%%%%%%%%%%%%%%%%%%
\begin{figure}
  \centering
  \includegraphics[width=0.45\textwidth]{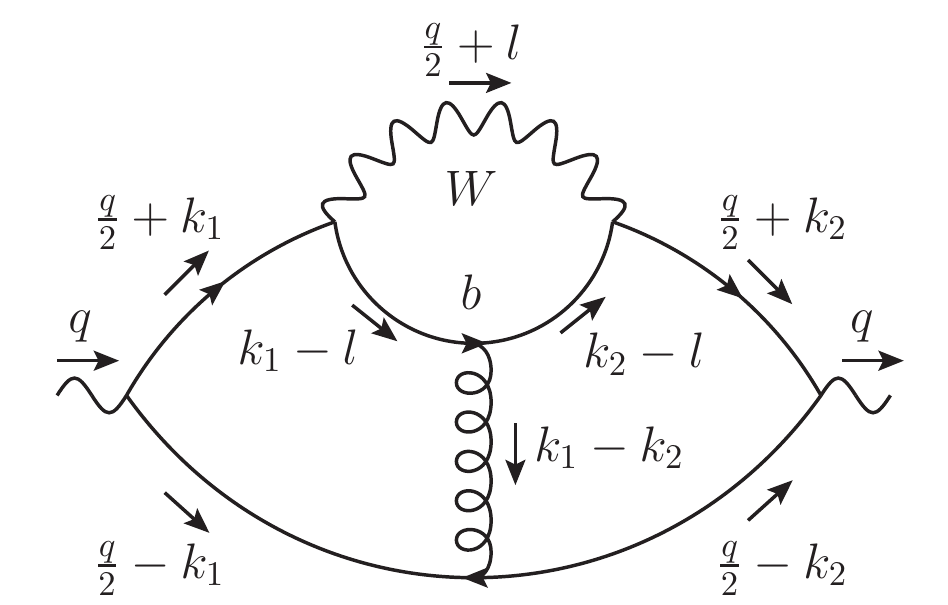}
  \caption{Forward-scattering diagram whose imaginary part is related to
  cut diagrams $h_{1b}$.\label{fig:h1b_extra_region}}
\end{figure}
%%%%%%%%%%%%%%%%%%%%%%%%%%%%%%%%%%%%%%%%%%%%%%%%%%%%%%%%%%%%%%%%%%%%%%%

The dominant terms in the $\rho$ expansion can also be obtained upon
application of the method of regions~\cite{Beneke:1997zp,Jantzen:2011nz}.
Let us consider the forward scattering diagram in
Figure~\ref{fig:h1b_extra_region}, whose imaginary part corresponds to the
sum of cut diagrams $h_{1b}$ and $g_5$ (plus left-right symmetric ones),
since no other cuts are kinematically possible.
The authors of~\cite{Penin:2011gg} discuss the contributions from the
regions that are obtained by replacing $v\to\rho^{1/2}$ in the hard,
soft, potential and ultrasoft region. However, for the case of the diagram
in Figure~\ref{fig:h1b_extra_region}, which is related to
$h_{1b}$ discussed above, it can be shown  that the leading order
contribution comes from an additional region with parametrically smaller
virtuality or order $\rho^2m_t^2$, that was not considered
in~\cite{Penin:2011gg}. With the momentum assignment of
Figure~\ref{fig:h1b_extra_region}, it
corresponds to $k_i^0\sim m_t\rho^2$, $\mathbf{k}_i\sim m_t\rho$ and
$l\sim m_t\rho$. We obtain for the relevant scalar integrals in this region:
\begin{eqnarray}
 I [\eta] & = &  \int\frac{d^dk_1}{i\pi^{d/2}}\int\frac{d^dk_2}{i\pi^{d/2}}
\int\frac{d^dl}{i\pi^{d/2}}\frac{1}{[2m_tk_1^0-\mathbf{k}_1^2]
[-2m_tk_1^0-\mathbf{k}_1^2][2m_tk_2^0-\mathbf{k}_2^2]
[-2m_tk_2^0-\mathbf{k}_2^2] }
\nonumber \\
& & \times\,\frac{1}{[2m_tl^0+2\rho m_t^2][(l^0)^2-(\mathbf{l}-\mathbf{k}_1)^2]
[(l^0)^2-(\mathbf{l}-\mathbf{k}_2)^2][-(\mathbf{k}_1-\mathbf{k}_2)^2]^{\eta}}
\nonumber\\[0.2cm]
& = & (-1)^\eta\,\frac{\pi(2\rho)^{1-2\eta-6\epsilon}}
{2m_t^{2+2\eta+6\epsilon}}
\,e^{6i\pi\epsilon}\, \frac{\Gamma\left(\frac12+\epsilon\right)
\Gamma\left(\frac12-\epsilon\right)\Gamma\left(\eta+2\epsilon\right)^2
\Gamma\left(1-\eta-2\epsilon\right)}{\Gamma\left(2-2\epsilon\right)
\Gamma\left(2\eta+4\epsilon\right)}
  \nonumber\\
 & & \times \,\Gamma\left(1-\eta-3\epsilon\right)
 \Gamma\left(-1+2\eta+6\epsilon\right) \,.
\end{eqnarray}
The relevant cases are $\eta=0,1$,
\begin{eqnarray}
I[0]
& = & [\text{real}]  - \frac{i\pi^3\rho}{\, m_t^2}\,
(4 e^{\gamma_E}\rho^2 m_t^2 )^{-3\epsilon}
\left[\frac{1}{\epsilon}+8+\dots\right],
\\
I[1]
 & = & [\text{real}] - \frac{i \pi^3}{4 \rho m_t^4}\,
(4 e^{\gamma_E}\rho^2 m_t^2 )^{-3\epsilon}
\left[\frac{1}{\epsilon}-2+\dots\right].
\end{eqnarray}
Both scalar integrals produce a contribution of order
$\Gamma_0\times \alpha_s/\rho$ to the cross section\footnote{
A factor $\alpha/s_w^2 \times \rho^2\propto \Gamma_0/m_t$, where the
$\rho^2$ term arises from the bottom propagators, appears in the numerator
that comes along $I[1]$. The numerator of the $I[0]$ term has no
$\rho^2$ suppression. Hence when $\Gamma_0$ is extracted as in
(\ref{eq:h1ares})--(\ref{eq:Nnorm}) below, both terms contribute
at order $1/\rho$.} and the imaginary part contains a $1/\epsilon$
divergence, which are the properties that are needed to cancel the
finite-width divergence at the leading order in the $\rho$ expansion discussed
above. We also note that a non-vanishing contribution from the region
$k_i^0\sim m_t\rho^2$, $\mathbf{k}_i\sim m_t\rho$ and $l\sim m_t\rho$, is
consistent with the findings of \cite{Ruiz-Femenia:2014ava},
which identified that the leading-order term in $\rho$ of diagram $h_{1b}$
originates from the region in the $t$-integration
$(1-t)\sim(1-x_W)^2\simeq \rho^2$, where $(1-t)\sim \mathbf{k}_2^2 /m_t^2$
adopting the momentum assignment of Figure~\ref{fig:h1b_extra_region}.

For completeness, we finally provide results for the individual contributions
to the cross section from diagrams $h_{1a}$ and $h_{1b}$ and their resonant
counterparts at the leading order in $\rho$:
\begin{eqnarray}
\sigma^{{\cal O}(\alpha_s/\rho)}_{h_{1a},\text{res}} & = &
%\sigma_0 \frac{24\pi N_c}{s}\,(C_0^{(v)^2} + C_0^{(a)^2})\,\frac{\alpha_sC_F}{4\pi}\,\frac{m_t\Gamma_0^{(d=4)}}{\rho} \nonumber\\
N \left[-\frac{1}{\epsilon} - \frac73 + 2 \ln2
- \ln\left(\frac{\mu_w^2}{m_t^2\rho^2}\right) -
2 \ln\left(\frac{\mu_w^2}{4m_t|E+i\Gamma|}\right)\right],
\label{eq:h1ares} \\[2mm]
\sigma^{{\cal O}(\alpha_s/\rho)}_{h_{1a},\text{non-res}} & = &
  %\sigma_0 \frac{24\pi N_c}{s}\,(C_0^{(v)^2} + C_0^{(a)^2})\,\frac{\alpha_sC_F}{4\pi}\,\frac{m_t\Gamma_0^{(d=4)}}{\rho} \nonumber\\
  N \left[\frac{1}{\epsilon} + \frac73 - 2 \ln2 + 2\ln\left(\frac{\mu_w^2}{m_t^2\rho^2}\right) + \ln\left(\frac{\mu_w^2}{m_t^2}\right)\right],
\label{eq:h1anonres} \\[2mm]
  \sigma^{{\cal O}(\alpha_s/\rho)}_{h_{1b},\text{res}} & = & -\frac{1}{2}  \sigma^{{\cal O}(\alpha_s/\rho)}_{h_{1a},\text{res}} \,,
  \label{eq:h1bres}\\[2mm]
  \sigma^{{\cal O}(\alpha_s/\rho)}_{h_{1b},\text{non-res}} & = &
  %\sigma_0 \frac{24\pi N_c}{s}\,(C_0^{(v)^2} + C_0^{(a)^2})\,\frac{\alpha_sC_F}{4\pi}\,\frac{m_t\Gamma_0^{(d=4)}}{\rho} \nonumber\\
  N \left[-\frac{1}{2\epsilon} -\frac76 + 3 \ln2 - \frac32 \ln\left(\frac{\mu_w^2}{m_t^2\rho^2}\right)\right] \,,\label{eq:h1bnonres}
\end{eqnarray}
with
\begin{equation}
N= \sigma_0 \frac{24\pi N_c}{s}\,\Big[C_0^{(v)^2} + C_0^{(a)^2}\Big]\,
\frac{m_t\Gamma_0}{\rho} \,\frac{\alpha_sC_F}{4\pi}\,.
\label{eq:Nnorm}
\end{equation}
The sum of the non-resonant contributions
$\sigma^{{\cal O}(\alpha_s/\rho)}_{h_{1a},\text{non-res}}$ and
$ \sigma^{{\cal O}(\alpha_s/\rho)}_{h_{1b},\text{non-res}}$ agrees with the
result given in \cite{Ruiz-Femenia:2014ava} if one takes into account that the
leptonic tensor was treated in $d$ dimensions there, which
introduces a factor $(1-\epsilon)$ compared to the $d=4$ result used in the
present work.\footnote{We take the opportunity to correct a typo in
\cite{Ruiz-Femenia:2014ava}:  a factor $(3-2\epsilon)/3$ was missed
in the hadronic tensor $H_{1a}$ written in Eq.~(3) therein.}

%%%%%%%%%%%%%%%%%%%%%%%%%%%%%%%%%%%%%%%%%%%%%%%%%%%%%%%%%%%%%%%%%%%%%%%%
%%%%%%%%%%%%%%%%%%%%%%%%%    Bibliography      %%%%%%%%%%%%%%%%%%%%%%%%%
%%%%%%%%%%%%%%%%%%%%%%%%%%%%%%%%%%%%%%%%%%%%%%%%%%%%%%%%%%%%%%%%%%%%%%%%

\end{document}